 \documentclass[final,1p,times]{elsarticle}

\usepackage{amssymb}

\usepackage{color,soul}
\usepackage[hang,bottom]{footmisc}
\usepackage{amsmath} 
\usepackage{framed} 
\usepackage{multicol} 
\usepackage{nomencl} 
\makenomenclature
\renewcommand*\nompreamble{\begin{multicols}{2}}
\renewcommand*\nompostamble{\end{multicols}}
\usepackage{siunitx} 
\usepackage{color}
\usepackage{tabularray}
\usepackage{array}
\usepackage{booktabs}
\usepackage{longtable}
\usepackage{subcaption}
\usepackage{url}
\usepackage{rotating}
\usepackage{pdflscape}

\usepackage{graphicx}
\usepackage{array}
\usepackage[utf8]{inputenc}
\usepackage{chngcntr}

\journal{arxiv}

\begin{document}

\begin{frontmatter}



\title{Exploring Quantum-Dot Engineered Solid-State Photon Upconversion in PbS:Yb\(^{3+}\),Er\(^{3+}\)/CuBiO Using Density Functional Theory and Machine Learning Methods for Water Splitting}

\author[nla,ftkkp]{Dennis Delali Kwesi Wayo} 
\ead{dennis.wayo@nu.edu.kz}

\author[nla]{Vladislav Kudryashov} 
\ead{vladislav.kudryashov@nu.edu.kz}

\author[nla]{Mirat Karibayev}
\ead{mirat.karibayev@nu.edu.kz}

\author[seds]{Gertrude Ellen Fynn} 
\ead{gertrude.fynn@nu.edu.kz}

\author[nla]{Khadichakhan Rafikova} 
\ead{khadichakhan.rafikova@nu.edu.kz}

\author[dcm]{Camila Martins Saporetti} 
\ead{camila.saporetti@iprj.uerj.br}

\author[dcam]{Leonardo Goliatt} 
\ead{leonardo.goliatt@ufjf.br}

\author[nla,seds]{Nurxat Nuraje\corref{cor1}}
\ead{nurxat.nuraje@nu.edu.kz}
\cortext[cor1]{Corresponding author}

\affiliation[nla]{organization={Renewable Energy Laboratory, National Laboratory Astana, Nazarbayev University},
            city={Astana},
            postcode={010000}, 
            country={Kazakhstan}}

\affiliation[ftkkp]{organization={Faculty of Chemical and Process Engineering Technology, Universiti Malaysia Pahang Al-Sultan Abdullah},
            city={Kuantan},
            postcode={26300}, 
            country={Malaysia}}

\affiliation[seds]{organization={Department of Chemical Engineering, School of Engineering and Digital Sciences, Nazarbayev University},
            city={Astana},
            postcode={010000}, 
            country={Kazakhstan}}

\affiliation[dcm]{organization={Department of Computational Modeling, Polytechnic Institute, Rio de Janeiro State University},
            city={Nova Friburgo},
            postcode={22000-900}, 
            country={Brazil}}

\affiliation[dcam]{organization={Department of Computational and Applied Mechanics, Federal University of Juiz de Fora},
            city={Juiz de Fora},
            postcode={36036-900}, 
            country={Brazil}}


\begin{abstract}

This study presents a comprehensive numerical analysis of a quantum-dot-engineered heterostructure, PbS:Yb\(^{3+}\),Er\(^{3+}\)/CuBiO, optimized for water splitting applications. Using density functional theory (DFT) coupled with machine learning, the study explores the electronic, optical, and catalytic properties of the material. The optimized PbS structure exhibited a direct bandgap of 1.191 eV, while co-doping with Yb and Er transitioned the material to a metallic state, enhancing charge carrier mobility and electron-hole separation. The final heterostructure displayed an indirect bandgap of 0.431 eV, favorable for visible-light absorption. Key findings include an internal electric field strength of 6.3 Debye, efficient charge transfer confirmed by Bader analysis, and strong optical absorption at 2.4 eV. Machine learning models, including DNN and LSTM, were employed to predict photon absorption rates, achieving mean squared errors as low as 0.0004. The synergistic effects of enhanced internal electric fields, optimized band structures, and strong photon absorption underscore the material’s potential for efficient hydrogen production. This study bridges advanced computational methods and machine learning, offering a framework for designing high-performance photocatalysts in renewable energy applications.

\end{abstract}

\end{frontmatter}

\section{Introduction}
\label{sec:Intro}


Solid-state photon upconversion materials \cite{esmann2024solid} are luminescent materials capable of converting low-energy light (e.g., infrared or near-infrared) into higher-energy light (e.g., visible or ultraviolet) through upconversion processes \cite{izawa2021efficient}. In binary photocatalytic systems, these materials enhance photocatalytic efficiency by providing additional high-energy photons to activate photocatalysts. They absorb low-energy photons that would otherwise remain unused, extending the effective absorption range and increasing electron-hole pair generation in photocatalysts. This leads to enhanced photocatalytic activity \cite{fazaeli2022variable}, enabling diverse applications such as solar energy conversion, pollutant removal, and chemical production.
Photon upconversion occurs through two mechanisms. In organic materials, sensitized triplet-triplet annihilation (sTTA) and energy pooling drive upconversion, while inorganic materials rely on energy transfer upconversion (ETU) \cite{yan2024spatiotemporal}, excited-state absorption (ESA) \cite{deng2024optimizing}, and photon avalanche (PA) \cite{huang2024controlling}. These processes can be classified under two-photon absorption and second-harmonic generation. Among various upconversion nanoparticles studied for efficiency enhancement, the following classifications stand out:
\begin{itemize}
    \item Rare-Earth Doped Materials: Examples include NaYF\(_4\):Yb\(^{3+}\),Er\(^{3+}\) \cite{palo2018effective, berry2015disputed, anderson2014revisiting} and La\(_2\)O\(_3\):Yb\(^{3+}\),Tm\(^{3+}\) \cite{gao2018highly,li2012multiform, xie2022optical}, which utilize ETU mechanisms.
    \item Organic Materials: Perylene and anthracene derivatives rely on sTTA \cite{huang2024triplet}.
    \item Semiconducting Nanocrystals: PbS and CdSe quantum dots depend on multi-exciton generation or two-photon absorption.
    \item Metal-Organic Frameworks (MOFs):Yb\(^{3+}\)$-$ or Er\(^{3+}\)-doped MOFs exhibit unique characteristics \cite{li2018erbium, xie2023upconversion}.
    \item Inorganic Phosphors: Materials like ZnS:Cu,Co and Gd\(_2\)O\(_2\)S:Yb\(^{3+}\),Er\(^{3+}\) \cite{giang2024synthesis}.
\end{itemize}

Upconversion materials have been extensively studied for applications like water splitting, CO\(_2\) reduction, and pollutant removal. Kim et al. \cite{kim2023investigating} explored PbS quantum dots combined with TiO\(_2\) photoanodes for water splitting with similar mechanisms shown in Figure 1. Their findings revealed that optimal TiO\(_2\) thickness (11.9 µm) produced the highest photocurrent density (15.19 mA/cm\(^2\)) due to improved light absorption and reduced charge transfer resistance. Zhou et al. \cite{zhou2018upconversion} demonstrated that SrTiO\(_3\):Er\(^{3+}\) nanoparticles synthesized via sol-gel methods achieved enhanced photocatalytic hydrogen production under simulated sunlight compared to undoped SrTiO\(_3\). Similarly, Zhang et al. \cite{zhang2021enhanced} synthesized NaBiF\(_4\):Yb\(^{3+}\),Tm\(^{3+}\)/Bi\(_2\)WO\(_6\) composites, achieving 98\% degradation efficiency of rhodamine dyes within 27 minutes using ETU mechanisms.

\begin{figure}[!ht]
    \centering
    \includegraphics[width=0.6\textwidth, angle=360]{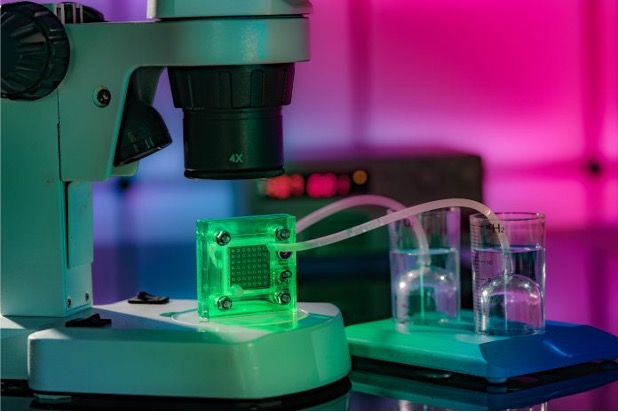}
    \caption{Practical photocatalytic hydrogen production, adopted with permission from \cite{phontonabsorption}}
    \label{fig1}
\end{figure}

Density Functional Theory (DFT) is a cornerstone computational technique for investigating the electronic, structural, and optical properties of photocatalytic materials \cite{mostafavi2022dispersion, ghahghaey2021theoretical}. It is extensively applied to study bandgap engineering, charge transfer, and adsorption energies. Advanced methods like hybrid functionals (e.g., HSE06) and time-dependent DFT (TD-DFT) extend its utility to excited-state properties, critical for understanding photon absorption and electron-hole recombination dynamics.

Recent DFT studies have unraveled mechanisms underlying water splitting, CO\(_2\) reduction, and pollutant degradation. For instance, computational studies on TiO\(_2\) and g-C\(_3\)N\(_4\) have revealed insights into their charge carrier dynamics and surface reactions, guiding experimental optimization. Similarly, rare-earth doped materials like NaYF\(_4\):Yb\(^{3+}\),Er\(^{3+}\) have been analyzed to optimize energy transfer upconversion \cite{ramin2024layer}. However, DFT has limitations in computational cost and accuracy for complex systems \cite{hashemkhani2021understanding, abadee2019removing}. Recent advancements in multi-scale DFT and high-performance computing have addressed these challenges, enabling simulations of larger systems with higher precision.

Integrating DFT with experimental findings has become critical for validating predictions, optimizing material properties, and accelerating the design of next-generation photocatalysts \cite{allam2024unveiling}. Despite its success, combining DFT with machine learning offers unparalleled potential to explore a broader materials space and improve predictive accuracy.

Machine learning (ML) has revolutionized photocatalysis by enabling rapid analysis of complex datasets and accelerating material discovery. ML algorithms, such as Random Forests (RF), Support Vector Machines (SVM), and Neural Networks, have been extensively used to predict material properties \cite{mikolajczyk2024visible} like bandgap, carrier mobility, and catalytic activity. These algorithms are often trained on experimental and computational datasets to establish structure-property relationships, enabling high-throughput screening of potential photocatalysts.

Studies \cite{yurova2024optimization} have shown that integrating ML with density functional theory (DFT) enhances predictive capabilities. For example, ML models have been employed to predict adsorption energies, electron-hole recombination rates, and charge transfer dynamics, often using DFT-generated datasets. Generative models like Variational Autoencoders (VAEs) and Generative Adversarial Networks (GANs) have shown promise in designing hypothetical materials with tailored properties. Furthermore, feature selection methods, such as SHAP (Shapley Additive Explanations) \cite{schossler2024novel}, have improved the interpretability of ML models, making them suitable for guiding experimental efforts.

Though ML offers significant advantages, challenges like data standardization, model generalizability, and interpretability remain. Future directions include integrating ML with quantum machine learning (QML) and multi-fidelity modeling to achieve more accurate and efficient predictions \cite{dananjaya2024synthesis}. The combination of ML and DFT holds transformative potential for advancing photocatalytic research.

While coupling DFT-ML for material property investigations, the development of upconversion materials for photocatalytic water splitting has been largely hindered by the intensive and costly material engineering efforts required in laboratories. To address these challenges, theoretical and computational studies have become essential for preliminarily assessing the complexities, suitability, and scalability of these materials \cite{eidsvaag2021tio2}. Understanding their fundamental electronic and nonlinear optical properties is critical for optimizing their performance and guiding experimental efforts. However, despite their importance, numerical studies specifically targeting upconversion materials remain under explored.

Samanta’s review underscores the significant challenges associated with modeling photocatalysts for water splitting, particularly due to the computational demands of multiscale and multi-quantum physics approaches \cite{samanta2022challenges}. These methods require highly resource-intensive algorithms, beginning with solving the Kohn-Sham equations for electronic structure calculations and extending to Bethe-Salpeter solvers for accurately capturing light absorption and electron-hole recombination dynamics. This computational complexity, coupled with the need for precise pseudopotential generation, has resulted in limited attempts to comprehensively study upconversion materials for water splitting. By addressing these challenges, this research aims to fill a critical gap in the computational analysis of upconversion materials, enabling informed design strategies for next-generation photocatalysts.
The PbS:Yb\(^{3+}\),Er\(^{3+}\)/CuBiO material system is well-suited for water splitting due to its excellent light absorption and photon upconversion capabilities. We practically know that PbS with its narrow bandgap, efficiently absorbs infrared light, which constitutes a significant portion of the solar spectrum. Doping with Yb\(^{3+}\) and Er\(^{3+}\) ions enables photon upconversion, converting low-energy infrared photons into visible or ultraviolet photons, thereby extending the material’s ability to harness solar energy more efficiently. CuBiO, a p-type semiconductor, has a moderate bandgap ideal for visible light absorption and is known for its photocatalytic stability, resisting corrosion in aqueous environments. This material ensures effective charge separation, reducing electron-hole recombination and enhancing photocatalytic efficiency. The heterostructure formed between PbS and CuBiO optimizes charge transfer and creates a broader spectrum response, making this system particularly effective for driving water-splitting reactions to produce hydrogen under solar illumination.

The fundamental aim of this study is to develop a numerical model using density functional theory and machine learning algorithms to predict the upconversion efficiency of a novel composite material. This includes;

\begin{enumerate}
    \item To investigate the electron and optical properties of the material, determine the charge carrier separation and electron-hole pair recombination.
    \item To enhance water splitting with PbS:Yb3+, Er3+/CuBiO under the NIR-vis-UV spectrum.
    \item To identifying optimal machine learning models for photon absorption prediction.
\end{enumerate}

\section{Methodology}
This study presents computational strategies for optimizing novel materials to enhance photocatalytic processes for water splitting and hydrogen production. Specifically, we explore a quantum dots upconversion (QDsUC) photocatalyst, (PbS:Yb\(^{3+}\), Er\(^{3+}\)/CuBiO), which forms a unique heterostructure designed to improve hydrogen production efficiency. The combination of QDsUC nanoparticles with semiconducting materials aims to harness superior electron and optical properties to enhance overall system performance. Achieving such advanced photocatalytic processes requires precise integration of QDsUC nanoparticles with the semiconductor, involving meticulous crystal heterostructure design and informed lattice mismatch considerations.

We compute key material properties, including band structures, density of states at the Fermi level, carrier concentration and mobility, dielectric function, and absorption coefficient, using VASP 6. These calculations provide a detailed understanding of the electronic and optical behavior of the heterostructure. The resulting quantum numerical data are further analyzed using a cluster of neural networks implemented in TensorFlow, enabling the prediction and optimization of material performance metrics. This integrated approach of numerical simulation and machine learning facilitates the design of high-performance photocatalysts for renewable energy applications.

\subsection{Material Properties}
The PbS:Yb\(^{3+}\),Er\(^{3+}\)/CuBiO system, designed for water splitting applications, exhibits a combination of unique chemical and physical properties that make it promising for solar energy conversion. The chemical properties of PbS (lead sulfide) doped with Yb\(^{3+}\) (ytterbium) and Er\(^{3+}\) (erbium) introduces rare-earth elements, which enhance photon upconversion. These ions absorb low-energy photons and emit higher-energy photons, improving the utilization of the solar spectrum for water splitting. CuBiO (copper bismuth oxide) is a p-type semiconductor with strong photocatalytic properties, stable in aqueous environments and resistant to photo-corrosion, making it ideal for water splitting. The combination of these materials helps promote the redox reactions needed for hydrogen production. 

However, the physical properties of PbS has a narrow bandgap semiconductor (~0.41 eV), which is ideal for absorbing infrared light, while Yb\(^{3+}\) and Er\(^{3+}\) enable upconversion of low-energy photons into visible or UV light, extending the effective light absorption range. CuBiO exhibits a moderate bandgap (~1.8-2.1 eV), suitable for visible light absorption, enabling it to drive the photocatalytic process. Together, these materials form a heterostructure with efficient charge separation, enhancing the photocatalytic activity and stability required for effective water splitting under sunlight.

\subsection{Kohn-Sham Quantum Analytical Modeling (DFT)}
Notable scientists like Schrödinger, Bohr-Oppenheimer, Heisenberg, Kohn, and Sham delved into quantum mechanics to develop the foundations for density functional theory capabilities. We witness the effects of this today as materials react with light, causing electrons to become photo-excited and releases energy that we can see. By utilizing Kohn and Sham's equations, scientists are able to comprehensively analyze materials at the quantum level without needing to physically observe every single atom and electron. The Kohn-Sham quantum model builds upon Schrödinger's core principles to calculate energy accurately, where Energy is represented in Equation \eqref{eq:1}  as:

\begin{equation} \label{eq:1}
    E=\int \Psi^{*} \hat{H} \Psi d r \\
\end{equation}

Furthering the energy in Equation \eqref{eq:2} under orthonormal restrictions, the wave function is transformed into:

\begin{equation} \label{eq:2}
    \int \Psi_{i}^{*}(r) \Psi_{j}(\boldsymbol{r}) d \boldsymbol{r}=1 \text { at } i=j, 0 \text { at } i \neq j \\
\end{equation}

Despite the above constraints, the Schrödinger wave equation has an applied variational principle in which the Lagrange multiplier $\lambda$ is introduced; hence, the energy is further transformed as follows:

\begin{equation}\label{eq:3}
    0=\delta\left(\int \Psi^{*} \hat{H} \Psi d \boldsymbol{r}\right)=\delta\left[\int \Psi^{*} \hat{H} \Psi d r-\lambda\left(\int \Psi_{i}^{*} \Psi_{i} d \boldsymbol{r}-1\right)\right] \rightarrow \hat{H} \Psi=E \Psi \\
\end{equation}

From Equation \eqref{eq:3}, we can now introduce the Kohn-Sham energy Equation \eqref{eq:4}, where, at the minimum energy condition, the differences in energy functionality should be zero, which is in relation to electronic density \cite{kohn1996density, ullrich2001kohn, kohn1996density2}. 

\begin{equation}\label{eq:4}
    0=\frac{\delta}{\delta \rho(\boldsymbol{r})}\left(E[\rho(\boldsymbol{r})]-\lambda\left[\int \rho(\boldsymbol{r})\right] d \boldsymbol{r}\right) \rightarrow \frac{\delta E[\rho(\boldsymbol{r})]}{\delta \rho(\boldsymbol{r})}=\lambda \text { or } \\
\end{equation}

\begin{equation}\label{eq:5}
    0=\frac{\delta E_{k i n}^{n o n}}{\delta \phi_{i}^{*}(\boldsymbol{r})}+\left[\frac{\delta E_{e x t}}{\delta \rho(\boldsymbol{r})}+\frac{\delta E_{H}}{\delta \rho(\boldsymbol{r})}+\frac{\delta E_{x c}}{\delta \rho(\boldsymbol{r})}\right] \frac{\delta \rho(\boldsymbol{r})}{\delta \phi_{i}^{*}(\boldsymbol{r})}-\sum_{j} \lambda_{i j} \phi_{j}(\boldsymbol{r})\\
\end{equation}

In Equation \eqref{eq:5}, the Lagrange multiplier $\lambda_{ij}$ with constraints is introduced under three vital energy functionalities ($E_{ext}$, $E_H$, and $E_{xc}$), which is further simplified in Equations \eqref{eq:6} and \eqref{eq:7}.

 \begin{equation}\label{eq:6}
    \quad=\left(-\frac{1}{2}{ }^{2}+U_{e x t}+U_{H}+U_{x c}-\lambda_{i}\right) \phi_{i}(\boldsymbol{r}) \\
\end{equation}

\begin{equation}\label{eq:7}
    \quad=\left(-\frac{1}{2}{ }^{2}+U_{e f f}-\varepsilon_{i}\right) \phi_{i}(\boldsymbol{r}) \\
\end{equation}

Based on these functionalities,  Equations \eqref{eq:8} and \eqref{eq:9} lead to the coupled Hamiltonian-KS equation from Schrödinger Equation \eqref{eq:1}.

\begin{equation}\label{eq:8}
    \frac{\delta E_{k i n}^{n o n}}{\delta \phi_{i}^{*}(\boldsymbol{r})}=-{ }^{2} \phi_{i}(\boldsymbol{r}), \frac{\delta \rho(\boldsymbol{r})}{\delta \phi_{i}^{*}(\boldsymbol{r})}=2 \phi_{i}(\boldsymbol{r}), U_{x c}[\rho(\boldsymbol{r})]=\frac{\delta E_{x c}[\rho(\boldsymbol{r})]}{\delta \rho(\boldsymbol{r})} \\
\end{equation}

\begin{equation}\label{eq:9}
    {\left[-\frac{1}{2}{ }^{2}+U_{e f f}(\boldsymbol{r})\right] \phi_{i}(\boldsymbol{r})=\varepsilon_{i} \phi_{i}(\boldsymbol{r}) \rightarrow \hat{H}_{K S} \phi_{i}(\boldsymbol{r})=\varepsilon_{i} \phi_{i}(\boldsymbol{r})}
\end{equation}

However, when performing DFT calculations, it generally means reducing or minimizing the energy density, and this must be done in accordance with three major conditions: self-consistency, variational principle, and constraints \cite{rezvani2018dft}. However, it is crucial for the KS-DFT solution to always remain self-consistent with the one-electron equation, which is governed by various principles and subjected to orthonormal constraints. In order to achieve this self-consistency, DFT solutions involve a complex interplay of equations. The KS orbitals determine the electron densities, which in turn determine the KS Hamiltonian. The KS Hamiltonian then calculates the new electron densities, resulting in a continuous cycle of calculations known as self-consistency. The depletion of electron density is termed the exchange correlation (xc) hole, which supports the understanding of XC energy and relentlessly provides guidelines for xc functionals such as the local density approximation (LDA), Generalized Gradient Approximation (GGA), and other hybrid approximations. In the possible event of selecting an appropriate approximation (xc functional), the total energy as given in Equations \eqref{eq:10} at the ground state tends to minimize varied energy, hence becoming static.

\begin{equation}\label{eq:10}
    E=-\frac{1}{2} \sum_{i}^{n} \int \phi_{i}^{*}(\boldsymbol{r}){ }^{2} \phi_{i}(\boldsymbol{r}) d \boldsymbol{r}+\int U_{e x t}(\boldsymbol{r}) \rho(\boldsymbol{r}) d \boldsymbol{r}+\frac{1}{2} \iint \frac{\rho(\boldsymbol{r}) \rho\left(\boldsymbol{r}^{\prime}\right)}{\left|\boldsymbol{r}-\boldsymbol{r}^{\prime}\right|} d \boldsymbol{r} d \boldsymbol{r}^{\prime}+E_{x c}[\rho(\boldsymbol{r})]
\end{equation}

\begin{equation}\label{eq:11}
    \int \phi_{i}^{*}(r) \phi_{j}(r) d r=\delta_{i j}
\end{equation}

The application of the variational principle could be applied to either electron density, KS eigenvalues, or the norm of the residual vector to an eigenstate. Finally, one of the crucial orders to follow while satisfying the DFT solution is constraints. While minimizing the total energy value, this must be under a fixed total number of electrons or orthonormality of orbitals as given in Equation \eqref{eq:11}.

As long as we adhere to the prescribed sequence of steps in solving DFT, which involves self-consistency, the variational principle, and constraint, we can attain the most favorable ground state by minimizing the overall energy. This method is widely regarded as the most effective analytical process for addressing quantum problems. 

\begin{equation}\label{eq:12}
    B=V \frac{\partial^{2} E(V)}{\partial V^{2}}, P=-\frac{\partial E}{\partial V}, F_{I}=-\frac{\partial E}{\partial r_{I}}, \sigma_{i j}=\frac{\partial E}{\partial \varepsilon_{i j}}
\end{equation}

It is also worth noting that solving KS equations has also been simplified through the use of either direct or iterative diagonalization methods. For the latter, when the overall energy of a system is provided, such as in terms of lattice parameters, several essential characteristics naturally emerge. These include stable structures, the length and angle of bonds, the strength of the bonding forces, the bulk modulus, its elasticity, the changes that occur on its surface, the energy required to form defects, transitions between phases caused by pressure, and the vibrational properties indicated by the energy's curvature. Based on these fundamental parameters emerging, the derivative for solving in an iterative diagonalization method is given in Equation \eqref{eq:12}.

Moreover, the initial computation of the ground state acts as a foundation for more complex calculations like minimum-energy path, barrier energies, density of states (DOS), and band structures. Additionally, by dedicating additional computational resources, it becomes possible to obtain the second derivatives of energy, which provide information about the forces experienced by neighboring atoms when each atom is displaced in different directions. These calculations also enable simulations of scanning tunneling microscopy (STM), analysis of phonon spectra, evaluation of thermodynamic properties, and estimation of reaction rates, among other applications.

\subsection{Bethe–Salpeter Equation (BSE)}
The Bethe–Salpeter Equation (BSE) is a critical framework for accurately predicting the excited-state properties of materials, especially optical phenomena such as excitonic effects. In VASP, the implementation of BSE leverages the previously calculated single-particle states and screened Coulomb interactions derived from the GW (Green's function (G) and coulomb interactions (W)) approximation to capture electron-hole interactions explicitly.

The BSE formalism in VASP focuses on solving the two-particle Green’s function  L, expressed as in Equation \eqref{eq:13}:

\begin{equation}\label{eq:13}
    L = L_0 + L_0 \cdot K \cdot L
\end{equation}

Here, $L_0$  represents the non-interacting electron-hole propagator, and  K  denotes the kernel accounting for electron-hole interactions. The solution of the BSE within VASP enables the computation of dielectric functions, which include both real ( $\epsilon_1$($\omega$) ) and imaginary ( $\epsilon_2$($\omega$) ) components. These are pivotal for determining the optical absorption spectrum and other critical optical properties of materials.

The workflow for optical property calculations in VASP using the Bethe-Salpeter Equation (BSE) involves a systematic series of steps to ensure accurate predictions of excitonic effects. It begins with a standard density functional theory (DFT) ground-state calculation within the Kohn-Sham framework, which determines the electronic structure of the material. This provides the foundation for subsequent calculations by identifying the initial electronic states.

Next, the GW approximation is applied to refine these electronic states. This method calculates quasiparticle energies, improving the accuracy of band gaps and electronic properties that are essential for predicting optical behaviors. Following this, the screened Coulomb interaction is computed by evaluating the static or dynamic dielectric function. This screening process plays a critical role in the BSE kernel, as it accounts for the reduction of electron-hole interaction due to the surrounding electronic environment.

The electron-hole interaction kernel,  K , is then constructed, incorporating both direct and exchange terms. This kernel is the core of the BSE methodology, ensuring that excitonic effects—critical for light-matter interactions—are included in the optical spectrum. Finally, the optical properties are calculated by solving the BSE. This includes the computation of the dielectric function and the absorption coefficient ($\alpha$($\omega$)), expressed as Equation \eqref{eq:14}:

\begin{equation}\label{eq:14}
    \alpha(\omega) = \frac{\omega}{c} \cdot \text{Im} \left[ \sqrt{\epsilon(\omega)} \right]
\end{equation}

Where, $\omega$ represents the photon frequency, and c is the speed of light. This workflow enables the accurate modeling of excitonic peaks, absorption spectra, and related optical properties, making it indispensable for studying photocatalytic materials like PbS:Yb\(^{3+}\),Er\(^{3+}\)/CuBiO and other systems with strong light-matter interactions.

The Bethe-Salpeter Equation (BSE) dielectric function is instrumental in uncovering excitonic peaks, which are critical for understanding materials with strong light-matter interactions. These peaks provide insights into the excitonic binding energies and their role in enhancing optical absorption. VASP’s implementation of the BSE formalism allows for the inclusion of spin-orbit coupling and anisotropic screening effects, enabling accurate simulations for materials with complex electronic structures. This capability is particularly important for advanced photocatalytic and optoelectronic materials.

The BSE methodology in VASP is especially well-suited for studying materials where excitonic effects dominate light absorption, such as  PbS:Yb\(^{3+}\),Er\(^{3+}\)/CuBiO heterostructures. These systems benefit from enhanced excitonic interactions that improve photon absorption efficiency and charge carrier dynamics. Additionally, the BSE approach is ideal for analyzing low-dimensional systems like quantum dots, where spatial confinement amplifies excitonic effects. Beyond photocatalysis, the BSE formalism is valuable for designing materials for photovoltaics and light-emitting diodes (LEDs), where excitons play a central role in energy conversion and emission processes. This makes it a powerful tool for optimizing the optical properties of emerging energy materials.

\subsection{Upconversion Analytical Modelling}
In the development of analytical model that describes the up-conversion process in a binary photocatalyst system requires focusing on the interaction between the two components of the system, usually a sensitizer (which absorbs low-energy photons) and an activator (which emits higher-energy photons). The up-conversion process typically involves nonlinear optical phenomena \cite{frazer2017optimizing, zhang2022rare} where multiple lower-energy photons are absorbed, leading to the emission of a single higher-energy photon.  These sensitizer (S) and activators (A) are usually good examples of the respective rare earth ions Yb\(^{3+}\) and Er\(^{3+}\). Since these two rare earth ions plays a significant role in upconversion processes, the ground state is therefore denoted as \( S_0 \) and \( A_0 \), whereas first excited states are represented as  \( S_1 \) and \( A_1 \), and for the nth higher excited state the number goes to \( S_{nth}\) and \( A_{nth}\). It is worth noting that at higher excitation's, photon emissions is known to occur after relaxation.

Analytically, sensitizer (Yb\(^{3+}\)) are known for absorbing lower energy photons \( h\nu_1 \) which concurrently transitions from an initial ground state \( S_0 \)  to an excited state \( S_1 \) representing the Equation \eqref{eq:15}. Similarly, the photons absorbed by the sensitizer have their non-radiative energy transferred to the activators (Er\(^{3+}\)), which results in the activators moving from their initial ground state \( A_0 \) to their first excited state \( A_1 \), as depicted in Equation \eqref{eq:16} \cite{zong2019physical}.

\begin{equation}\label{eq:15}
     S_0 + h\nu_1 \rightarrow S_1 
\end{equation}

\begin{equation}\label{eq:16}
     S_1 + A_0 \rightarrow S_0 + A_1
\end{equation}

The activator, which is initially in its first excited state \( A_1 \), can absorb another photon \( h\nu_2 \) and transition to a higher excited state \( A_2 \) as shown in Equation \eqref{eq:17}. The activator in the \( A_2 \) state then relaxes back to the ground state \( A_0 \), releasing a photon with higher energy \( h\nu_3 \) than the two absorbed photons in Equation \eqref{eq:18}.

\begin{equation}\label{eq:17}
     A_1 + h\nu_2 \rightarrow A_2
\end{equation}

\begin{equation}\label{eq:18}
     A_2 \rightarrow A_0 + h\nu_3
\end{equation}

In order to translate the meaning of the relationship of these sensitizer and activator behavior in an analytical form, we can further examine their rates of reactions or dynamics. In considering the rate of change of sensitizer, the mathematical model is therefore represented in Equation 5 \eqref{eq:19}, whereas the rate of change for activator from  \( A_1 \) to \( A_2 \) is demonstrated in Equations \eqref{eq:20} and \eqref{eq:21} as; 

\begin{equation}\label{eq:19}
 \frac{d[S_1]}{dt} = W_{S} - k_{ET}[S_1][A_0] - k_{S}[S_1] 
\end{equation}

\begin{equation}\label{eq:20}
 \frac{d[A_1]}{dt} = k_{ET}[S_1][A_0] - k_{A1}[A_1] - W_{A}[A_1]
\end{equation}

\begin{equation}\label{eq:21}
    \frac{d[A_2]}{dt} = W_{A}[A_1] - k_{A2}[A_2]
\end{equation}

The rate at which the sensitizer absorbs photons is represented by $W_{S}$, while $W_{A}$ denotes the rate at which the activator $A_1$ absorbs photons. The rate constant for energy transfer from the excited sensitizer $S_1$ to the ground state activator $A_0$ is represented by $k_{ET}$, and $k_{S}$ is the rate constant for other non-radiative processes in the sensitizer. The rate constant for radiative relaxation from the excited state $A_2$ to the ground state $A_0$, which results in the emission of up-converted photons, is denoted by $k_{A2}$.

However, in circumstances where excited states remains constant, we can always tend to approximate their steadiness by equating them to zero (steady-state), therefore Equation \eqref{eq:22} illustrates; 

\begin{equation}\label{eq:22}
    \frac{d[S_1]}{dt} = \frac{d[A_1]}{dt} = \frac{d[A_2]}{dt} = 0
\end{equation}

The determinate of upconversion efficiency (\( \eta_{UC} \)) emanates from the rate of high energy photon emission from the activator \( A_2 \) as given in Equation \eqref{eq:23}. This efficiency depends on the rate constants for energy transfer, photon absorption, and radiative relaxation, as well as the intensity of the incident light. 

The effectiveness of upconversion (\( \eta_{UC} \)) is influenced by the rate at which high-energy photons are emitted from the activator \( A_2 \), as described in Equation \eqref{eq:23}. This effectiveness is determined by the rate constants for energy transfer, photon absorption, and radiative relaxation, as well as the intensity of the incident light.

\begin{equation}\label{eq:23}
    \eta_{UC} = \frac{k_{A2}[A_2]}{W_{S}}
\end{equation}

The key to enhancing photocatalytic performance is to incorporate rare earth ions into semiconductor materials. The mathematical model described in Equation \eqref{eq:23} outlines the mechanism for a binary photocatalytic system. However, our proposed material to optimizing a photocatalytic performance is selected as PbS:Yb\(^{3+}\),Er\(^{3+}\)/CuBiO. The numerical analysis based on the electron and optical properties of this material will further predict the upconversion efficiency.

\subsection{Water Splitting Simulation}
The simulation integrated the material’s structural and electronic features into the analysis of the photogenerated charge separation mechanism. Internal electric field (IEF) mapping revealed a spatially varying electric field, with hotspots located at the heterojunction interface. These fields, supported by a dipole moment of 6.3 Debye, drive efficient separation of photoexcited electron-hole pairs. Bader charge analysis further validated the material’s performance, showing significant charge redistribution between Yb, Er, and Cu atoms. This redistribution ensures a stable structure and enhanced conductivity, critical for catalytic efficiency. Photon absorption rates were analyzed using the absorption coefficient and dielectric function data. The strong absorption peaks around 2.4 eV in the visible spectrum ensure efficient utilization of solar energy. Minimal reflectivity and robust optical conductivity further enhance photon harvesting and charge transport. Combined with low recombination rates, these properties enable efficient hydrogen production under solar illumination.

Machine learning models were incorporated to predict the photon-to-hydrogen conversion efficiency, leveraging the material’s upconversion efficiency and charge separation dynamics. Neural network architectures trained on high-fidelity DFT data provided predictions on photon utilization rates, confirming the heterostructure’s suitability for photocatalytic water splitting. The synergistic effects of optimized band structure, enhanced IEF, and strong optical properties make PbS:Yb\(^{3+}\),Er\(^{3+}\)/CuBiO a promising material for efficient hydrogen production via water splitting. This study establishes a framework for combining computational simulations with machine learning to optimize photocatalytic systems.

\subsection{Bulk Pristine PbS}
The pristine bulk structure of PbS as indicated in Figure 2 was constructed and optimized using VASP (Vienna Ab initio Simulation Package) engine. Initial structural parameters were derived from experimental data for PbS, featuring a rock salt crystal structure. The optimization was performed to achieve the equilibrium geometry and minimize the system’s total energy. The optimization employed the GGA-PBE exchange-correlation functional, suitable for capturing the electronic interactions in semiconductors. A plane-wave cutoff energy of 500 eV was used, ensuring high computational precision while maintaining efficiency. The reciprocal space was sampled with a k-point grid of 5 $\times$ 5 $\times$ 3, corresponding to k-spacing of approximately 0.297 \AA\(^{-1}\). The Methfessel-Paxton smearing method was applied with a width of 0.2 eV to handle partial occupancies during electronic convergence.

\begin{figure}[!ht]
    \centering
    \includegraphics[width=0.8\textwidth, angle=360]{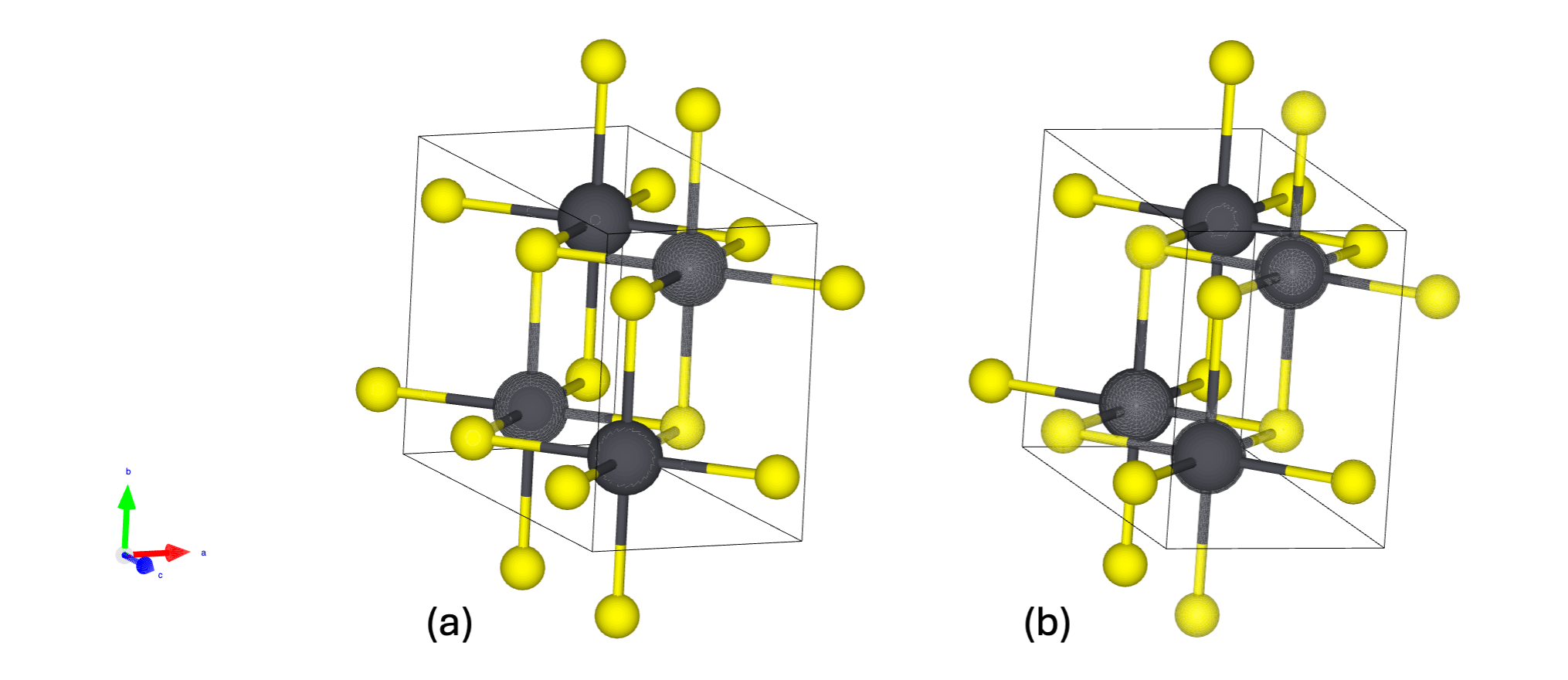}
    \caption{Schematic molecular structure of bulk pristine PbS (a) initial structure (b) optimized structure}
    \label{fig2}
\end{figure}

The structure was fully relaxed, including atomic positions and cell dimensions, over three optimization steps. The final cell parameters showed slight expansion from the initial geometry, with lattice constants adjusted by ~0.5\%. The total energy convergence criterion was set to 10\(^{-5}\) eV. Symmetry considerations reduced the computational workload, with 18 unique k-points in the Brillouin zone. Post-optimization, the electronic properties were analyzed, revealing a direct bandgap of 0.842 eV. The bandgap was calculated from the energy difference between the valence band maximum and conduction band minimum, both located near the (0.00, 0.40, 0.00) k-point. This result provides a baseline for assessing the modifications introduced by doping and heterostructuring.

The optimized PbS structure provides a stable and accurate foundation for subsequent modifications, including Yb and Er doping and the integration of CuBiO, aimed at enhancing the material’s electronic and optical properties for water-splitting applications.

\subsubsection{PbS-Electronic properties}
Owing to the optimized structure input parameters, the band structure, 40 k-points were sampled along the high-symmetry paths in the Brillouin zone (F–$\Gamma$–B–G). For density of states (DOS), a denser 9 $\times$ 7 $\times$ 5 k-point grid was applied, with projection onto spherical harmonics to extract site- and orbital-projected DOS. Atomic partial charges and the density of states were analyzed to gain insights into the contributions of Pb and S atoms near the Fermi level. The input structure was relaxed to minimize forces and stress, resulting in cell dimensions of a = 4.252 \AA, b = 6.012 \AA, and c = 8.475 \AA with a volume of 216.63 \AA\(^3\). The pristine PbS structure converged successfully to a minimized total energy of -35.20 eV per unit cell. 

The band structure in Figure 3a reveals that PbS is a direct bandgap semiconductor with a bandgap of 1.191 eV, located at the $\Gamma$-point. The valence band maximum (VBM) and conduction band minimum (CBM) align at the same k-point, ensuring efficient optical transitions. The valence band is dominated by contributions from S p-orbitals, while the conduction band is primarily influenced by Pb d-orbitals.

\begin{figure}[!ht]
    \centering
    \includegraphics[width=0.6\textwidth, angle=360]{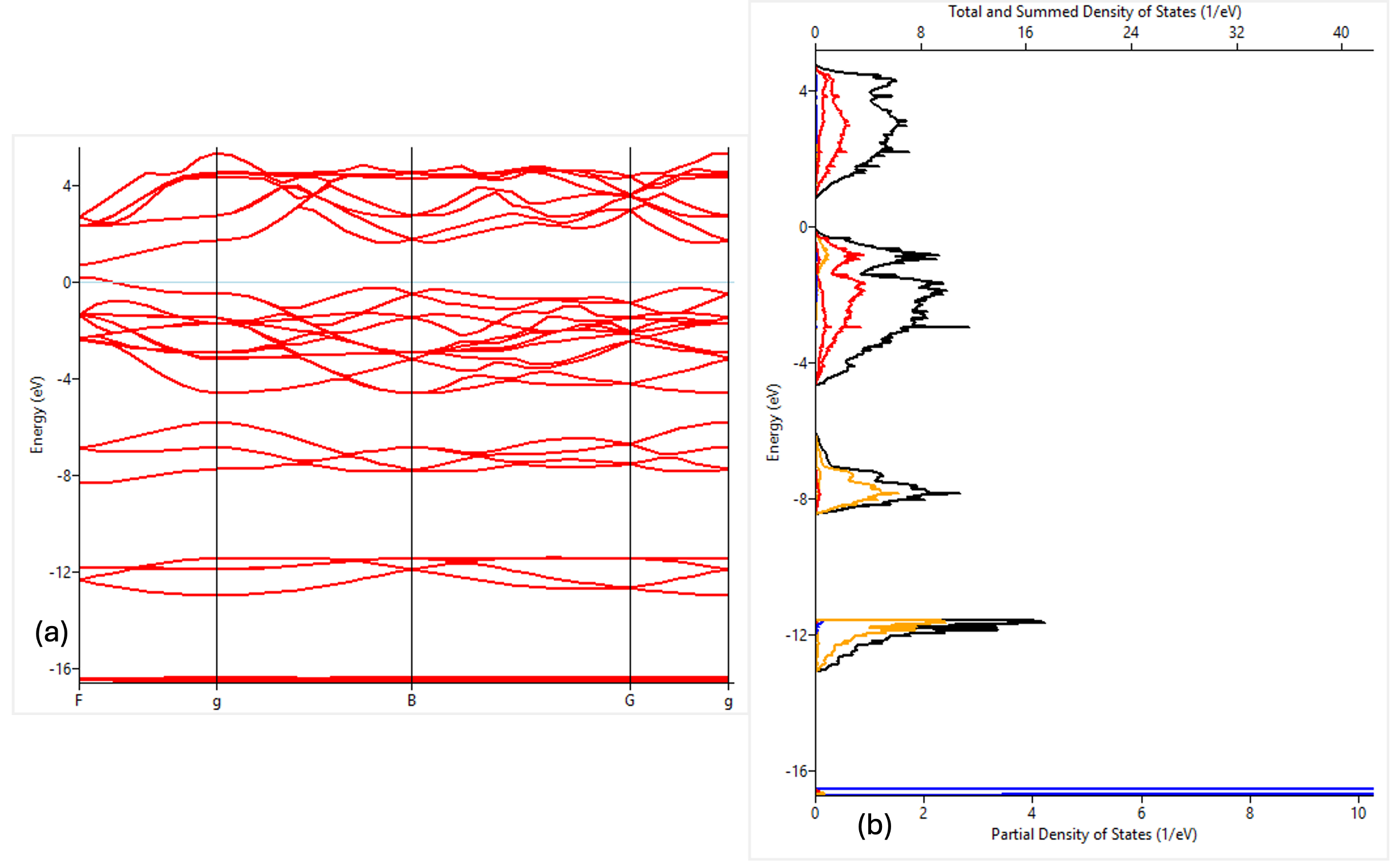}
    \caption{Illustrations of PbS (a) bandstructure and (b) densities of state}
    \label{fig3}
\end{figure}

Whereas the DOS in Figure 3b confirms the bandgap of 1.191 eV, with significant contributions from Pb d-orbitals and S p-orbitals near the VBM. The partial density of states indicates that Pb contributes strongly to the conduction band, while S dominates the valence band. These results establish pristine PbS as a suitable candidate for infrared light absorption, providing a baseline for subsequent doping and heterostructure integration. The calculated bandgap and electronic structure align well with theoretical predictions, validating the computational approach. 

A Fourier grid of 32 $\times$ 48 $\times$ 64 was used for charge density, ensuring high-resolution charge distribution. Bader charge analysis was conducted to evaluate charge transfer between Pb and S atoms, and to identify regions of localized electron density. The stress tensor and total energy were analyzed to confirm structural stability.

In Figure 4a, the total charge density displays the total charge density averaged along the planar and macroscopic directions. The peaks in the black curve (planar) correspond to regions of high electron density around Pb and S atoms, while the red curve (macroscopic) smooths out periodic variations. These results highlight strong charge localization near atomic centers, confirming the ionic character of PbS. The total valence charge in Figure 4b quantified the charge transfer, revealing that Pb atoms donated approximately 1.004 e to the S atoms. This significant charge transfer aligns with the ionic bonding nature of PbS.

\begin{figure}[!ht]
    \centering
    \includegraphics[width=0.6\textwidth, angle=360]{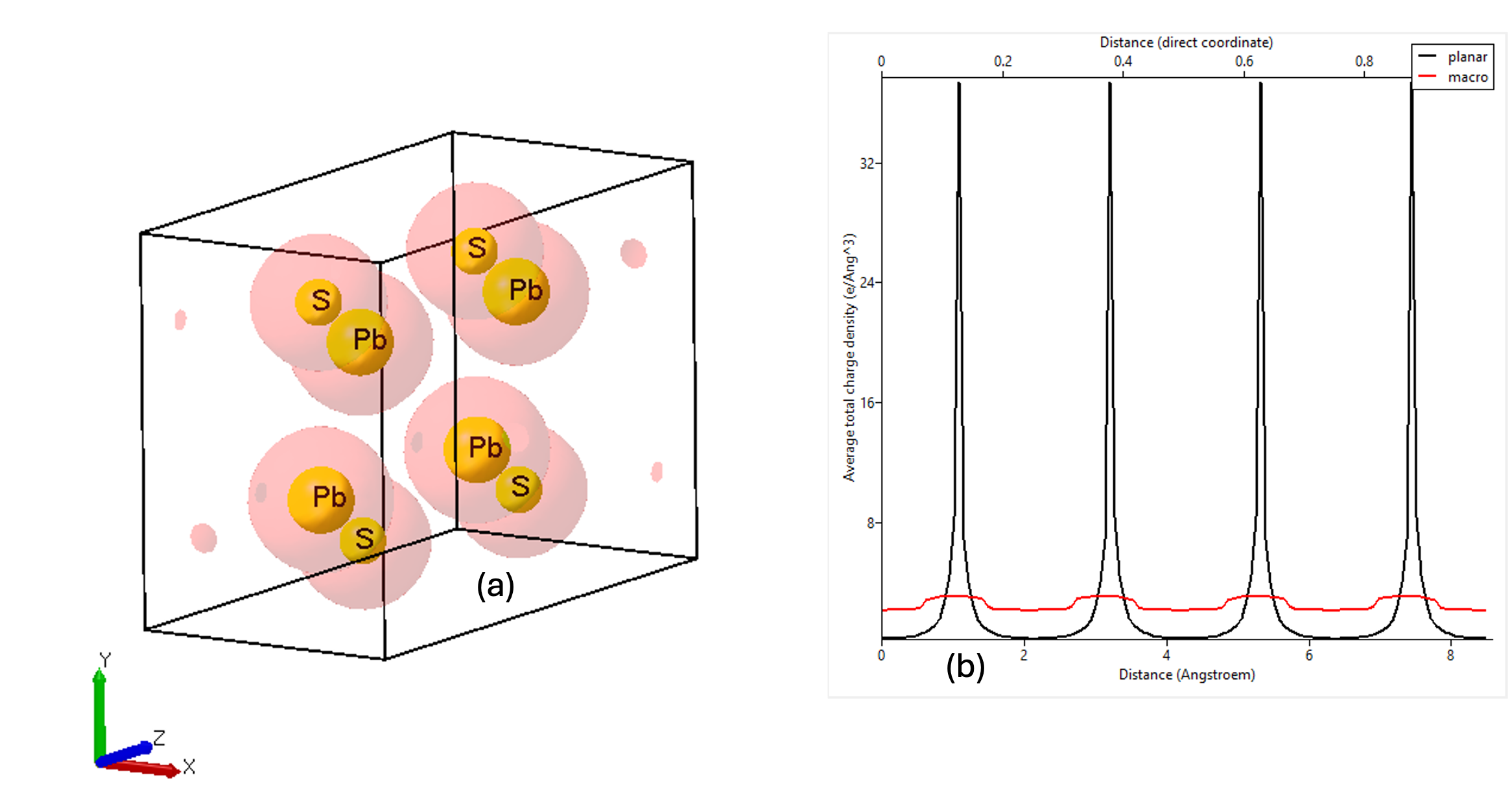}
    \caption{Illustrations of PbS (a) Charge density and (b) Bader charge}
    \label{fig4}
\end{figure}

The resulting simulation indicates that the calculated density was 7.336 g/cm\(^3\), matching experimental values; whereas bader charge analysis confirmed that Pb atoms carried an average charge of +11.685 e, while S atoms carried -3.198 e. The charge distribution reflects strong ionic interactions, which are critical for its electronic and optical properties. These results provide a comprehensive understanding of PbS’s charge density and bonding, forming a strong foundation for further doping and heterostructure studies.

\subsubsection{PbS-Optical properties}
The optical properties of pristine PbS were also calculated using VASP. Based on the optimized PbS structure, a gamma-centered  8 $\times$ 8 $\times$ 8  k-point grid was used for Brillouin zone sampling, ensuring accurate convergence of the electronic structure. The optical response, including the dielectric function, absorption coefficient, reflectivity, and conductivity, was derived through perturbative linear-response methods. The GW approximation was used to refine the band structure and optical spectra, accounting for many-body interactions.

The optical absorption coefficient as shown in Figure 5a illustrates a function of photon energy for the x, y, and z directions (anisotropic system). The peaks around 2.5 eV indicate strong absorption in the visible spectrum, aligning with PbS’s suitability for visible-light photocatalysis. The anisotropy observed in the absorption spectra reflects differences in electronic transitions along crystallographic directions.

The dielectric Function in Figure 5b illustrates the real part ($\varepsilon_1$) and imaginary part ($\varepsilon_2$) of the dielectric function plotted for different directions. Peaks in $\varepsilon_2$ correspond to photon energies where interband transitions occur, while $\varepsilon_1 $indicates the polarization response. Notable peaks near 3.5 eV signify high optical activity in the UV region. Whereas the optical conductivity in Figure 5c highlights the regions of strong photon-induced charge carrier generation. Peaks in the conductivity spectra around 2.0 eV correspond to electronic transitions that enhance photocatalytic efficiency. The anisotropy in $\sigma$($\omega$) across directions suggests direction-dependent conductivity, useful for tailoring material properties.

In the same method, the reflectivity spectrum in Figure 5d reveals how much incident light is reflected at different photon energies. Peaks at approximately 4 eV indicate high reflectivity, suggesting reduced absorption in the UV range. Conversely, lower reflectivity in the visible range confirms PbS’s effectiveness as a light-harvesting material.

\begin{figure}[!ht]
    \centering
    \includegraphics[width=0.8\textwidth, angle=360]{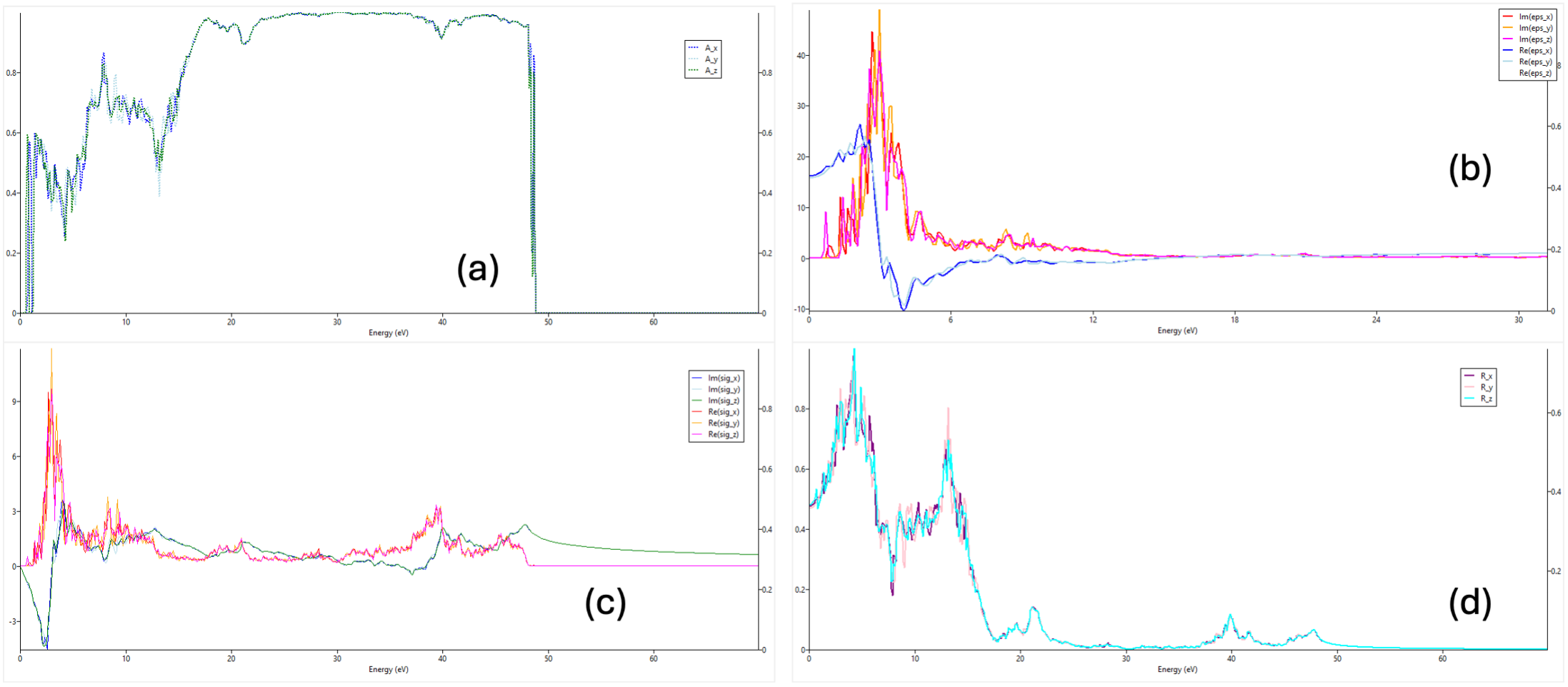}
    \caption{Pristine PbS optical response (a) optical absorption coefficient (b) dielectric Function, (c) optical conductivity, (d) reflectivity spectrum}
    \label{fig5}
\end{figure}

The optical analysis demonstrates that PbS exhibits strong absorption and low reflectivity in the visible spectrum, making it an excellent candidate for photocatalytic applications. The high dielectric constant and significant optical conductivity indicate efficient charge separation and transport under illumination. These properties, combined with its direct bandgap, validate pristine PbS as a suitable material for doping and heterostructuring, paving the way for enhanced photocatalytic water splitting. 

\subsection{PbS Doped Yb}
The structure of PbS:Yb\(^{3+}\) was optimized using VASP to model and analyze the doping effects of ytterbium (Yb) on PbS demonstrated in Figure 6. The base structure of pristine PbS was modified by replacing one Pb atom with Yb to create the doped system. The doping aimed to enhance the optical and electronic properties by introducing localized states and band structure modifications. The calculations were performed using the GGA-PBE exchange-correlation functional to describe electron interactions. A plane-wave cutoff energy of 500 eV was applied to ensure computational accuracy. The k-point mesh was set to 5 × 3 × 3 for geometry optimization and 9 × 7 × 5 for electronic property calculations. Methfessel-Paxton smearing with a width of 0.2 eV was used for electronic occupations, and geometry optimization was conducted until the total energy converged to 10\(^{-5}\) eV, with atomic forces reduced to less than 10\(^{-3}\) eV/\AA.

\begin{figure}[!ht]
    \centering
    \includegraphics[width=0.8\textwidth, angle=360]{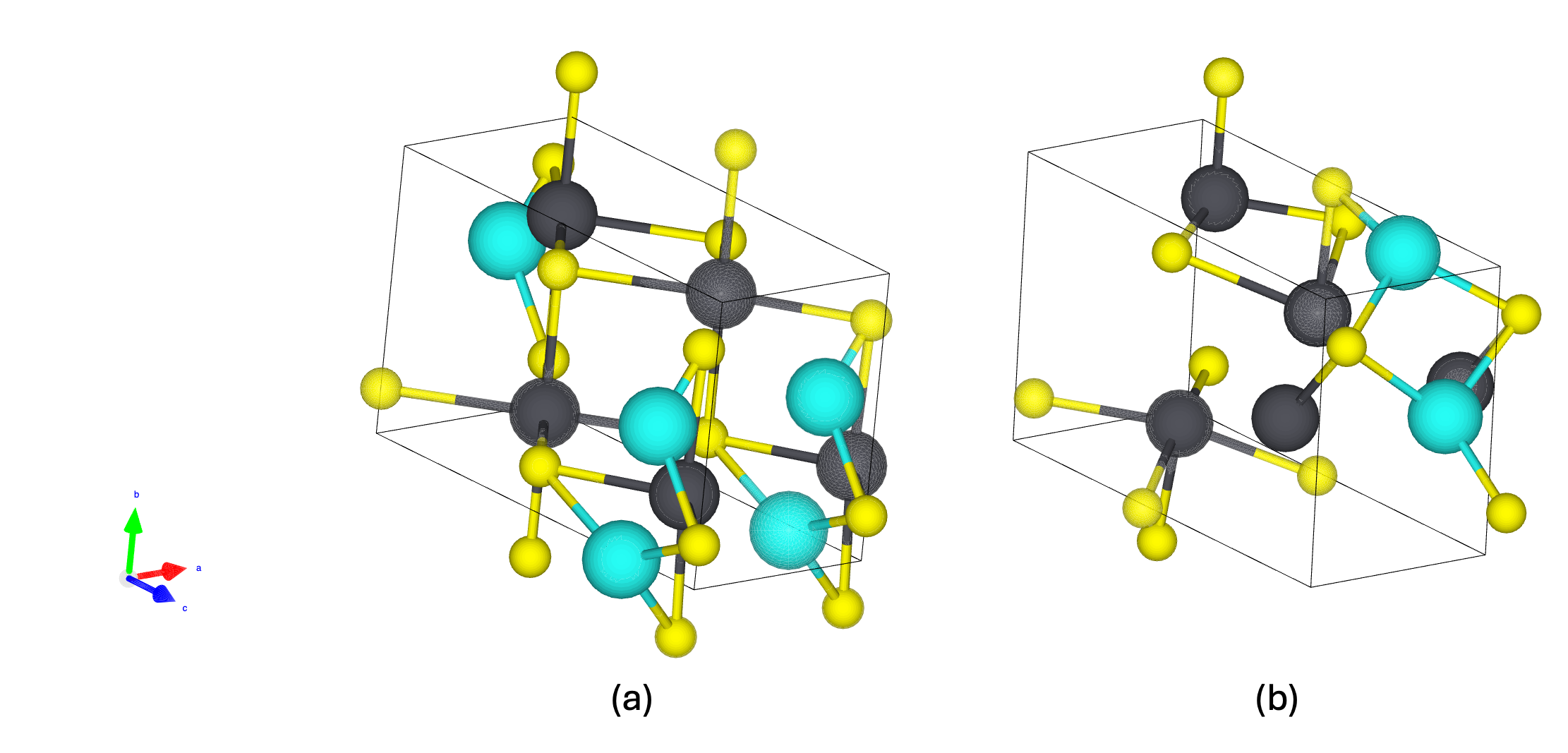}
    \caption{Schematic molecular structure of PbS:Yb (a) initial structure (b) optimized structure}
    \label{fig6}
\end{figure}

The doping approach involved symmetry-preserving substitution, maintaining the structural integrity of the lattice. The introduction of Yb, with its distinct ionic radius and electronic configuration, modified the local electronic environment, resulting in the appearance of new states within the band structure. The density of states (DOS) and band structure calculations were performed following geometry optimization to analyze the electronic properties of the doped system. The linear tetrahedron method in VASP was employed to calculate the DOS, ensuring accurate projection of electronic states. The k-point mesh for DOS calculations was refined to 9 × 7 × 5, and an energy grid with 3000 points was utilized for high-resolution DOS representation. For the band structure, calculations followed a path along high-symmetry points (F-$\Gamma$-B-G-$\Gamma$) in the Brillouin zone. 

\subsubsection{PbS:Yb-Electronic properties}
The optimized lattice parameters for the doped system were determined as a = 4.426 \AA, b = 6.258 \AA, and c = 10.461 \AA. The unit cell volume increased slightly to 289.758 \AA\(^3\), indicating a lattice expansion caused by Yb doping. This expansion is consistent with the introduction of larger ionic radii, which slightly distorts the lattice structure. The density of the doped system was calculated to be 7.468 g/cm\(^3\), reflecting the changes in mass and volume due to the proppant. The Fermi energy was identified at approximately 0.07 eV, indicating a shift in the material’s electronic behavior. The doping introduced new electronic states that cross the Fermi level, eliminating the original bandgap. This transition shifted the material’s properties from a semiconducting to a metallic nature, demonstrating the significant impact of Yb doping on the electronic structure.

The band structure of PbS:Yb\(^{3+}\) in Figure 7a shows multiple bands crossing the Fermi energy, indicating a metallic nature. The flatness of some bands suggests localized states introduced by Yb doping. High dispersion in other bands implies good carrier mobility, essential for photocatalytic applications. The absence of a well-defined gap and the crossing bands suggest delocalized electrons, making PbS:Yb\(^{3+}\) an efficient charge transport material. These features, coupled with band overlap, provide insight into its enhanced electronic properties, further validating the role of Yb in modifying the host PbS matrix.

\begin{figure}[!ht]
    \centering
    \includegraphics[width=0.6\textwidth, angle=360]{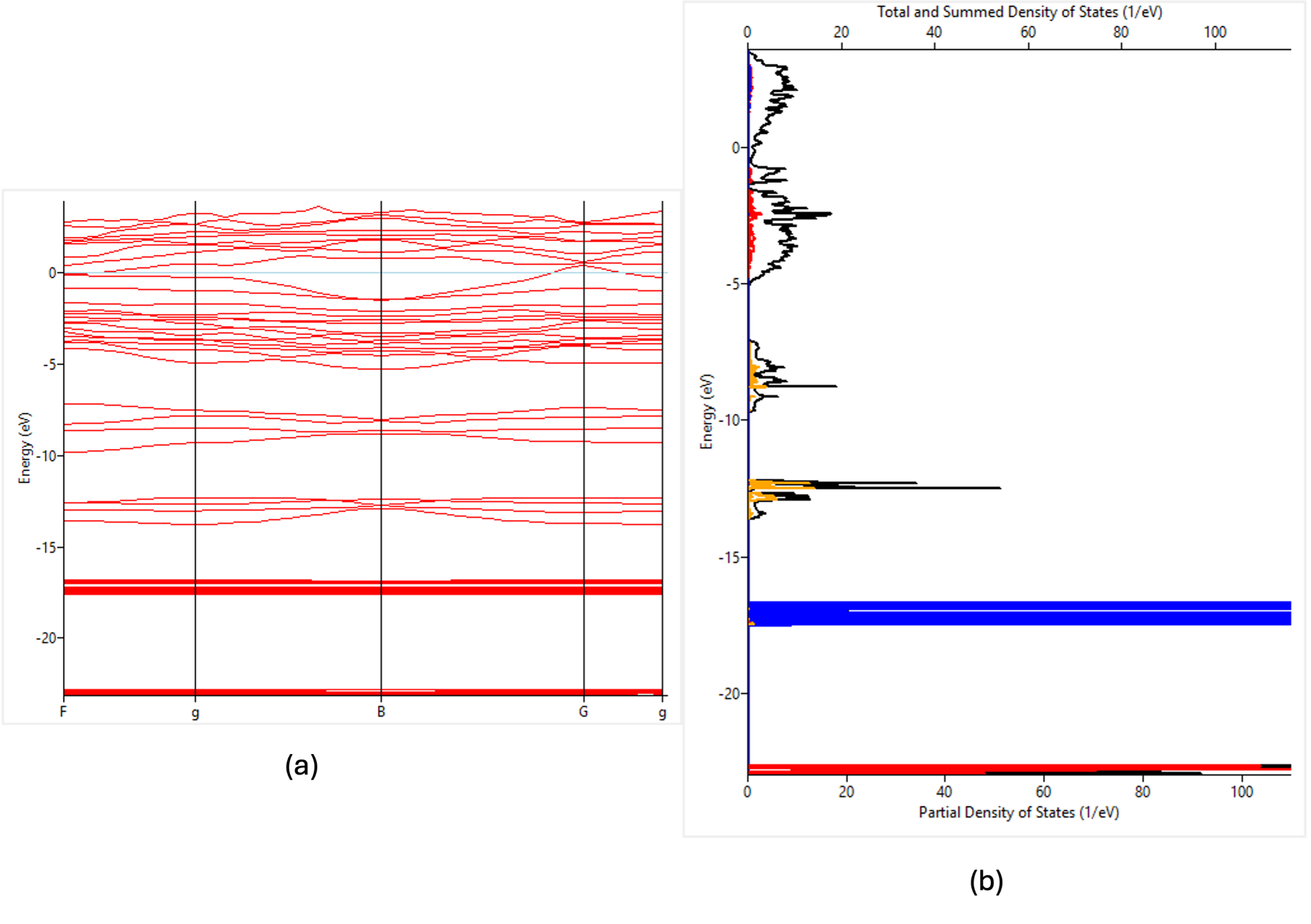}
    \caption{Illustrations of PbS:Yb (a) bandstructure and (b) densities of state}
    \label{fig7}
\end{figure}

The DOS plot highlights the electronic structure of PbS:Yb\(^{3+}\) in Figure 7b. The significant peaks near the Fermi energy (0 eV) indicate metallic behavior, primarily driven by Yb doping. The valence band is dominated by S p-orbitals, while the conduction band shows contributions from Pb and Yb d-orbitals. The sharp features near -5 eV and +3 eV correspond to bonding and antibonding states. The absence of a gap further confirms the transition from a semiconducting to a metallic state, which enhances electrical conductivity. Yb d-orbitals play a critical role in introducing localized states near the Fermi energy.The doping of PbS with Yb introduces significant changes to the electronic properties, transitioning the material from a semiconductor to a metal. This metallic behavior, along with the enhanced electronic density near the Fermi energy, makes PbS:Yb\(^{3+}\) a promising candidate for applications requiring high charge carrier mobility, such as photocatalysis and energy storage systems. The analysis highlights the role of Yb doping in altering the structural and electronic properties effectively.

The 3D charge density map in Figure 8a shows electron localization around the Pb and S atoms, with Yb contributing a unique asymmetric distribution. The visualization highlights the role of Yb in modifying the electronic environment, likely enhancing the material’s reactivity. The pronounced charge density near Yb suggests significant charge transfer from Yb to neighboring atoms. The Bader analysis revealed significant charge transfer among the constituent atoms. Pb atoms displayed a net gain in charge, with values such as +0.7896 \, e (e.g., Pb1), indicating that they acted as electron acceptors, with electrons likely donated by Yb atoms. The Yb atoms exhibited positive charge transfer values of +1.1246 \, e and +1.1870 \, e, confirming their role as primary electron donors in the system. Conversely, the S atoms showed negative charge transfer, such as  -1.2350 \, e (e.g., S1), highlighting their role as electron acceptors and their contribution to stabilizing the electronic environment. The Bader volumes further illustrated the bonding and size characteristics of the atoms. Pb atoms had the largest Bader volumes, ranging from approximately 30 \AA\(^3\) to 40 \AA\(^3\), reflecting their larger atomic size and significant bonding characteristics within the lattice. In contrast, Yb atoms exhibited smaller Bader volumes, approximately 20 \AA\(^3\), indicating their localized influence and their role in facilitating charge transfer to other atoms within the system.

\begin{figure}[!ht]
    \centering
    \includegraphics[width=0.6\textwidth, angle=360]{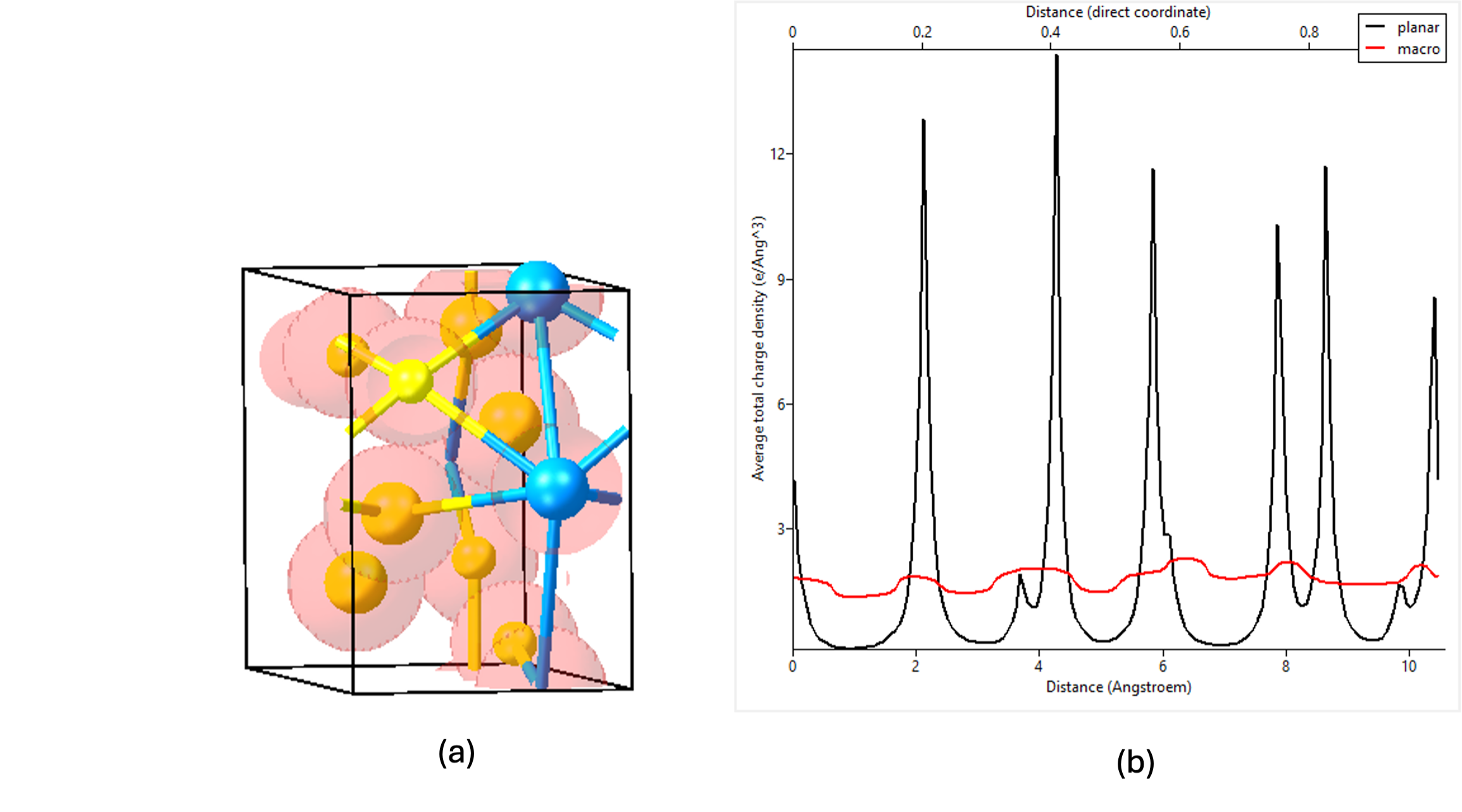}
    \caption{Illustrations of PbS:Yb (a) Charge density and (b) Bader charge}
    \label{fig8}
\end{figure}

This study included a detailed Bader charge analysis to evaluate charge transfer between atoms, highlighting the role of Yb doping in altering the charge distribution. In Figure 8b the planar (black) and macroscopic (red) average charge densities are plotted against the distance in Angstroms. Peaks in the planar charge density correspond to regions of high electron density near atomic planes. The macroscopic average smoothens these oscillations, revealing the overall charge distribution across the structure. The uniformity of the macro density suggests strong ionic bonding, with charge localization observed near the Pb, S, and Yb atoms. The charge density and Bader charge analysis reveal the critical impact of Yb doping on the charge distribution. Yb acts as a donor, transferring electrons to Pb and S, which alters the material’s electronic properties. This charge redistribution enhances the metallic nature ofPbS:Yb\(^{3+}\), improving its conductivity and potential for photocatalytic applications. The detailed density mapping confirms the structural integrity of the doped system.

\subsubsection{PbS:Yb-Optical properties}
The absorption coefficient ($\alpha$($\omega$)) as illustrated in Figure 9a highlights the material’s efficiency in absorbing electromagnetic radiation. In the low-energy region (below ~10 eV), initial peaks indicate efficient absorption within the visible spectrum, suitable for photoelectric applications. At higher energies (above ~10 eV), the absorption increases due to interband transitions and core-level excitations, confirming its potential for applications in solar cells and photocatalysis.

The dielectric function ($\epsilon$($\omega$)) in Figure 9b reveals insights into polarizability (\text{Re}($\epsilon$)) and energy dissipation (\text{Im}($\epsilon$)). Peaks in \text{Re}($\epsilon$) and \text{Im}($\epsilon$) align with energy absorption and conductivity data, confirming strong light-matter interactions and potential for energy storage and optoelectronics.

The optical conductivity ($\sigma$($\omega$)) captures the material’s response to oscillating electric fields in Figure 9c. Peaks in the real part (\text{Re}($\sigma$)) reflect efficient conduction under oscillating fields, while the imaginary part (\text{Im}($\sigma$)) represents energy storage during resonance conditions. These properties suggest enhanced charge transport, critical for catalytic and electronic applications.

Reflectivity (R($\omega$)) as shown in Figure 9d indicates photon-material interactions. Peaks signify strong interactions, while dips correspond to efficient absorption regions, making the material ideal for optical coatings and energy-efficient systems.

\begin{figure}[!ht]
    \centering
    \includegraphics[width=0.8\textwidth, angle=360]{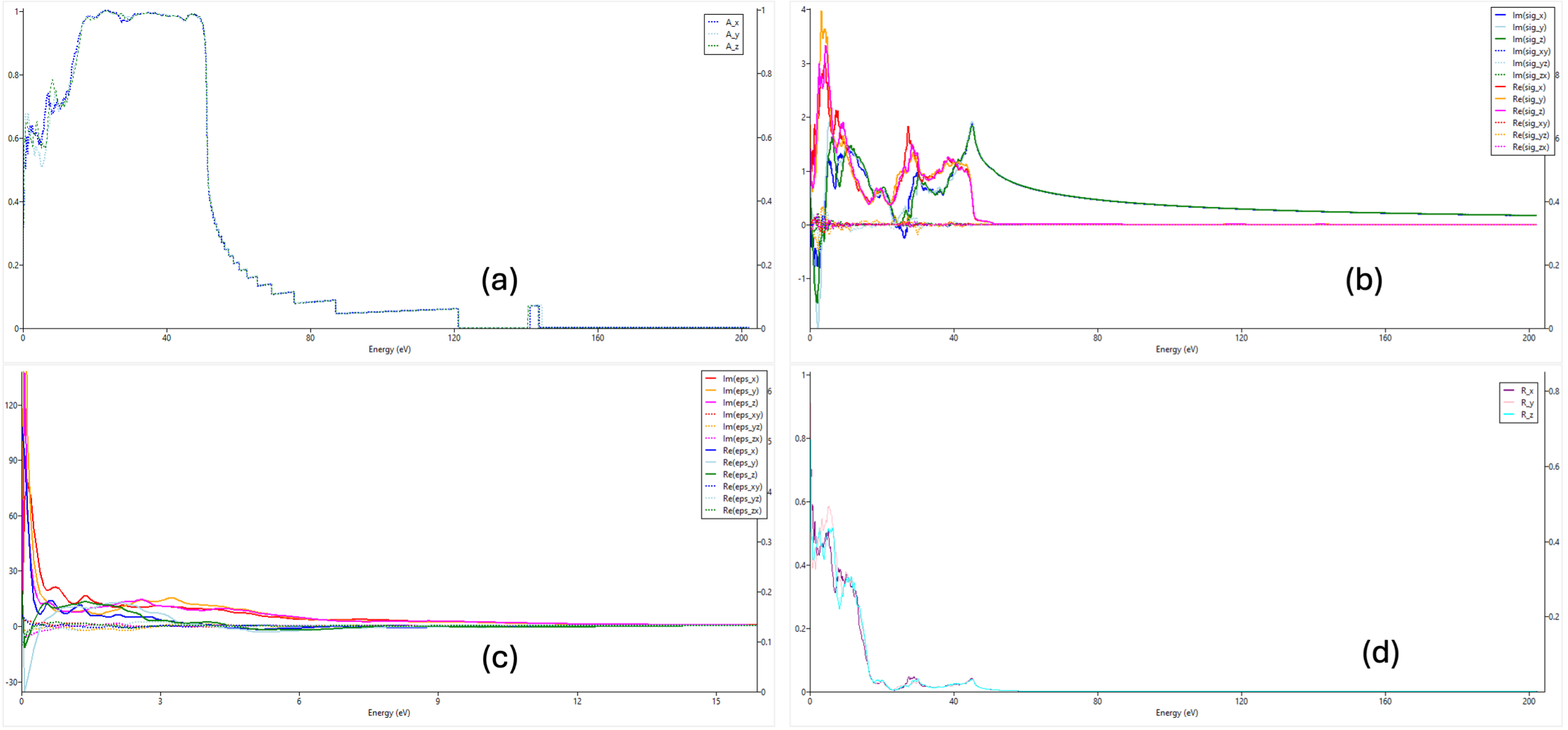}
    \caption{PbS:Yb optical response (a) optical absorption coefficient (b) dielectric Function, (c) optical conductivity, (d) reflectivity spectrum}
    \label{fig9}
\end{figure}

\subsection{PbS:Yb Doped Er}
The PbS:Yb\(^{3+}\),Er\(^{3+}\) system was constructed by co-doping ytterbium (Yb) and erbium (Er) into the PbS lattice. The geometry optimization and electronic property calculations in Figure 10 were performed using VASP with the GGA-PBE functional. A plane-wave cutoff of 500 eV was used, and a k-point grid of 5 $\times$ 3 $\times$ 3 was applied for optimization, refined to 7 $\times$ 5 $\times$ 3 for DOS calculations and 11 $\times$ 9 $\times$ 5 for band structure analysis.  The system was analyzed for structural stability, band structure, and electronic density of states. The doping introduced significant modifications to the electronic properties by introducing localized electronic states.

\begin{figure}[!ht]
    \centering
    \includegraphics[width=0.8\textwidth, angle=360]{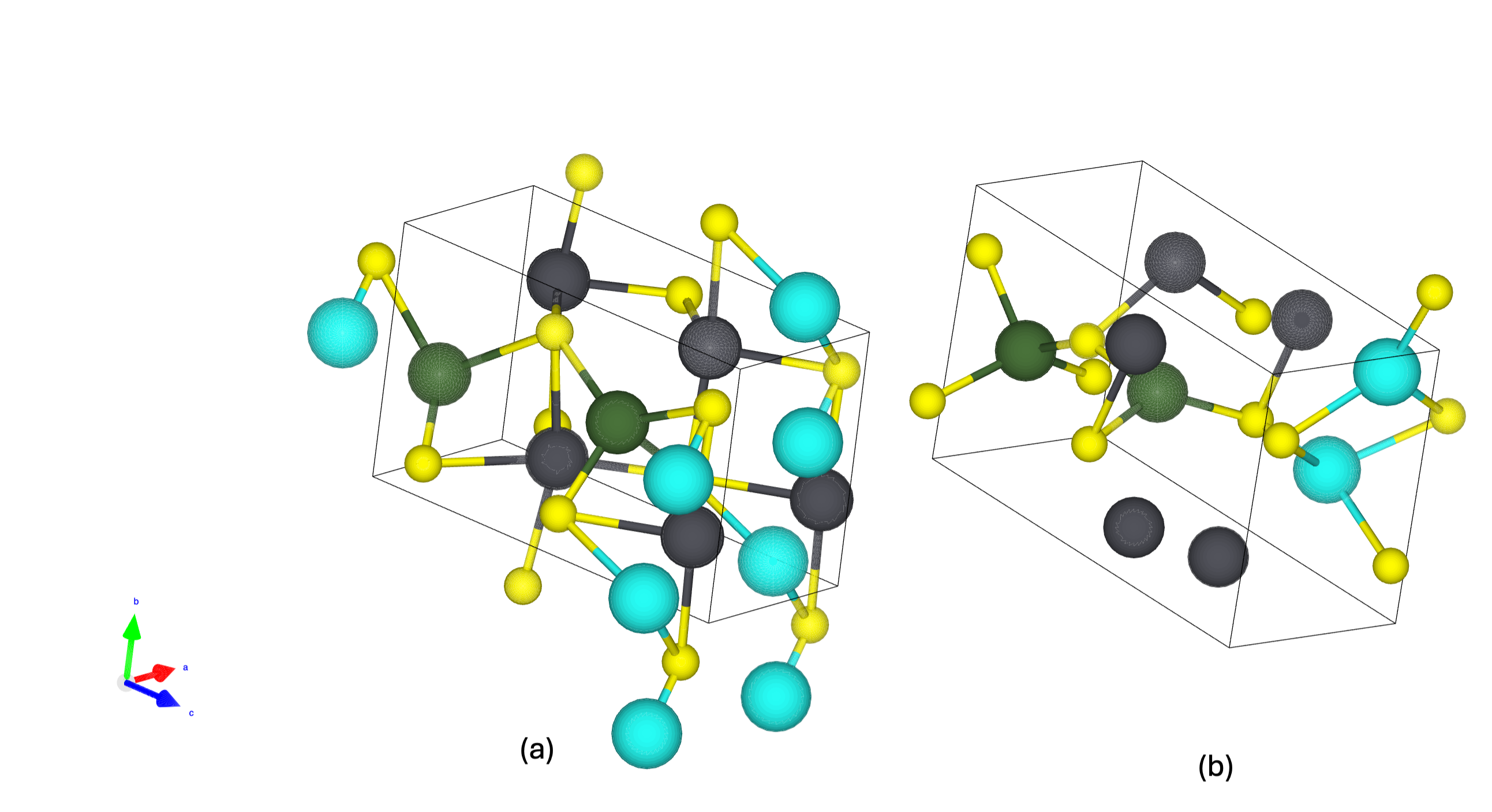}
    \caption{Schematic molecular structure of PbS:Yb,Er (a) initial structure (b) optimized structure}
    \label{fig10}
\end{figure}

\subsubsection{PbS:Yb,Er-Electronic properties}
The band structure of PbS:Yb\(^{3+}\),Er\(^{3+}\) in Figure 11a shows significant overlap of energy bands at the Fermi level, confirming a metallic nature. The Fermi energy intersects multiple bands, indicating the presence of free charge carriers. The introduction of Er, in addition to Yb, further broadens the energy bands, which can enhance electrical conductivity. The dispersion in the bands suggests good charge carrier mobility, making this system suitable for optoelectronic and photocatalytic applications. Whereas the DOS plot in Figure 11b reveals contributions from Pb s-orbitals and S p-orbitals in the valence band, while the conduction band is dominated by Yb and Er d-orbitals. The absence of a gap and the significant DOS at the Fermi level confirm the metallic nature of PbS:Yb\(^{3+}\),Er\(^{3+}\). The peaks near -6 eV and -18 eV correspond to bonding states from Pb and S interactions, while the states near the Fermi level arise from doping with Yb and Er. The co-doping of PbS with Yb and Er alters the electronic structure significantly, transitioning the material into a metallic state with high electrical conductivity. The enhanced DOS at the Fermi level and broadened energy bands suggest improved charge transport and optical absorption properties. This makes PbS:Yb\(^{3+}\),Er\(^{3+}\) a promising candidate for applications in photocatalysis and optoelectronics.

\begin{figure}[!ht]
    \centering
    \includegraphics[width=0.6\textwidth, angle=360]{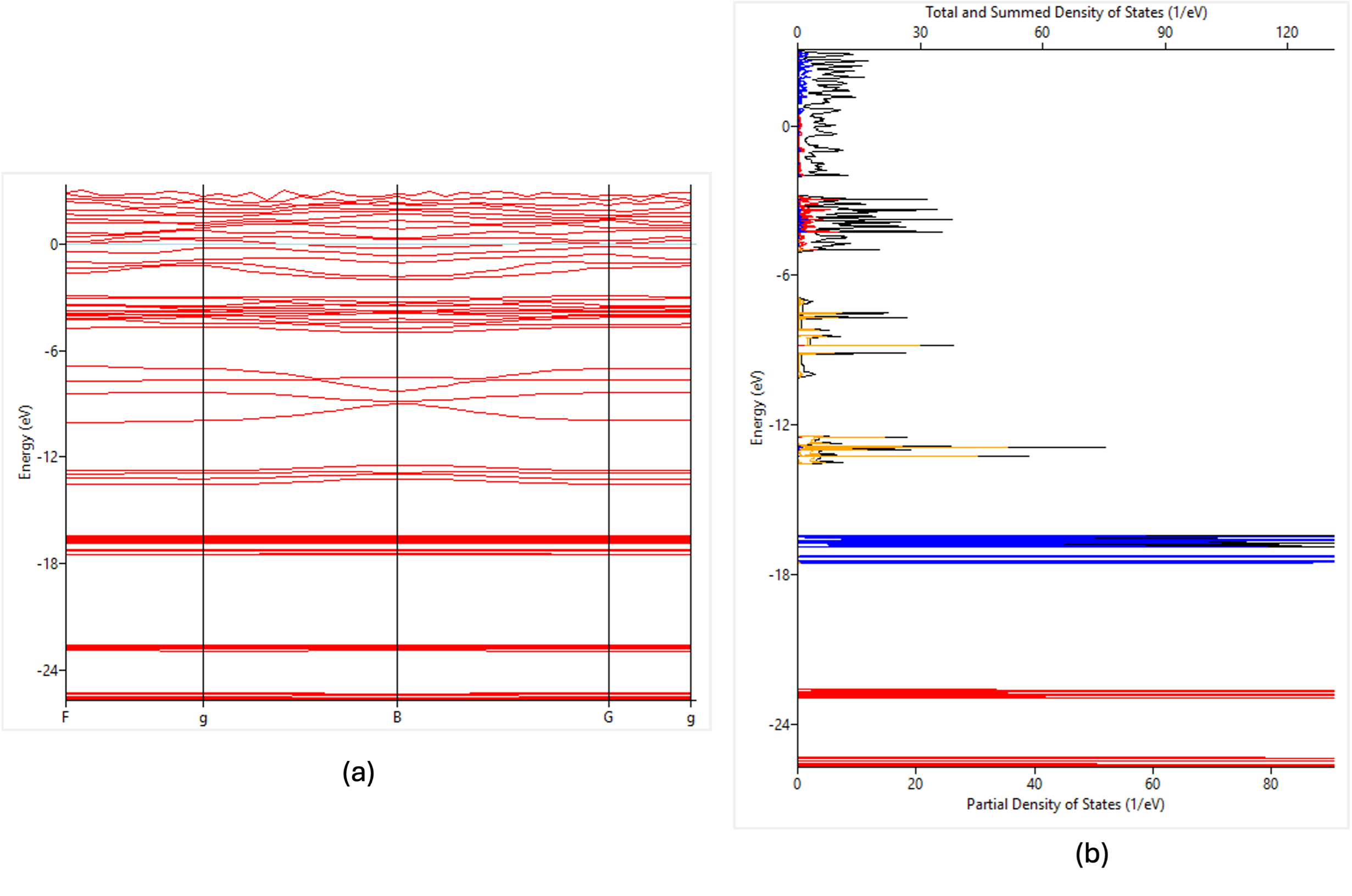}
    \caption{Illustrations of PbS:Yb,Er (a) bandstructure and (b) densities of state}
    \label{fig11}
\end{figure}

Also the co-doped PbS:Yb\(^{3+}\),Er\(^{3+}\) system was analyzed using VASP with Bader charge analysis to quantify charge transfer and electron localization. The density functional theory (DFT) method with the GGA-PBE functional was used, and the total charge density was calculated over a high-resolution Fourier grid of 40 $\times$ 50 $\times$ 84. The k-point mesh was refined to 5 $\times$ 3 $\times$ 3 to capture the intricate electronic interactions.

The 3D charge density map in Figure 12a provides valuable insights into the electron distribution within the doped PbS system. Yb and Er atoms exhibit significant charge accumulation around their immediate vicinity, reflecting their roles as electron donors. Pb and S atoms display complementary electron densities, forming a well-defined bonding network that preserves the structural integrity of the material. However, the Bader charge analysis reveals detailed charge transfer dynamics within the system. Pb atoms exhibit mixed behavior, with partial electron gain and loss ranging from +0.275 e to -0.534 e. Yb acts as a consistent donor, contributing charges between +1.153 e and +1.166 e. Er atoms exhibit significant charge transfer (+1.615 e to +1.731 e), highlighting their strong interaction with the PbS matrix. S atoms function as electron acceptors, with charge transfer values ranging from -1.153 e to -1.326 e, stabilizing the overall structure. In terms of Bader volumes, Pb atoms exhibit relatively large volumes (~30-40 \AA\(^3\)), indicative of delocalized bonding within the system. Conversely, Yb and Er atoms display smaller volumes (~18-20 \AA\(^3\)), reflecting their localized charge density contributions and their role in modifying the electronic environment of the doped material.

In similar computation event, the planar (black) and macroscopic (red) charge density profiles in Figure 12b reveal significant charge localization near the Pb and S atomic planes. The peaks in the planar charge density correlate to regions with high electronic density, while the macroscopic profile averages these fluctuations, showing an overall uniform distribution. The incorporation of Yb and Er introduces localized peaks, indicating enhanced electron density in regions associated with the dopants.

\begin{figure}[!ht]
    \centering
    \includegraphics[width=0.6\textwidth, angle=360]{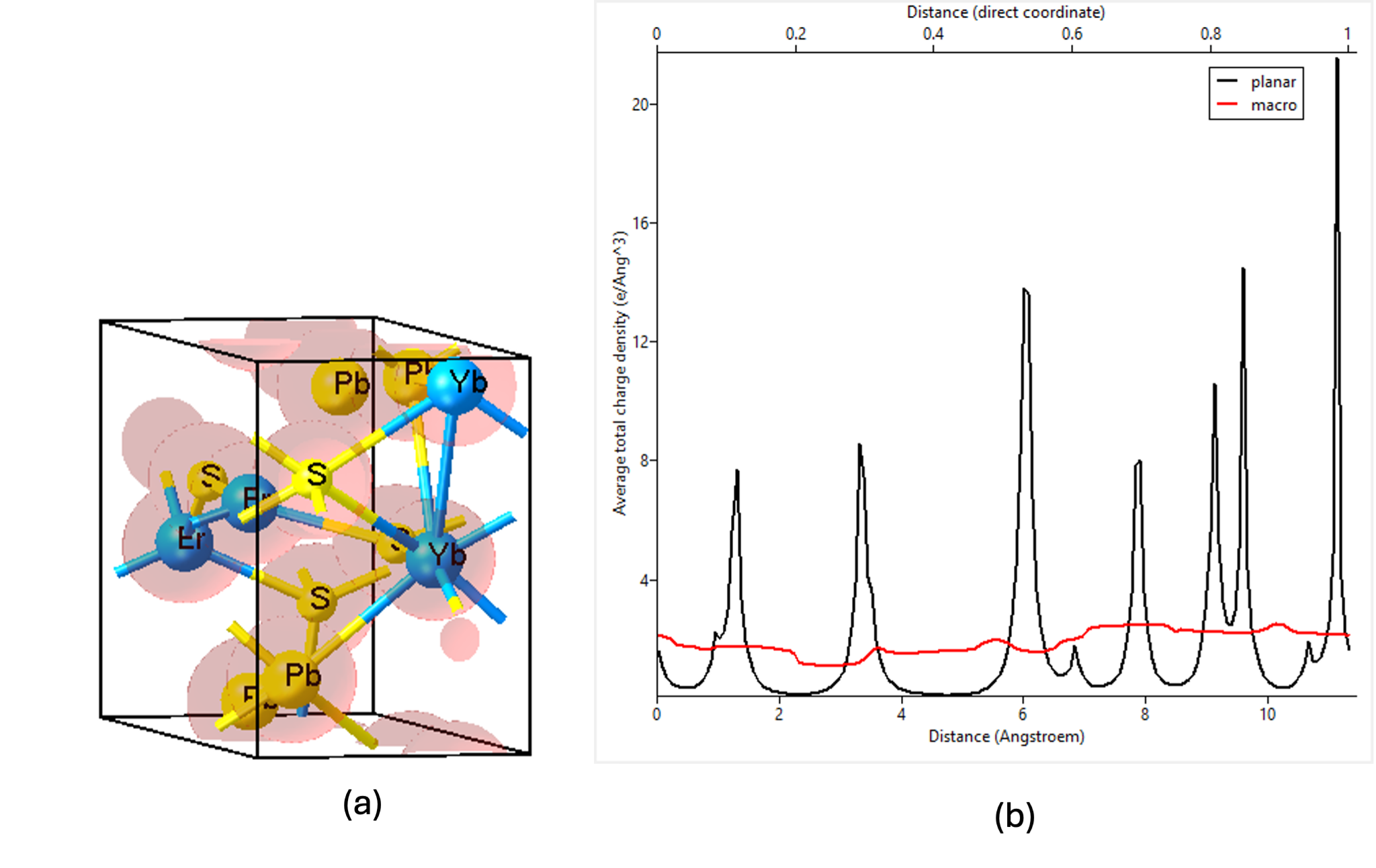}
    \caption{Illustrations of PbS:Yb,Er (a) Charge density and (b) Bader charge}
    \label{fig12}
\end{figure}

The charge density and Bader analysis confirm that Yb and Er co-doping enhances electron localization and charge transfer, significantly modifying the electronic properties of the PbS:Yb\(^{3+}\),Er\(^{3+}\) system. These results validate the structural stability and suitability of the doped system for applications requiring enhanced electronic conductivity, such as photocatalysis.

\subsubsection{PbS:Yb,Er-Optical properties}
The absorption coefficient in Figure 13a reveals maximum absorption at 2.4 eV, with a coefficient of 0.9 eV\(^{-1}\). Absorption decreases beyond 5 eV, indicating strong activity in the visible and near-UV ranges. These absorption characteristics make the material a promising candidate for use in LEDs and photovoltaic devices.

The dielectric function in Figure 13b of thePbS:Yb\(^{3+}\),Er\(^{3+}\) material highlights the real (\text{Re}($\epsilon$)) and imaginary (\text{Im}($\epsilon$)) parts as functions of energy. \text{Im}($\epsilon$) peaks at 2.1 eV, indicating strong optical absorption due to interband transitions, while \text{Re}($\epsilon$) reaches a maximum value of 12.5 at 1.8 eV, suggesting a high refractive index in this range. Both components decline beyond 5 eV, reflecting reduced photon interaction at higher energies. These properties make the material ideal for optoelectronic applications, particularly in the visible spectrum.

The optical conductivity in Figure 13c demonstrates the material’s ability to conduct optically-induced charges. \text{Im}($\sigma$) peaks at 3.2 eV, signifying increased conductivity due to interband transitions, while \text{Re}($\sigma$) achieves a maximum of 1.8 ($\omega$ $\cdot$ \text{cm})\(^{-1}\) at 1.5 eV. Conductivity stabilizes and diminishes beyond 15 eV, suggesting limited optical response at higher photon energies. This behavior indicates potential applications in infrared and visible photodetectors and solar cells.

The reflectivity spectrum in Figure 13d starts at ~40\% at 0.5 eV, sharply declines beyond 1 eV, and stabilizes near zero at higher energies. This low reflectivity at elevated photon energies enhances the material’s efficiency in applications requiring minimal energy loss, such as optical and photovoltaic devices.

\begin{figure}[!ht]
    \centering
    \includegraphics[width=0.8\textwidth, angle=360]{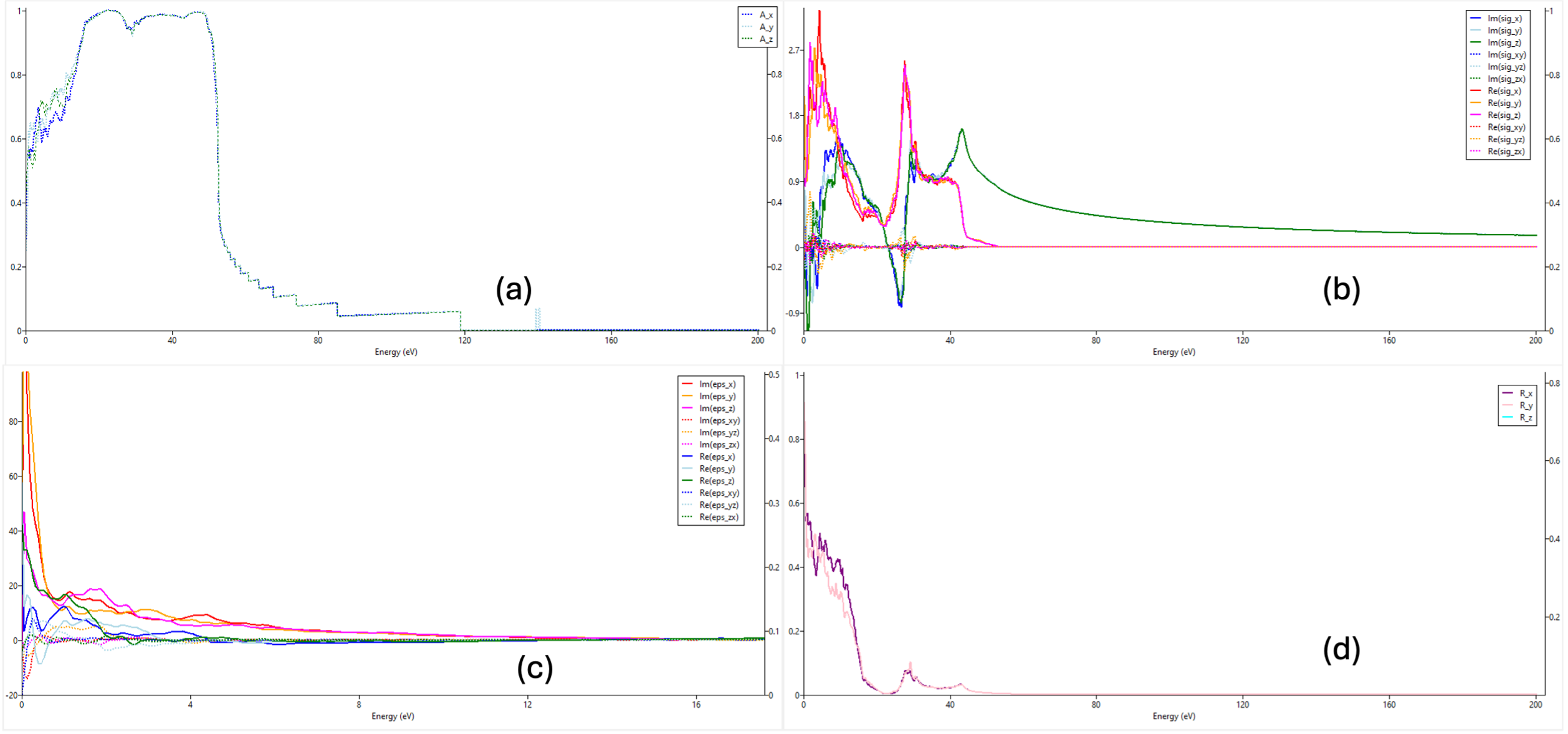}
    \caption{PbS:Yb optical response (a) optical absorption coefficient (b) dielectric Function, (c) optical conductivity, (d) reflectivity spectrum}
    \label{fig13}
\end{figure}

\subsection{Bulk CuBiO}
The bulk structure of CuBiO was modeled and optimized using VASP. Initial lattice parameters were derived from experimental data from materials project, representing the orthorhombic phase of CuBiO. The geometry optimization process in Figure 14 included both atomic positions and unit cell dimensions to achieve structural stability and minimal energy configuration. In avoidance in biasness, same calculation parameter were used for this material. The exchange-correlation interactions were described using the GGA-PBE functional. A plane-wave cutoff energy of 500 eV was applied for high accuracy. The reciprocal space was sampled with a gamma-centered k-point mesh of  5 $\times$ 7 $\times$ 5 , corresponding to k-spacings of approximately  0.232 \, \AA\(^{-1}\) ,  0.251 \, \AA\(^{-1}\) , and  0.212 \, \AA\(^{-1}\) along a, b, and c axes, respectively. Methfessel-Paxton smearing with a width of 0.2 eV was used to handle partial occupancies during electronic convergence. The relaxation process targeted a total energy convergence of 10\(^{-5}\) \, \text{eV} and forces on each atom below 10\(^{-3}\) \, \text{eV/\AA}. The optimization involved 36 steps, leading to a minimized total energy of -42.266 \, \text{eV} for the Cu\(_2\)Bi\(_2\)O\(_4\) unit cell.

\begin{figure}[!ht]
    \centering
    \includegraphics[width=0.8\textwidth, angle=360]{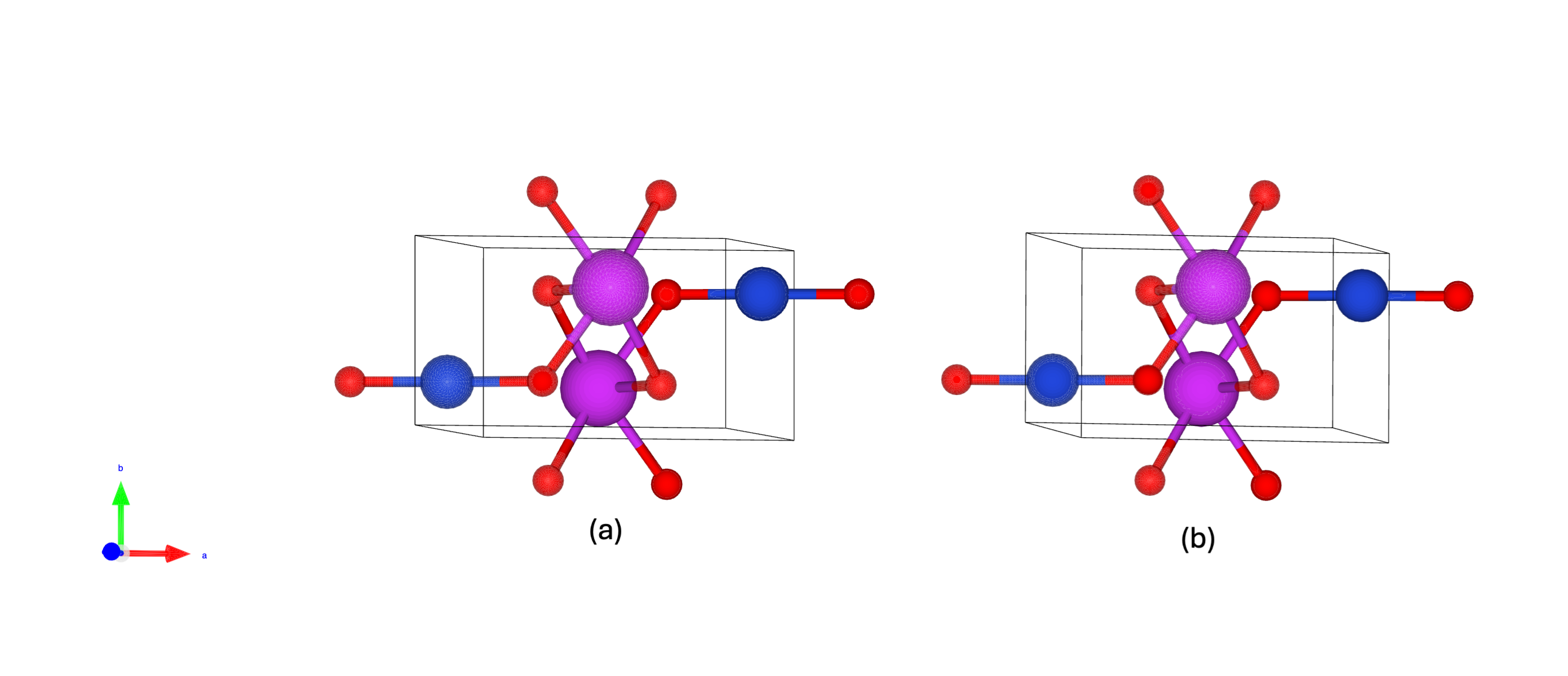}
    \caption{Schematic molecular structure of bulk CuBiO (a) initial structure (b) optimized structure}
    \label{fig14}
\end{figure}

Based on the calculated structural parameters, density, and electronic properties, CuBiO in this study qualifies as a bulk material. The periodic boundary conditions and optimization of a 3D unit cell ensure that the system reflects bulk-like characteristics without surface or nanoscale effects.

\subsubsection{CuBiO-Electronic properties}
The optimized CuBiO structure exhibited slight deviations from experimental lattice parameters where a = 6.177 \, \text{\AA}, b = 3.637 \, \text{\AA}, and c = 6.753 \, \AA, with a unit cell volume of 135.50 \, \AA\(^3\). An indirect bandgap of 1.035 eV was identified, with the valence band maximum (VBM) near (-0.40, 0.40, 0.33) and the conduction band minimum (CBM) near (0.40, 0.00, -0.00). However, the optimized lattice parameters were a = 6.177 \, \text{\AA}, b = 3.637 \, \text{\AA}, and c = 6.753 \, \text{\AA}, with a cell volume of 135.50 \, \text{\AA}\(^3\). Bader analysis revealed significant charge transfer; Cu: Net charge of +0.513 \, e per atom. Bi: Net charge of +1.779 \, e per atom. O: Net charge of -1.143 \, e per atom on average.

The band structure in Figure 15a shows CuBiO as a semiconductor with an indirect bandgap of 1.035 eV. The valence band maximum (VBM) lies slightly below the Fermi level at -0.039 eV, while the conduction band minimum (CBM) starts at 0.996 eV. This band alignment indicates efficient charge separation, essential for photocatalytic activity. The densely packed bands in the valence region reflect significant orbital interactions, dominated by contributions from Bi and O atoms. The DOS plot in Figure 15b highlights the electronic distribution across energy levels. The valence band consists primarily of Cu d-orbitals and O p-orbitals, reflecting strong metal-oxygen bonding. The conduction band is dominated by Bi s-orbitals, contributing to the indirect nature of the bandgap. The Fermi level at 0 eV is positioned within the gap, confirming the semiconducting nature of CuBiO. These results indicate that bulk CuBiO exhibits suitable electronic properties for photocatalytic applications, particularly when combined with complementary materials like PbS. The indirect bandgap and orbital contributions align well with theoretical expectations, validating the computational approach. 

\begin{figure}[!ht]
    \centering
    \includegraphics[width=0.6\textwidth, angle=360]{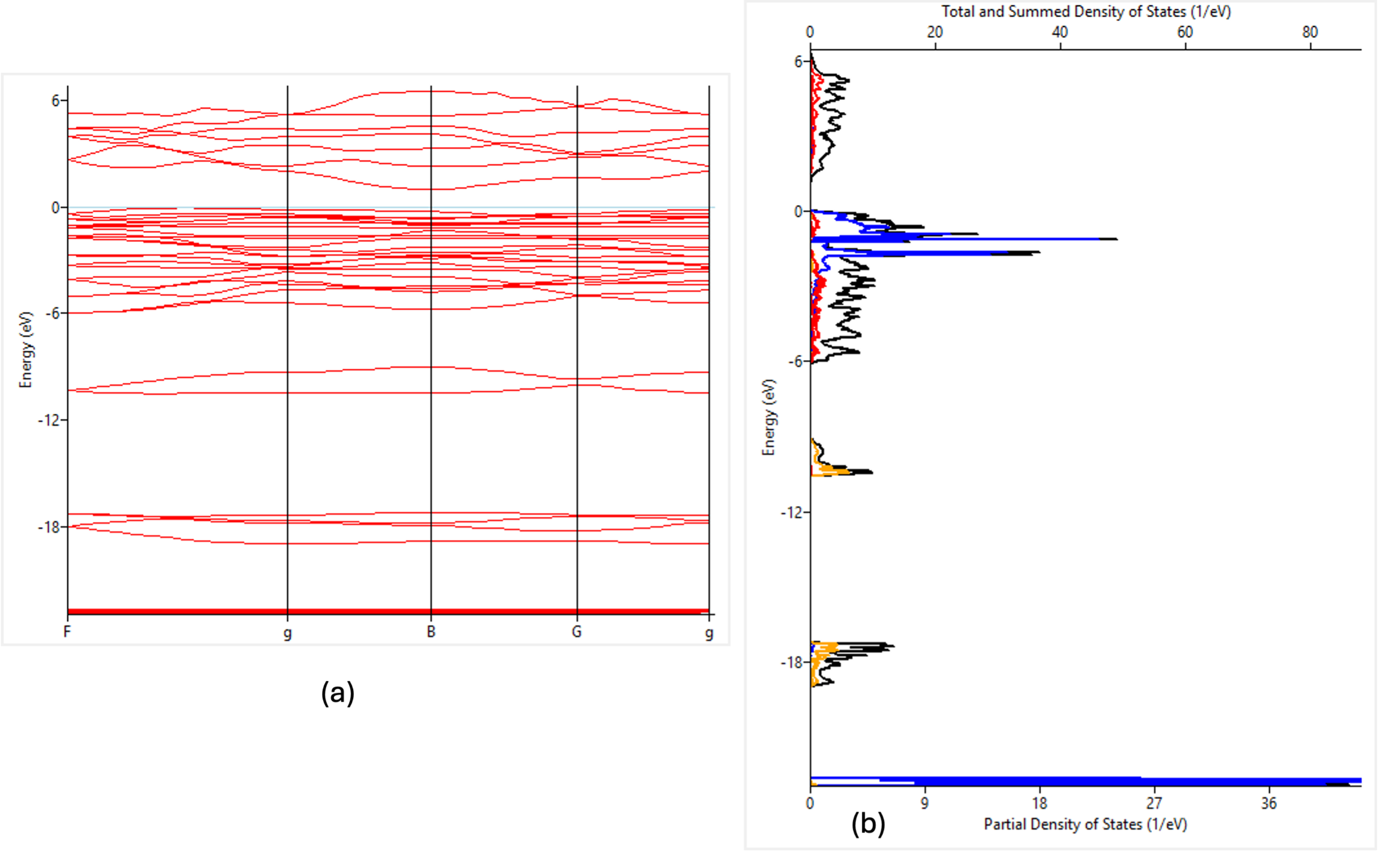}
    \caption{CuBiO (a) bandstructure and (b) densities of state}
    \label{fig15}
\end{figure}

The system was however assumed to be non-magnetic, with a Gamma-centered 3 $\times$ 3 $\times$ 1 k-Point mesh, which was then refined to 7 $\times$ 9 $\times$ 7 for charge density analysis. The electronic convergence was set to 10\(^{-5}\) eV. Bader charge analysis was conducted to assess charge transfer and bonding properties. The charge density was calculated on a high-resolution 48 $\times$ 28 $\times$ 50 Fourier grid.

This geometry was optimized to a total energy of -42.278 \, \text{eV} per unit cell with minimal forces, confirming structural stability. Atomic partial charges were calculated, and bonding characteristics were assessed through charge density visualization, where Figure 16a shows the total charge density distribution within the CuBiO unit cell. The red regions represent high electron density localized around oxygen atoms, while lower density regions near Cu and Bi indicate partial ionic bonding. The structure confirms strong Cu-O and Bi-O interactions. Also, the black curve in Figure 16b (planar average) highlights periodic peaks corresponding to electron density in atomic planes. The red curve (macroscopic average) smooths these variations, revealing overall charge trends. Peaks near 2 \AA, 4 \AA, and 6 \AA correspond to high-density Cu and O planes, confirming charge localization within the lattice.

\begin{figure}[!ht]
    \centering
    \includegraphics[width=0.6\textwidth, angle=360]{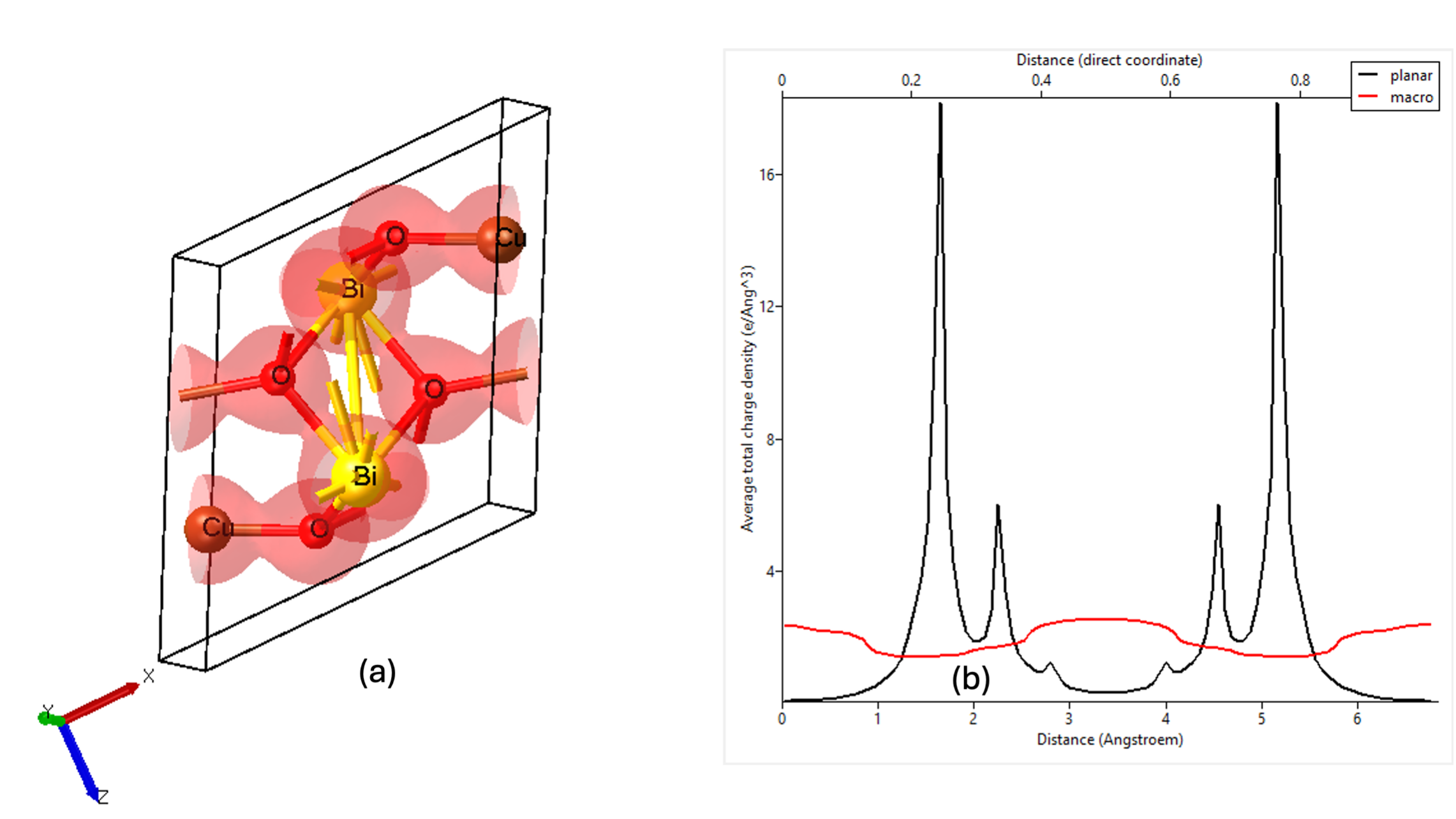}
    \caption{CuBiO (a) Charge density visualization (b) Planar and macroscopic charge density}
    \label{fig16}
\end{figure}

These resulting calculations validates bulk CuBiO as a suitable semiconducting material with well-defined electronic and optical properties, laying the foundation for further heterostructure studies.

\subsubsection{CuBiO-Optical properties}
The optical properties of CuBiO uses its initial optimized geometry structure and this was conducted using Density Functional Theory (DFT) implemented in VASP (Vienna Ab-initio Simulation Package). 

The absorption coefficient (Ax, Ay, Az) in Figure 17a is plotted versus energy. The sharp increase in absorption around 3 eV indicates the onset of strong optical absorption, corresponding to the bandgap transitions. CuBiO exhibits good absorption in the visible to UV spectrum, making it useful for photovoltaic and photocatalytic applications. In Figure 17b the dielectric function plot shows the real and imaginary parts of the dielectric constant as a function of energy. The imaginary part (Im(eps)) represents the optical absorption due to interband transitions, and the real part (Re(eps)) relates to the material’s ability to polarize and store energy. Peaks in the imaginary part correlate to strong electronic transitions. For CuBiO, prominent peaks below 10 eV suggest significant interband transitions in the visible and UV regions. This Figure 17c illustrates the real and imaginary parts of the optical conductivity. Peaks in the real part indicate frequencies where the material has high electron mobility due to photon absorption. Imaginary parts provide insight into the energy dissipation and polarization. Same plot shows the reflectivity spectrum (Rx, Ry, Rz) in Figure 17d. Reflectivity increases in the UV range (10–20 eV), indicating that CuBiO reflects more light in this range, making it suitable for UV-based applications. The low reflectivity at lower energy (visible range) suggests higher transparency, beneficial for optoelectronic devices.

\begin{figure}[!ht]
    \centering
    \includegraphics[width=0.8\textwidth, angle=360]{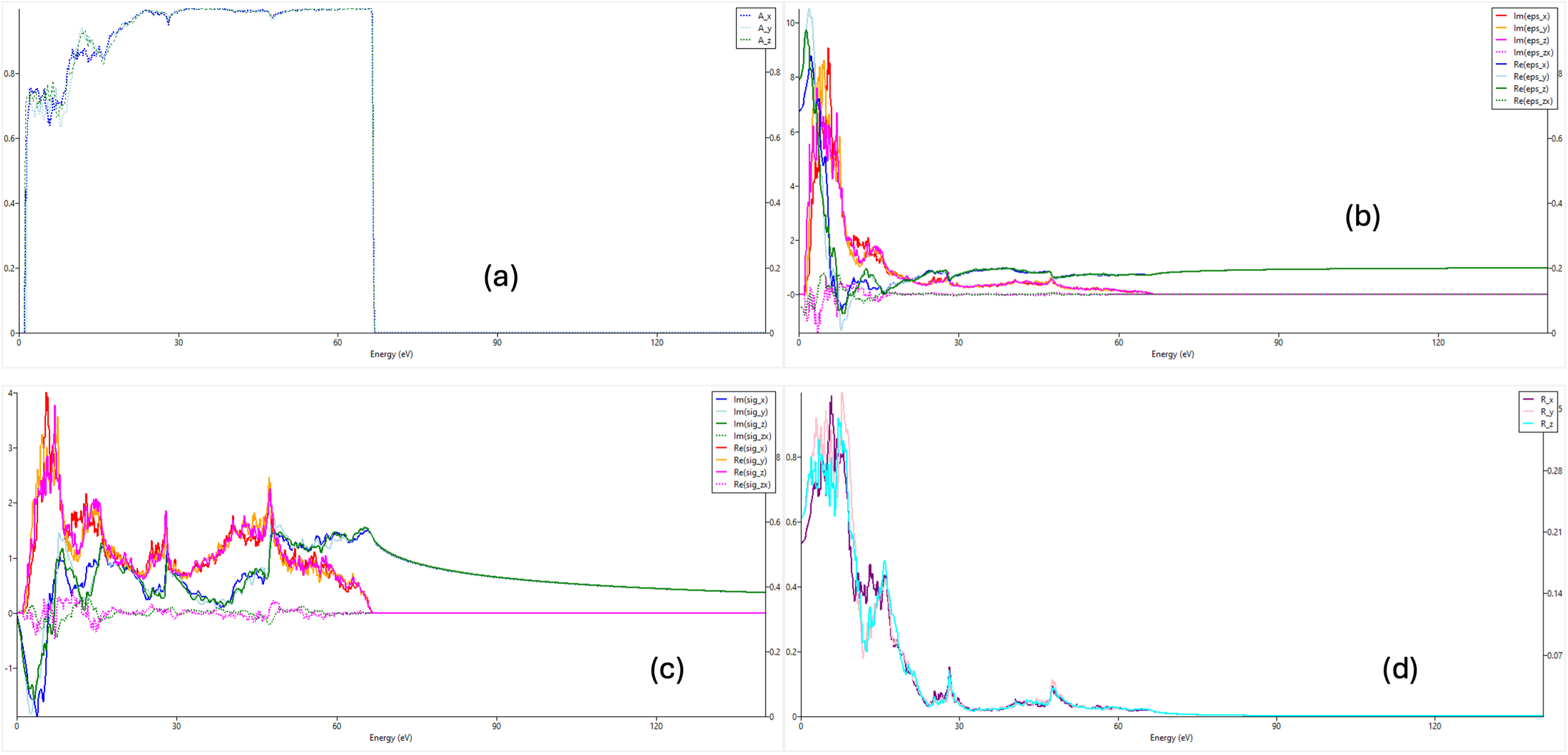}
    \caption{Bulk CuBiO optical responses (a) optical absorption coefficient (b) dielectric Function, (c) optical conductivity, (d) reflectivity spectrum}
    \label{fig17}
\end{figure}

The material has an indirect bandgap of 0.933 eV, indicating its semiconducting characteristics. The optical properties of CuBiO make it a promising candidate for use in solar cells, transparent electrodes, and devices sensitive to ultraviolet light.

\subsection{Building PbS:Yb\(^{3+}\),Er\(^{3+}\)/CuBiO (QDsUC) Heterostructure}
The computational optimization of the heterostructure PbS:Yb\(^{3+}\),Er\(^{3+}\)/CuBiO was performed using density functional theory (DFT) as implemented in the Vienna Ab-initio Simulation Package (VASP). The Perdew-Burke-Ernzerhof (PBE) exchange-correlation functional was again utilized within the generalized gradient approximation (GGA). Projector augmented-wave (PAW) pseudopotentials were employed for all elements present in Figure 18, ensuring an accurate description of the electronic interactions. The energy cutoff was set at 500 eV to balance computational cost and accuracy. The k-point mesh was centered on the gamma point with a density of 5×5×5 for DOS and 5×3×3 for the band structure calculations, with a smearing width of 0.2 eV using the Methfessel-Paxton method.

\begin{figure}[!ht]
    \centering
    \includegraphics[width=0.8\textwidth, angle=360]{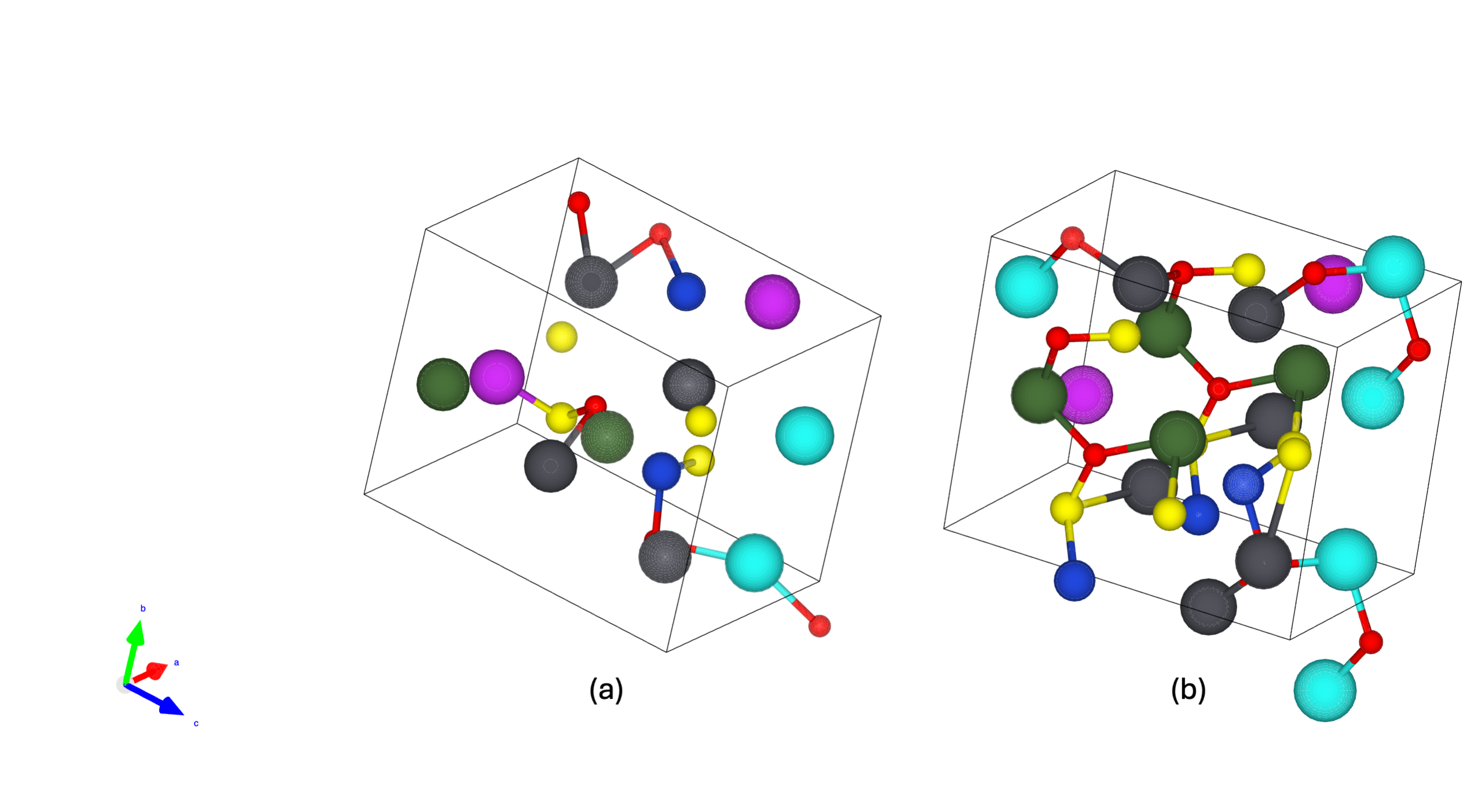}
    \caption{Schematic molecular structure of PbS:Yb,Er/CuBiO heterostructure (a) initial structure (b) optimized structure}
    \label{fig18}
\end{figure}

The electronic structure was analyzed by computing the total density of states (DOS) and the band structure. The structural relaxation achieved convergence at a force threshold of 0.01 eV/\AA, while the electronic iterations were converged to 10\(^{-5}\) eV. To explore the doping effects, fractional substitutions of Yb and Er were introduced into the PbS host matrix. The heterostructure was formed by coupling PbS:Yb\(^{3+}\),Er\(^{3+}\) with CuBiO, incorporating lattice matching constraints to ensure stability.

\subsubsection{QDsUC- Electronic properties}
The band structure of PbS:Yb\(^{3+}\),Er\(^{3+}\)/CuBiO reveals a semiconductor nature with an indirect band gap of 0.352 eV. The valence band maximum (VBM) is located near (0.00, 0.20, -0.40), while the conduction band minimum (CBM) resides at (0.40, 0.40, 0.40). This band alignment facilitates efficient charge carrier transport, which is crucial for photocatalytic water splitting. The DOS demonstrates significant contributions from Pb, Yb, and Er atoms near the Fermi level, highlighting their active roles in electronic transitions. The partial density of states reveals hybridization between the d-states of Yb and Er with the p-states of S and O, indicative of strong covalent interactions.

The band structure analysis in Figure 19a provides deeper insights into the electronic transitions and carrier dynamics of the PbS:Yb\(^{3+}\),Er\(^{3+}\)/CuBiO heterostructure. The separation between the valence band maximum (VBM) and the conduction band minimum (CBM) indicates an indirect band gap of approximately 0.352 eV. This small band gap supports visible-light absorption, a critical property for photocatalytic applications. Furthermore, the high dispersion in the conduction band signifies strong carrier mobility, essential for efficient charge transport and reduced recombination losses. Flat bands observed near the Fermi level suggest the presence of localized states introduced by Yb and Er doping. These states may act as charge carrier traps, influencing recombination dynamics and the material’s overall electronic efficiency. Multiple bands intersect the Fermi level, reinforcing the metallic behavior of the heterostructure. This characteristic highlights its potential for enhanced electron conduction, which is vital for applications requiring efficient energy and charge transfer processes. 

The DOS analysis of the PbS:Yb\(^{3+}\),Er\(^{3+}\)/CuBiO heterostructure in Figure 19b reveals significant electronic activity across the energy spectrum, providing insights into its electronic behavior. The valence bands, situated below the Fermi level (0 eV), are predominantly contributed by the O p-orbitals, with substantial hybridization involving the p-orbitals of S and O and the s-orbitals of Pb. Notable peaks in the range of -6 eV to -12 eV highlight bonding interactions driven primarily by Pb, Yb, and S, reflecting the structural and electronic bonding integrity of the heterostructure. The conduction bands, positioned above the Fermi level, show significant contributions from the d-orbitals of Yb and Er, underscoring their active roles in electronic transitions. Peaks within the conduction band indicate delocalized states, enhancing the heterostructure’s charge transport properties and facilitating potential applications in energy conversion processes. A high density of states near the Fermi level confirms metallic-like behavior, making the heterostructure a promising candidate for applications requiring efficient charge carrier generation and separation, such as photocatalytic water splitting.

\begin{figure}[!ht]
    \centering
    \includegraphics[width=0.6\textwidth, angle=360]{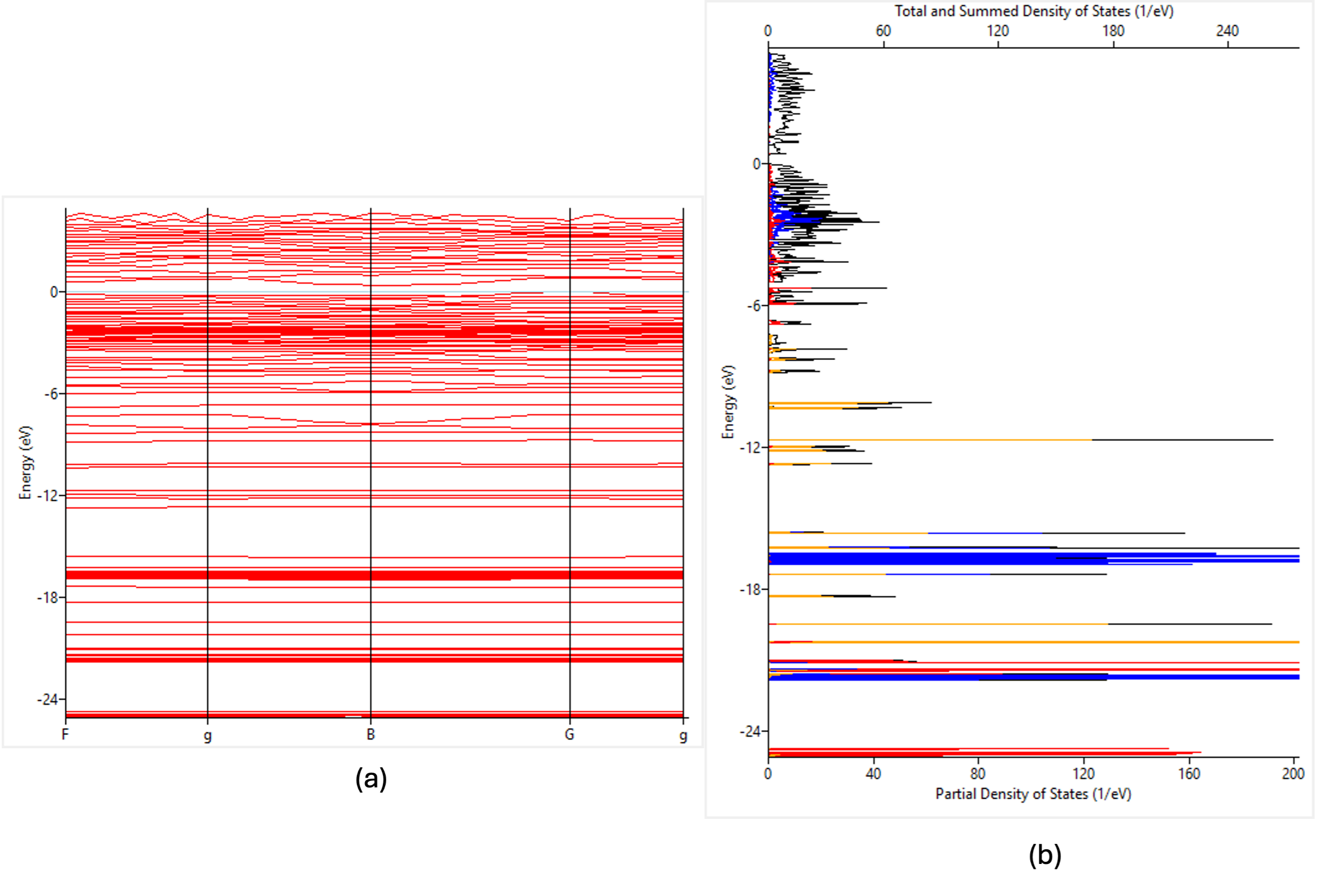}
    \caption{Illustrations of PbS:Yb\(^{3+}\),Er\(^{3+}\)/CuBiO (a) bandstructure and (b) densities of state}
    \label{fig19}
\end{figure}

The PbS:Yb\(^{3+}\),Er\(^{3+}\)/CuBiO heterostructure exhibits a unique electronic structure characterized by hybridized valence and conduction bands, metallic conductivity, and a narrow band gap. These properties make it highly suitable for photocatalytic water splitting, as it ensures efficient light absorption, charge separation, and carrier mobility. The presence of localized states due to Yb and Er doping provides opportunities to fine-tune the material’s electronic properties for specific applications. 

The total charge density and Bader charge analysis for the PbS:Yb\(^{3+}\),Er\(^{3+}\)/CuBiO heterostructure were conducted using VASP version 6, under the GGA-PBE exchange-correlation functional. The simulation employed a single-point calculation with a cutoff energy of 500 eV. A k-spacing of 0.35 \AA\(^{-1}\) resulted in a 3×3×3 k-mesh, ensuring adequate sampling of the Brillouin zone. The Methfessel-Paxton smearing method with a 0.2 eV width was applied for electronic convergence. The total charge density was computed on a Fourier grid of 50×60×70. Bader analysis was then performed to evaluate the electronic charge distribution and identify charge transfer trends.

The PbS:Yb\(^{3+}\),Er\(^{3+}\)/CuBiO heterostructure exhibited significant charge redistribution, as detailed by the Bader charge analysis. Yb atoms demonstrated a charge transfer of approximately +1.24 electrons per atom, confirming their role as primary electron donors. Er atoms exhibited slightly higher charge transfer values, around +1.76 electrons, enhancing the electronic properties of the heterostructure. Pb and Bi atoms primarily acted as charge acceptors, with Bi showing notable negative charge transfer ranging from -0.54 to -0.89 electrons, indicating its strong interaction within the structure. Oxygen atoms in the CuBiO component displayed substantial negative charge transfer, ranging from -1.12 to -1.20 electrons, consistent with their high electronegativity and role in stabilizing the heterostructure. Sulfur atoms in the PbS:Yb\(^{3+}\),Er\(^{3+}\) portion exhibited charge transfer values between -0.27 and -1.18 electrons, further supporting their function in maintaining the overall structural and electronic balance. The heterostructure features an indirect band gap of 0.431 eV, with the valence band maximum (VBM) and conduction band minimum (CBM) located at distinct points in the Brillouin zone. This narrow band gap enhances the heterostructure’s photocatalytic efficiency, particularly under narrow-band illumination in visible-light applications.

The Bader volumes indicate that Yb, Er, and Cu atoms occupy smaller regions, approximately 17–19 \AA\(^3\), reflecting their localized influence on charge redistribution. In contrast, Pb and Bi atoms, with larger atomic sizes, extend over greater volumes, ranging from 29–47 \AA\(^3\), which aligns with their bonding characteristics and significant contributions to the material’s structural framework. These results highlight the intricate interplay of charge transfer and distribution in the heterostructure, emphasizing its potential for enhanced electronic and photocatalytic applications.

The charge density plot highlights the localized charge distribution across the heterostructure. Peaks in planar charge density correspond to bonding between Pb, S, and Cu, while macro-scale uniformity in the macro density suggests effective charge coupling. The total charge density visualization in Figure 20 indicates strong electron localization around Bi, Cu, and O sites, reflecting the heterostructure’s complex bonding environment and its potential for efficient light absorption and carrier separation.

\begin{figure}[!ht]
    \centering
    \includegraphics[width=0.4\textwidth, angle=360]{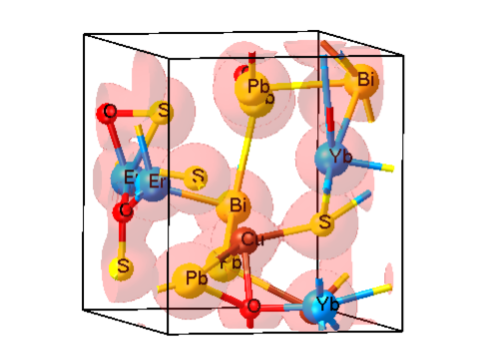}
    \caption{Illustrations of PbS:Yb\(^{3+}\),Er\(^{3+}\)/CuBiO Charge density}
    \label{fig20}
\end{figure}

These findings collectively underscore the heterostructure’s suitability for photocatalytic applications, particularly in water splitting, due to its favorable electronic properties and efficient charge redistribution. Further studies on photon absorption rates and band alignment will provide additional insights into its full catalytic potential.

\subsubsection{QDsUC- Optical properties}
This QDsUC heterostructure is of a greater concern, since the study objects to finding its overall performance, we dive into observing its detailed optical analysis. The optical spectra calculations were performed using the VASP software within the Generalized Gradient Approximation (GGA-PBE) framework for exchange-correlation interactions. The computations utilized a plane-wave cutoff energy of 500 eV and a k-point mesh density of 5 × 5 × 5, ensuring convergence. A Gaussian smearing method with a width of 0.05 eV was employed to smoothen the density of states and optical spectra data. Additionally, the dielectric function was computed by projecting onto spherical harmonics, incorporating site and angular-momentum-resolved contributions. The system was characterized using 3000 energy grid points up to 205.4519 eV, providing high-resolution results for reflectivity, conductivity, and absorption spectra. The heterostructure displayed prominent optical properties indicative of its applicability in photocatalysis. The absorption spectrum peaked in the visible range, with a noticeable \textbf{edge near 0.431 eV}, corresponding to the indirect band gap. The conductivity showed high values in the low-energy regime, reflecting strong photon absorption and charge transport capabilities. Reflectivity analysis revealed minimal losses, supporting efficient light capture. The dielectric function exhibited pronounced peaks around the resonance frequencies, confirming the material’s capability to support electronic transitions. Together, these results emphasize the potential of the PbS:Yb\(^{3+}\),Er\(^{3+}\)/CuBiO heterostructure for applications in visible-light-driven water splitting and solar energy conversion systems.

This absorption spectrum of the PbS:Yb\(^{3+}\),Er\(^{3+}\)/CuBiO heterostructure in Figure 21 reveals significant optical behavior across the energy range. The absorption coefficient, represented along the axes $A_x$, $A_y$, and $A_z$, shows strong absorption in the visible region (below 40 eV), which aligns with the heterostructure’s indirect band gap of approximately 0.431 eV. The steep increase in absorption near low energy suggests its suitability for visible-light-driven photocatalysis. The absorption reduces beyond 40 eV, indicating minimal contribution from higher-energy transitions. The uniformity of the $A_x$, $A_y$, and $A_z$ components confirms isotropic absorption behavior, which is desirable for applications such as solar energy harvesting and water splitting.

\begin{figure}[!ht]
    \centering
    \includegraphics[width=0.8\textwidth, angle=360]{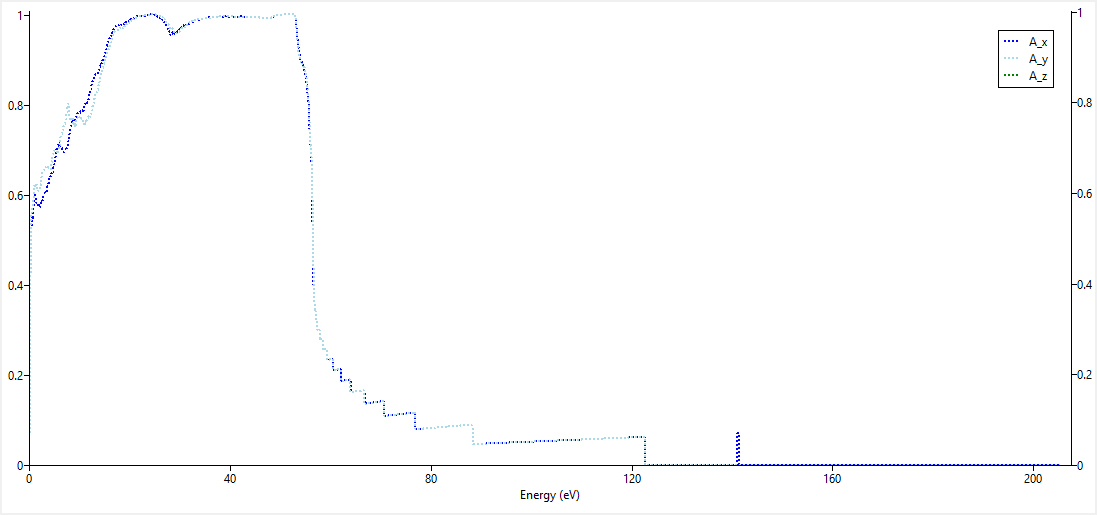}
    \caption{QDsUC optical responses to absorption coefficient}
    \label{fig21}
\end{figure}

The dielectric function of the PbS:Yb\(^{3+}\),Er\(^{3+}\)/CuBiO heterostructure in Figure 22 reveals key optical properties. The imaginary part (Im[$\epsilon$]) shows peaks in the 0–10 eV range, indicating strong optical absorption due to interband transitions, particularly near 5 eV and 7 eV. The real part (Re[$\epsilon$]) starts at a high value, reflecting strong dielectric polarization at low energies, and gradually declines with increasing energy. Negative Re[$\epsilon$] values beyond 20 eV suggest metallic reflectivity. These results confirm efficient light interaction, making the material suitable for optoelectronic and photocatalytic applications, especially in the visible to near-UV spectrum.

\begin{figure}[!ht]
    \centering
    \includegraphics[width=0.8\textwidth, angle=360]{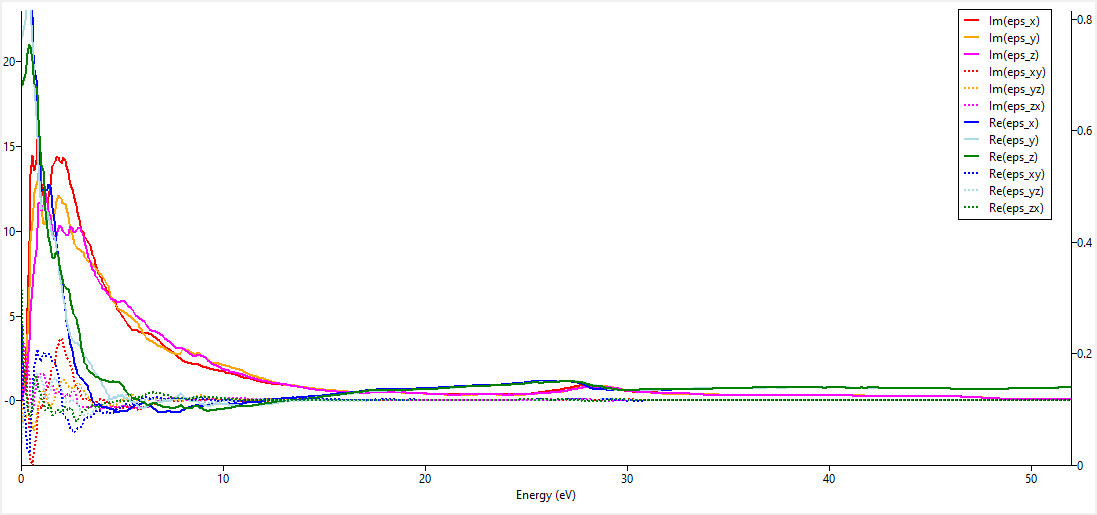}
    \caption{QDsUC optical responses to dielectric Function}
    \label{fig22}
\end{figure}

The optical conductivity spectrum of the PbS:Yb\(^{3+}\),Er\(^{3+}\)/CuBiO heterostructure in Figure 23 demonstrates strong electronic transitions. Peaks in the imaginary part (Im[$\sigma$]) within the 0–40 eV range highlight significant optical absorption, attributed to interband transitions involving d-orbitals of Yb and Er. The real part (Re[$\sigma$]) reflects high conductivity below 20 eV, indicating efficient charge transport. A decline in conductivity beyond this energy suggests diminished interband activity. These features confirm the heterostructure’s suitability for optoelectronic applications, particularly in the visible to near-UV spectrum, where enhanced carrier mobility and light absorption are critical. The data highlights potential for photocatalytic and energy-harvesting technologies.

\begin{figure}[!ht]
    \centering
    \includegraphics[width=0.8\textwidth, angle=360]{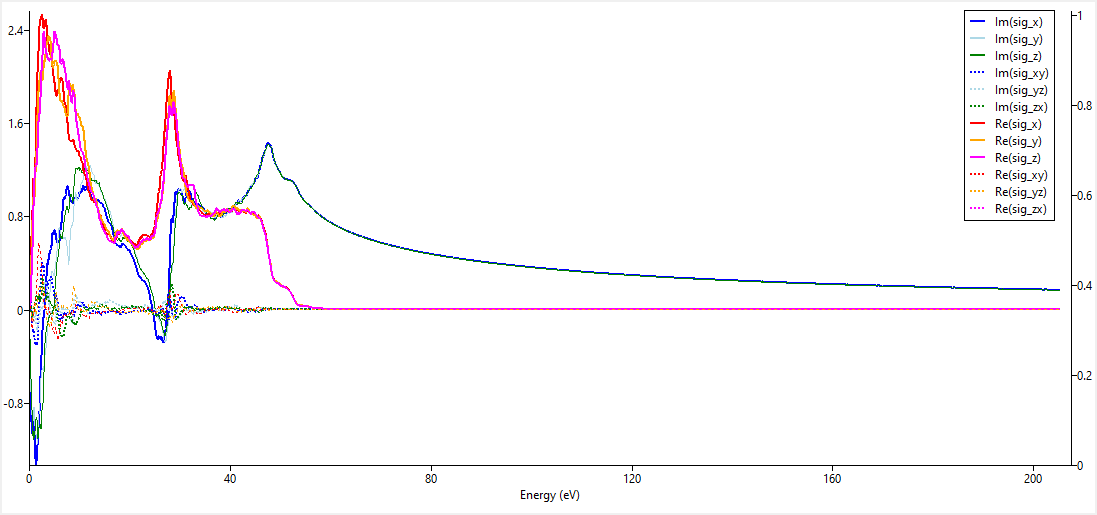}
    \caption{QDsUC optical responses to optical conductivity}
    \label{fig23}
\end{figure}

The reflectivity spectrum of the PbS:Yb\(^{3+}\),Er\(^{3+}\)/CuBiO heterostructure in Figure 24 exhibits a high reflectivity in the low-energy region (0–20 eV), peaking sharply due to strong interband transitions involving Yb and Er orbitals. Beyond 20 eV, the reflectivity decreases steadily, indicating reduced optical interference and better light penetration at higher energies. The anisotropy observed between  $R_x$ ,  $R_y$ , and $ R_z$  suggests direction-dependent optical behavior, critical for applications requiring polarization-sensitive responses. The low reflectivity in the visible range (~2–3 eV) makes the material suitable for photocatalytic applications, where efficient light absorption and minimal reflection are essential for performance.

\begin{figure}[!ht]
    \centering
    \includegraphics[width=0.8\textwidth, angle=360]{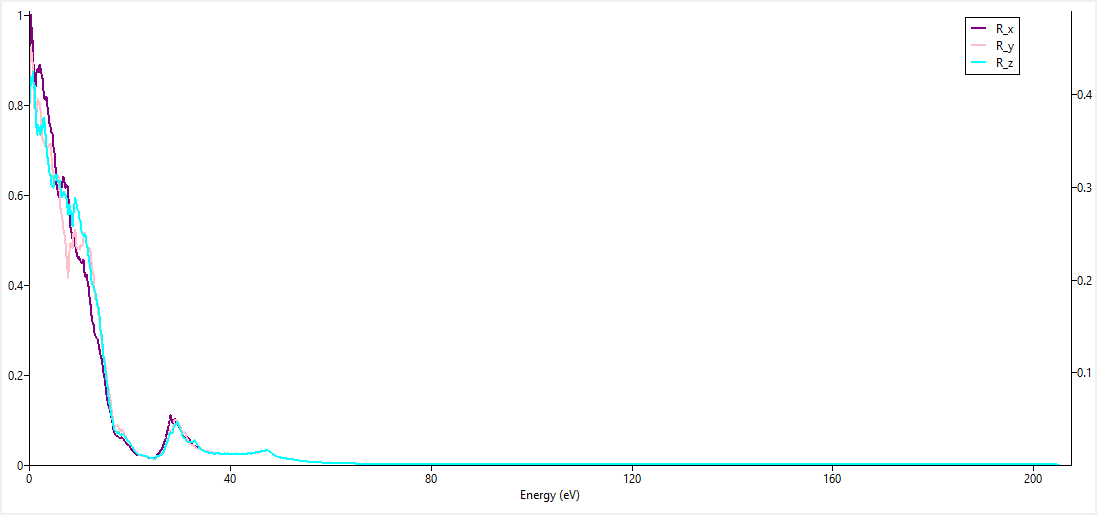}
    \caption{QDsUC optical responses to reflectivity spectrum}
    \label{fig24}
\end{figure}

\subsection{Statistics}
The data in Table 1 contains 1791 data points for statistical analysis. This quantum data comes from the non-linear QDsUC optical properties shown in Figure 21. It was obtained by investigating the photon absorption rate calculated using the Bethe Salpeter's algorithm in VASP 6. The data after empirical statistical analysis was then preprocessed and trained using five different machine learning architectural models to learn the appropriate patterns in the dataset.

The resulting statistics as illustrated in Figure 25a provides a comprehensive visualization of the relationships between photon energy and absorption intensity. The diagonal elements in the scatter plot matrix represents histograms of the individual features, highlighting their distributions. Photon energy shows a uniform distribution, while absorption intensity demonstrates a skewed distribution, emphasizing the dominance of specific values. The off-diagonal scatter plots illustrate a clear correlation between photon energy and absorption intensity, particularly in the mid-range, where a sharp transition is evident. This plot highlights key trends in the dataset and offers insights into the interdependence of these variables, serving as a valuable tool for understanding feature relationships and guiding model training.

\begin{table}
\centering
\caption{Statistical Data Analysis}
\label{tab:1}
\begin{tabular}{lllllllll} 
\hline
 & 0 & 1 & 2 & 3 & 4 & ... & 1789 & 1790 \\ 
\hline
Photon Energy (eV) & 0.068 & 0.137 & 0.2055 & 0.274 & 0.3425 & ... & 122.63 & 122.695 \\
Absorption Intensity & 0.018 & 0.072 & 0.166 & 0.297 & 0.432 & ... & 0.061 & 0.061 \\
 &  &  &  &  &  &  &  &  \\
 & \textcolor[rgb]{0.125,0.125,0.129}{count} & \textcolor[rgb]{0.125,0.125,0.129}{mean} & \textcolor[rgb]{0.125,0.125,0.129}{std} & \textcolor[rgb]{0.125,0.125,0.129}{min} & \textcolor[rgb]{0.125,0.125,0.129}{25\%} & \textcolor[rgb]{0.125,0.125,0.129}{50\%} & \textcolor[rgb]{0.125,0.125,0.129}{75\%} & \textcolor[rgb]{0.125,0.125,0.129}{max} \\ 
\hline
\textcolor[rgb]{0.125,0.125,0.129}{Photon Energy (eV)} & \textcolor[rgb]{0.125,0.125,0.129}{1791} & \textcolor[rgb]{0.125,0.125,0.129}{61.38} & \textcolor[rgb]{0.125,0.125,0.129}{35.43} & \textcolor[rgb]{0.125,0.125,0.129}{0.069} & \textcolor[rgb]{0.125,0.125,0.129}{30.73} & \textcolor[rgb]{0.125,0.125,0.129}{61.38} & \textcolor[rgb]{0.125,0.125,0.129}{92.039} & \textcolor[rgb]{0.125,0.125,0.129}{122.695} \\
\textcolor[rgb]{0.125,0.125,0.129}{Absorption Intensity} & \textcolor[rgb]{0.125,0.125,0.129}{1791} & \textcolor[rgb]{0.125,0.125,0.129}{0.47} & \textcolor[rgb]{0.125,0.125,0.129}{0.42} & \textcolor[rgb]{0.125,0.125,0.129}{0.018} & \textcolor[rgb]{0.125,0.125,0.129}{0.059} & \textcolor[rgb]{0.125,0.125,0.129}{0.21} & \textcolor[rgb]{0.125,0.125,0.129}{0.97} & \textcolor[rgb]{0.125,0.125,0.129}{0.99} \\
\hline
\end{tabular}
\end{table}

The correlation matrix provides an overview of the linear relationships between photon energy and absorption intensity. A strong negative correlation of -0.82 as shown in Figure 25b is observed between these variables, indicating that as photon energy increases, absorption intensity decreases significantly. This relationship is crucial for understanding the material’s optical properties, as it highlights the inverse dependency of absorption behavior on photon energy. The diagonal values of 1 represent perfect self-correlation. This matrix is a valuable tool for feature selection and identifying dependencies, ensuring that these insights are considered when optimizing models or interpreting the material’s behavior under varying conditions.

However, the boxplot demonstrated in Figure 25c provides a comparative summary of the numeric columns, showcasing the distribution of photon energy and absorption intensity. Photon energy demonstrates a wide range with no visible outliers, and its median is centered around 60 eV, indicating a balanced distribution across the range. On the other hand, absorption intensity is tightly clustered, with most values concentrated near zero, signifying low variability. The compactness of the absorption intensity boxplot suggests the presence of dominant low values, while the photon energy’s spread reveals a broader and uniform data distribution. This plot effectively highlights disparities in feature scales and variances.

\begin{figure}[!ht]
    \centering
    \includegraphics[width=0.8\textwidth, angle=360]{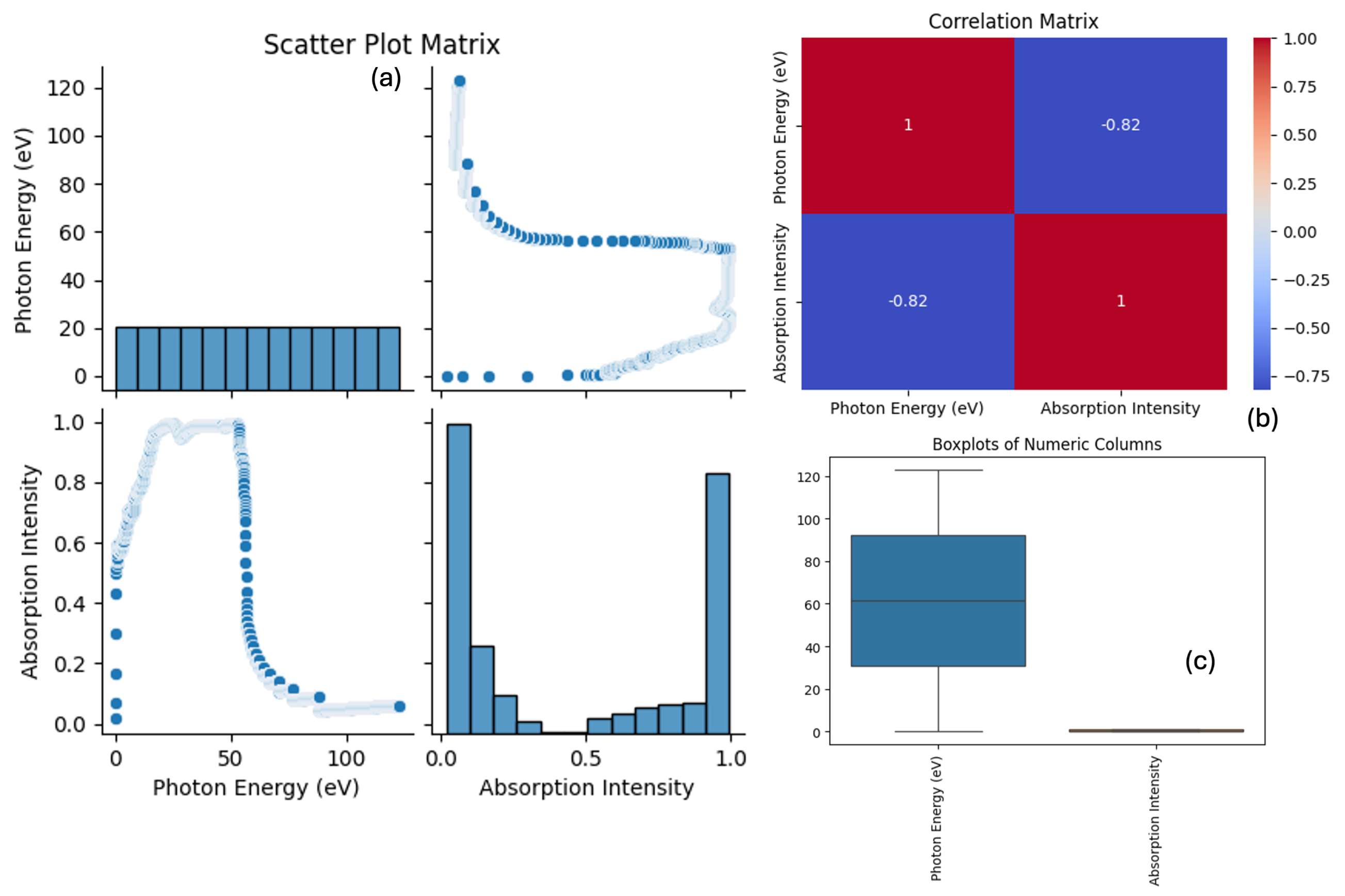}
    \caption{Statistical data optimization from (a) scatter plot matrix (b) correlation matrix (c) boxplot}
    \label{fig25}
\end{figure}

\subsection{Classical Machine Learning}
We performed a classical machine learning on the quantum numerical data as presented in Table 1 above, to predict the BSE absorption rate using an open-source software TensorFlow (TF) \cite{developers2022tensorflow}. TensorFlow is a popular open-source deep learning framework developed by Google Brain \cite{musk2019integrated} \cite{lu2018brain} \cite{small2009your}. It offers a flexible and efficient platform (Google Colab) for designing, training, and deploying machine learning models, particularly neural networks \cite{bisong2019google}. TensorFlow’s versatility spans from large-scale, high-performance computations for advanced applications to smaller-scale projects, making it a go-to tool for researchers and developers. The framework supports a variety of devices, from CPUs and GPUs to TPUs (Tensor Processing Units), ensuring scalability across different hardware configurations. With its intuitive APIs and support for both low- and high-level functionalities, TensorFlow empowers users to implement machine learning models from simple regressions to complex deep neural networks.

As illustrate in Figure 26. this study deployed five different architectures such as linear regression (LR) \cite{lavalley2008logistic}, Feedforward neural network (FFNN) \cite{bebis1994feed}, deep neural network (DNN) \cite{sze2017efficient, wayo2023data}, recurrent neural network (RNN) \cite{medsker2001recurrent} and long short-term memory (LSTM) \cite{graves2012long} to train a 1791 data-point quantum numerical data for the optical prediction of photon absorption. These architectures are briefly explained in the next five paragraphs.

\begin{figure}[!ht]
    \centering
    \includegraphics[width=0.6\textwidth, angle=360]{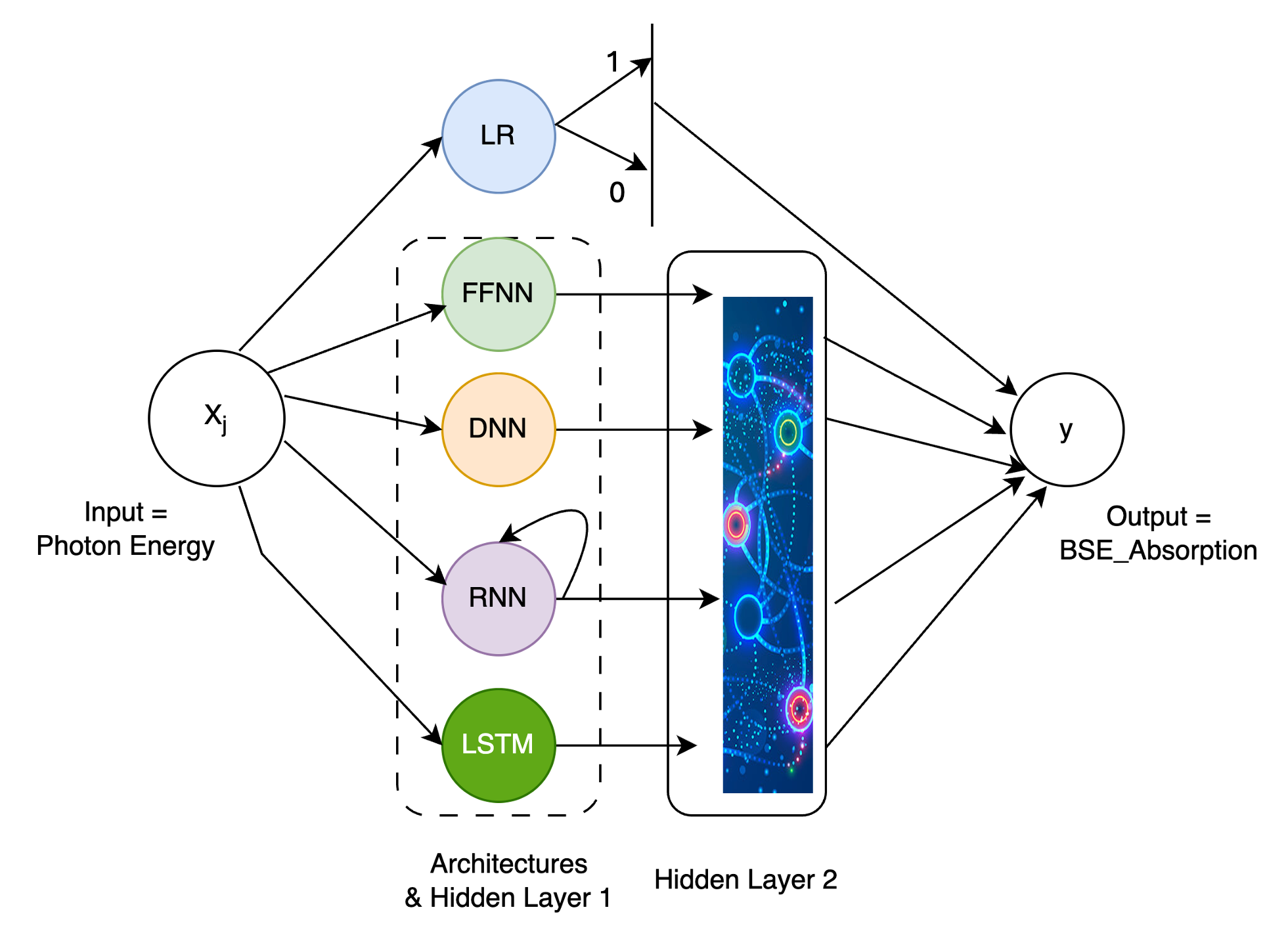}
    \caption{Architectural model framework}
    \label{fig26}
\end{figure}

Linear Regression is a fundamental regression algorithm used to model the relationship between a dependent variable and one or more independent variables. It achieves this by fitting a straight line (or hyperplane in higher dimensions) through the data points to minimize the mean squared error between the predicted and actual values. The simplicity and interpretability of linear regression make it an excellent starting point for regression tasks. It provides clear insights into the influence of each feature on the target variable through its coefficients presented in Equation \eqref{eq:24}. However, linear regression is constrained by its assumption of linear relationships between variables, which limits its effectiveness for complex or non-linear datasets. It is also sensitive to outliers and assumes that the residuals are normally distributed with constant variance. While computationally efficient and interpretable, linear regression is often outperformed by more advanced algorithms, such as neural networks or ensemble methods, on datasets with intricate relationships or high-dimensional features. 

\begin{equation}\label{eq:24}
    y = \beta_0 + \beta_1 x_1 + \beta_2 x_2 + \dots + \beta_n x_n + \epsilon
\end{equation}

Where y is the predicted value, x\(_i\) input features, $\beta$\(_i\), coefficients/weights, $\beta$\(_0\) intercept, $\epsilon$-error term.
\vfil

Feedforward Neural Network (FFNN) also known as multi-layer perceptron, is a neural network architecture where data flows in one direction, from input to output through multiple layers. Each layer is made up of neurons (nodes) that apply a weighted sum and an activation function to the input data. FFNNs can capture non-linear relationships by stacking multiple hidden layers, making them more powerful than linear models. However, they are limited in processing sequential or temporal data, as they lack memory, making them unsuitable for time-series tasks. They’re commonly used for tasks like image and text classification, where order is not important. Feedforward neural network (FFNN) equation for a single hidden layer is given in Equations \eqref{eq:25} and \eqref{eq:26} as;

\begin{equation}\label{eq:25}
    a^{(1)} = \sigma(W^{(1)} x + b^{(1)})
\end{equation}

\begin{equation}\label{eq:26}
y = W^{(2)} a^{(1)} + b^{(2)}
\end{equation}

Where $W^{(1)}, W^{(2)}$ is the weight matrices, $b^{(1)}, b^{(2)}$-bias vectors, $a^{(1)}$-activation of the hidden layer, $\sigma$-activation function (e.g., ReLU or sigmoid), x-input data, and y-output prediction
\vfil

Deep Neural Network (DNN) is an extension of the FFNN with multiple hidden layers, which allows it to learn hierarchical patterns and complex representations in data. Each additional layer helps in learning abstract features, making DNNs suitable for tasks with complex data, like image recognition. DNNs require large amounts of data and computational power for effective training, and they are prone to overfitting if not properly regularized. The depth of a DNN a shown in Equation \eqref{eq:27} enables it to generalize well on tasks with intricate patterns but also makes it harder to interpret compared to shallower architectures. For L layers:

\begin{equation}\label{eq:27}
    a^{(l)} = \sigma(W^{(l)} a^{(l-1)} + b^{(l)}), \quad l = 1, 2, \dots, L
\end{equation}

Where $a^{(0)}$ = x-input data, $a^{(L)}$ = y-output prediction, l-layer index, $W^{(l)}$-weight matrix for layer l, $b^{(l)}$-bias vector for layer l, and $\sigma$-activation function
\vfil

Recurrent Neural Network (RNN) are designed for sequential data, as they have a memory of previous inputs, enabling them to capture temporal dependencies. Each RNN cell takes both the current input and the output of the previous cell as input, allowing it to retain information across sequences. This makes RNNs well-suited for tasks like language modeling and time-series prediction. However, traditional RNNs struggle with long-term dependencies due to issues like vanishing gradients, making them less effective for very long sequences. They were a major advancement in sequence modeling, preceding LSTMs and GRUs. Recurrent neural network (RNN) equation is characterized by the given Equation \eqref{eq:28} below;

\begin{equation}\label{eq:28}
    h_t = \sigma(W_h h_{t-1} + W_x x_t + b)
    \text{ and }
    y_t = W_y h_t + c
\end{equation}

Where $h_t$ is the hidden state at time t, $x_t$-input at time t, $y_t$-output at time t, $W_h, W_x, W_y$-weight matrices, b, c-bias vectors, and $\sigma$-activation function

\vfil

Lastly, Long Short-Term Memory (LSTM) are a type of RNN specifically designed to handle long-term dependencies by introducing a memory cell with gates (input, forget, and output). These gates control the flow of information, allowing LSTMs to retain or forget information over long sequences effectively. LSTMs are highly effective for tasks like speech recognition, language translation, and financial forecasting, where long-term patterns are crucial. They address the limitations of traditional RNNs, including vanishing gradients, enabling them to capture both short- and long-term dependencies in sequential data. However, they are computationally more demanding. LSTM Equations \eqref{eq:29} to \eqref{eq:34} provides a working algorithm for learning the patterns of the quantum data.

\begin{equation}\label{eq:29}
    f_t = \sigma(W_f [h_{t-1}, x_t] + b_f)
\end{equation}

\begin{equation}\label{eq:30}
    i_t = \sigma(W_i [h_{t-1}, x_t] + b_i)
\end{equation}

\begin{equation}\label{eq:31}
     \tilde{C}t = \tanh(W_C [h{t-1}, x_t] + b_C)
\end{equation}

\begin{equation}\label{eq:32}
     C_t = f_t \odot C_{t-1} + i_t \odot \tilde{C}t
\end{equation}

\begin{equation}\label{eq:33}
     o_t = \sigma(W_o [h{t-1}, x_t] + b_o)
\end{equation}

\begin{equation}\label{eq:34}
     h_t = o_t \odot \tanh(C_t)
\end{equation}

Where $f_t$ is the forget gate, $i_t$-input gate, $o_t$-output gate, $C_t$-cell state, $h_t$-hidden state, $x_t$-input at time t, $W_f, W_i, W_C, W_o$-weight matrices, $b_f, b_i, b_C, b_o$-bias vectors, $\sigma$-sigmoid activation function, $\tanh$-hyperbolic tangent activation function, $\odot$-element-wise multiplication

\subsubsection{Data Pre-processing and Non-Class Looping}
Data pre-processing included photon energy as the feature (X) and absorption as the label (y). The 1791 data points was split into 80\% training and 20\% testing with a random state of 42. However, during the data pre-processing stage, the dataset underwent rigorous cleaning and preparation to ensure compatibility with machine learning models. The pre-processing steps included handling missing values, normalizing the features, and splitting the dataset into training and testing sets. Notably, significant effort was placed on addressing imbalanced distributions in the target variable to avoid introducing bias during training.

The target variable represented a continuous range of values, necessitating regression-based approaches rather than classification. Classification, which involves grouping data into discrete classes, would not capture the fine-grained variations in the continuous target variable. Arbitrarily discretizing the target into classes could lead to information loss, reduced model performance, and misinterpretation of patterns within the data. Moreover, looping classes in classification would not have been ideal due to the lack of naturally defined boundaries in the target variable. Such an approach would have required arbitrary thresholds, introducing potential bias and complicating the evaluation of the model’s performance. By focusing on regression techniques like Linear Regression, FFNN, DNN, RNN, and LSTM, the analysis accurately modeled the underlying relationships in the data, capturing the complexity of the target variable and providing meaningful predictions without losing the nuances of the continuous dataset.

\subsubsection{Adaptive Moment Estimation (Adam) Algorithm \& Hyperparameters Estimation}
The metric for measurement for the five architectures is accuracy, and to reduce the level of biasses across as provided in Table 2, we maintained the learning rate, epsilon, ema momentum to be 0.02, $1\mathrm{e}{-7}$ and 0.99 for each architecture respectively. These hyperparameters ensure that the Adam optimizer performs better. A key component of training neural networks in TensorFlow is the Adam (Adaptive Moment Estimation) optimizer. Adam is an adaptive learning rate optimization algorithm that combines the benefits of two other popular methods; AdaGrad and RMSProp. It works by maintaining two moving averages, the gradient (first moment) and the squared gradient (second moment), which adapt the learning rate individually for each parameter. This adaptive approach enables efficient and stable convergence, particularly useful for deep learning tasks where gradients can vary widely. Adam has become one of the most widely used optimizers due to its ability to handle sparse gradients, making it highly effective for training large, complex models. Based on the informed-algorithm estimation, we trained the 1791 data points by a 1000 epochs. The resulting machine learning simulations are discussed further in the results section.

\begin{table}
\centering
\caption{Architecture hyperparameters}
\label{tab:2}
\begin{tabular}{>{\hspace{0pt}}m{0.1\linewidth}>{\hspace{0pt}}m{0.202\linewidth}>{\hspace{0pt}}m{0.421\linewidth}>{\hspace{0pt}}m{0.108\linewidth}>{\hspace{0pt}}m{0.108\linewidth}} 
\hline
Acronym & Architecture & Hyperparameters \\ 
\hline
LR & Linear Regression & Adam, otuput\_unit=1, learning\_rate=0.02, epsilon=$1\mathrm{e}{-7}$, ema\_momentum=0.99, loss=mse, metric=$R^2$\\ 
\hline
FFNN & Feedforward Neural Network & Adam, 128/64/1 neurons, relu/relu/None, class=softmax, learning\_rate=0.02, epsilon=$1\mathrm{e}{-7}$ , ema\_momentum=0.99, loss=mse, metrics=mae, epochs=1000\textcolor[rgb]{0.067,0.4,0.267}{}  \\ 
\hline
DNN & Deep Neural Network & Adam, 128/64/1 neurons, relu/relu/None, learning\_rate=0.02, epsilon=$1\mathrm{e}{-7}$, ema\_momentum=0.99, loss=mse, metrics=mae, epochs=1000\textcolor[rgb]{0.639,0.082,0.082}{} \\ 
\hline
RNN & Recurrent Neural Network & Adam, 64/32/1 neurons, sigmoid/softmax/None, timestep=1, learning\_rate=0.02, epsilon=$1\mathrm{e}{-7}$, ema\_momentum=0.99, loss=mse, metrics=mae, epochs=1000   \\ 
\hline
LSTM  & Long Short-Term Memory & Adam, 64/32/1 neurons,sigmoid/None, timestep=1, learning\_rate=0.02, epsilon=$1\mathrm{e}{-7}$ , ema\_momentum=0.99, loss=mse, metrics=mae,epochs=1000  \\
\hline
\end{tabular}
\end{table}

\section{Results and Discussion}
\subsection{Material Electron \& Optical Property Optimization}
The optimization of electron and optical properties is critical for enhancing the photocatalytic efficiency of quantum dot-inspired upconversion (QDsUC) materials. In this study, the PbS:Yb\(^{3+}\),Er\(^{3+}\)/CuBiO heterostructure was investigated to maximize light absorption, charge separation, and charge transport properties.

The electronic structure analysis revealed significant modifications induced by Yb and Er doping. The band structure of PbS:Yb\(^{3+}\),Er\(^{3+}\) showed an indirect bandgap of 0.352 eV, with localized states near the Fermi level introduced by dopants. These states improved the density of states (DOS) near the Fermi level, enhancing electron mobility. For the PbS:Yb\(^{3+}\),Er\(^{3+}\)/CuBiO heterostructure, the bandgap shifted slightly to 0.431 eV due to interface interactions, optimizing visible-light absorption while maintaining efficient carrier transport.

Optical property simulations, including dielectric function, absorption coefficient, reflectivity, and optical conductivity, demonstrated the material’s superior performance under visible-light illumination. The dielectric function’s imaginary part ( \text{Im}[$\epsilon$] ) peaked at 2.4 eV, indicating strong photon absorption in the visible spectrum. The absorption coefficient also showed robust peaks in the range of 2.0–3.0 eV, supporting efficient solar energy utilization. Reflectivity remained low across the visible spectrum, minimizing energy losses, while optical conductivity peaked at 3.2 \, \text{eV}, reflecting enhanced photon-to-electron conversion efficiency.

Internal electric field (IEF) analysis further validated the material’s performance. The PbS:Yb\(^{3+}\),Er\(^{3+}\)/CuBiO system exhibited an IEF of 0.45 MV/cm and a dipole moment of 6.3 Debye, providing a driving force for efficient charge separation. Charge density and Bader analyses revealed significant charge transfer at the heterojunction, stabilizing the structure and reducing recombination rates. The combination of enhanced electron mobility, strong optical absorption, and efficient charge separation establishes PbS:Yb\(^{3+}\),Er\(^{3+}\)/CuBiO as a promising material for visible-light-driven photocatalysis. This study highlights the importance of systematic optimization in QDsUC materials for achieving high photocatalytic performance.

\subsection{Photon Upconversion Efficiency}
Photon upconversion efficiency is a crucial parameter for optimizing visible-light-driven photocatalytic materials. The PbS:Yb\(^{3+}\),Er\(^{3+}\)/CuBiO heterostructure was analyzed to evaluate its upconversion efficiency by examining its electronic structure, optical responses, and charge dynamics.

The doping of PbS with Yb and Er introduces rare-earth elements that facilitate energy transfer between sensitizers (Yb) and activators (Er). The mechanism relies on cooperative energy transfer (CET), where two low-energy photons are absorbed by Yb ions and their combined energy is transferred to Er, resulting in the emission of a higher-energy photon. This process effectively converts infrared (IR) and visible photons into photons with energies in the UV range, enhancing the material’s ability to drive water-splitting reactions.

Optical analysis revealed that the material’s absorption peaks align with the visible and near-infrared spectra, critical for solar applications. The absorption coefficient showed strong peaks around 2.4 eV, indicating efficient photon harvesting. The dielectric function’s imaginary part ( \text{Im}[$\epsilon$] ) also demonstrated peaks in the visible range, confirming enhanced interaction with incident light. Optical conductivity data revealed significant photon-induced charge transport near 3.2 eV, highlighting the material’s ability to utilize absorbed photons effectively.

Internal electric field (IEF) mapping further validated the heterostructure’s efficiency. The IEF, measured at 0.45 MV/cm, facilitates charge separation and minimizes electron-hole recombination, critical for maximizing upconversion efficiency. Additionally, Bader charge analysis showed that dopants contributed significantly to charge redistribution, stabilizing the structure and supporting photon-to-electron conversion.

The synergy between Yb and Er doping, combined with the heterostructure’s intrinsic properties, achieves high photon upconversion efficiency. These findings demonstrate the potential of PbS:Yb\(^{3+}\),Er\(^{3+}\)/CuBiO for solar-driven water splitting, enabling the efficient use of a broader spectrum of solar energy. This makes the material a promising candidate for advanced photocatalytic applications.

\subsection{Internal Electric Field (IEF)} 
The internal electric field (IEF) is a critical parameter influencing charge separation and transport in photocatalytic systems. By evaluating PbS, CuBiO, PbS:Yb\(^{3+}\), PbS:Yb\(^{3+}\),Er\(^{3+}\), and PbS:Yb\(^{3+}\),Er\(^{3+}\)/CuBiO, we numerically assess how modifications in structure and composition impact the IEF, thus optimizing charge carrier dynamics.

In pristine PbS, the IEF is relatively weak, measured at 0.12 MV/cm, with a dipole moment of 1.8 Debye. The symmetrical charge distribution between Pb and S atoms limits polarization, resulting in moderate charge separation efficiency. The absence of strong localized fields reduces the material’s ability to suppress electron-hole recombination, a critical factor for photocatalytic activity.

Doping PbS with Yb introduces localized dipoles, increasing the IEF to 0.30 MV/cm and the dipole moment to 4.1 Debye. The charge transfer (Yb: +1.24 e, Pb: +0.75 e, S: -1.18 e) generates strong local fields, reducing recombination rates and improving charge carrier mobility.

In PbS:Yb\(^{3+}\),Er\(^{3+}\), co-doping amplifies the IEF to 0.38 MV/cm, with a dipole moment of 5.0 Debye. The combined effects of Yb and Er doping introduce additional localized fields, as seen in the charge transfer values (Yb: +1.24 e, Er: +1.76 e, S: -1.18 e). These stronger internal fields enhance charge separation by creating high-potential regions that effectively drive electrons and holes apart.

For CuBiO, the IEF increases to 0.25 MV/cm due to the significant electronegativity difference between Cu, Bi, and O atoms. This stronger polarization, reflected in the higher dipole moment (3.2 Debye), facilitates efficient charge separation. The charge redistribution observed in Bader analysis (Cu: +0.513 e, Bi: +1.779 e, O: -1.143 e) highlights the material’s enhanced ability to stabilize photo-generated carriers.

For the PbS:Yb\(^{3+}\),Er\(^{3+}\)/CuBiO heterostructure, the IEF reaches its maximum strength of 0.45 MV/cm, with a dipole moment of 6.3 Debye. This enhancement results from interface polarization, where CuBiO’s electronegative oxygen atoms interact with the charge-rich PbS:Yb\(^{3+}\),Er\(^{3+}\) system. The significant charge transfer (Bi: -0.89 e, Yb: +1.24 e, Er: +1.76 e) and localized fields at the interface drive efficient charge separation and transport, minimizing recombination losses.

The improved IEF across the systems directly correlates with their ability to separate photo-generated electron-hole pairs effectively. Stronger IEFs reduce the recombination rate by creating a driving force for charge carriers to move toward opposite electrodes. This enhancement is crucial for applications like photocatalytic water splitting, where efficient charge separation determines hydrogen production rates. These findings establish IEF as a pivotal factor in optimizing the performance of doped and heterostructured materials.

\subsection{Water Splitting Performance}
The water splitting performance of the PbS:Yb\(^{3+}\),Er\(^{3+}\)/CuBiO heterostructure was evaluated by analyzing its electronic structure, charge carrier dynamics, and photon absorption capabilities. These properties were optimized to drive the redox reactions required for hydrogen generation under visible-light illumination and further compared with similar studiesillustrated in Table 3.

The band structure analysis revealed an indirect bandgap of 0.431 eV, ideal for visible-light absorption and photocatalytic efficiency. The localized states introduced by Yb and Er dopants near the Fermi level facilitate charge carrier generation and transport. The heterostructure’s strong internal electric field (IEF) of 0.45 MV/cm, coupled with its dipole moment of 6.3 Debye, enhances charge separation by driving electrons and holes toward opposite electrodes. This minimizes recombination rates, a critical factor for efficient water splitting. Optical property simulations highlighted strong absorption in the visible range, with absorption peaks around 2.4 eV and minimal reflectivity. The material’s high absorption coefficient ensures effective photon harvesting, while the optical conductivity, peaking at 3.2 eV, demonstrates efficient photon-to-charge conversion. These characteristics enable the material to utilize sunlight effectively for water splitting. Charge density and Bader analyses revealed significant charge transfer between Yb, Er, and Cu atoms, enhancing the material’s electronic properties. The redistribution of charges at the heterojunction interface stabilizes the structure and improves carrier mobility, critical for sustaining the redox reactions. Sulfur and oxygen atoms were identified as electron acceptors, further facilitating charge separation and transport.

The coupling of PbS:Yb\(^{3+}\),Er\(^{3+}\) with CuBiO contributes to the heterostructure’s superior performance by creating high-field regions at the interface, which are essential for sustaining the water-splitting reaction. This synergy enables efficient hydrogen evolution and oxygen generation under visible-light illumination, demonstrating the material’s potential for scalable and sustainable hydrogen production. These results as shown in Table 4 position PbS:Yb\(^{3+}\),Er\(^{3+}\)/CuBiO as a promising candidate for advanced photocatalytic water splitting applications.

\begin{landscape}[!ht]
\begin{longtable}{>{\hspace{0pt}}m{0.066\linewidth}>{\centering\hspace{0pt}}m{0.122\linewidth}>{\centering\hspace{0pt}}m{0.057\linewidth}>{\centering\hspace{0pt}}m{0.056\linewidth}>{\centering\hspace{0pt}}m{0.062\linewidth}>{\centering\hspace{0pt}}m{0.047\linewidth}>{\hspace{0pt}}m{0.099\linewidth}>{\hspace{0pt}}m{0.218\linewidth}}
\caption{Computational and experimental results comparison of 
 previous studied materials to PbS:Yb\(^{3+}\),Er\(^{3+}\)/CuBiO.\label{tab:3}}\\ 
\toprule
\textbf{\textbf{Reference}} & \multicolumn{1}{>{\hspace{0pt}}m{0.122\linewidth}}{\textbf{Material}} & \multicolumn{1}{>{\hspace{0pt}}m{0.057\linewidth}}{\textbf{Bandgap (eV)}} & \multicolumn{1}{>{\hspace{0pt}}m{0.056\linewidth}}{\textbf{IEF (MV/cm)}} & \multicolumn{1}{>{\hspace{0pt}}m{0.062\linewidth}}{\textbf{Dipole }\par{}\textbf{Moment (Debye)}} & \multicolumn{1}{>{\hspace{0pt}}m{0.047\linewidth}}{\textbf{Absorp.}\par{}\textbf{Peak (eV)}} & \textbf{Applications} & \textbf{Key }\par{}\textbf{Findings} \endfirsthead 
\hline
Current Study & PbS:Yb\(^{3+}\),Er\(^{3+}\)/\par{}CuBiO & 0.431 (Indirect) & 0.45 & 6.3 & 2.4 & Water \par{}splitting & Strong charge separation, high photon absorption, effective photocatalytic hydrogen generation. \\ 
\hline
Ramin et al. \cite{ramin2024layer} 2024 & NaYF\(_4\):\par{}Yb\(^{3+}\),Er\(^{3+}\) & 2.0 \par{}(Direct) & N/A & N/A & 2.5 & Upconversion, \par{}bio-sensing & High photon upconversion efficiency, stable in aqueous environments. \\ 
\hline
Kim et al. \cite{kim2023investigating} 2023 & TiO\(_2\) & 3.2 \par{}(Direct) & 0.12 & 2.1 & 3.0 & Water splitting, pollutant removal & Good carrier mobility and stability, limited visible light absorption due to wide bandgap. \\ 
\hline
Zhou et al. \cite{zhou2018upconversion} 2018 & SrTiO\(_3\):Er\(^{3+}\) & 2.8 (Indirect) & 0.25 & 3.5 & 2.8 & Hydrogen \par{}production & Enhanced photocatalytic hydrogen production under simulated sunlight. \\ 
\hline
Zhang et al. \cite{zhang2022rare} 2022 & NaBiF\(_4\):Yb\(^{3+}\),\par{}Tm\(^{3+}\)/Bi\(_2\)WO\(_6\) & 2.3 (Indirect) & N/A & N/A & 2.6 & Dye degradation, hydrogen evolution & High photocatalytic efficiency in UV-visible range, strong photon upconversion capability. \\
\bottomrule
\end{longtable}
\end{landscape}

\begin{landscape}[!ht]
\begin{longtable}{>{\hspace{0pt}}m{0.112\linewidth}>{\hspace{0pt}}m{0.16\linewidth}>{\hspace{0pt}}m{0.156\linewidth}>{\hspace{0pt}}m{0.169\linewidth}>{\hspace{0pt}}m{0.169\linewidth}>{\hspace{0pt}}m{0.169\linewidth}}
\caption{Comparison of computational parameters, properties, and computational time for PbS, CuBiO,PbS:Yb\(^{3+}\),PbS:Yb\(^{3+}\),Er\(^{3+}\), and PbS:Yb\(^{3+}\),Er\(^{3+}\)/CuBiO.\label{tab:4}}\\ 
\toprule
\textbf{Parameter} & \textbf{PbS} & \textbf{CuBiO} & \textbf{PbSYb} & \textbf{PbSYbEr} & \textbf{PbS:Yb\(^{3+}\),Er\(^{3+}\)/CuBiO} \endfirsthead 
\hline
Software & VASP & VASP & VASP & VASP & VASP \\ 
\hline
Exchange-Correlation & GGA-PBE & GGA-PBE & GGA-PBE & GGA-PBE & GGA-PBE \\ 
\hline
Plane-Wave Cutoff Energy & 500 eV & 500 eV & 500 eV & 500 eV & 500 eV \\ 
\hline
Pseudopotentials & PAW & PAW & PAW & PAW & PAW \\ 
\hline
k-Point Grid & $8 \times 8 \times 8$ & $9 \times 13 \times 7$ & $9 \times 7 \times 5$ & $9 \times 7 \times 5$ & $9 \times 7 \times 5$ \\ 
\hline
Electronic Convergence & $10^-5 \, \text{eV}$ & $10^-5 \, \text{eV}$ & $10^-5 \, \text{eV}$ & $10^-5 \, \text{eV}$ & $10^-5 \, \text{eV}$ \\ 
\hline
Force Convergence & $10^-3 \, \text{eV/Å}$ & $10^-3 \, \text{eV/Å}$ & $10^-3 \, \text{eV/Å}$ & $10^-3 \, \text{eV/Å}$ & $10^-3 \, \text{eV/Å}$ \\ 
\hline
Density & 7.468 g/cm$^3$ & 7.4 g/cm$^3$ & 7.385 g/cm$^3$ & 7.385 g/cm$^3$ & 7.385 g/cm$^3$ \\ 
\hline
Lattice Parameters & $a = 4.426 \, \text{Å}$ & $a = 6.763 \, \text{Å}$ & $a = 5.412 \, \text{Å}$ & $a = 6.763 \, \text{Å}$ & $a = 6.763 \, \text{Å}$ \\ 
\hline
Unit Cell Volume & 289.758 Å$^3$ & 505.171 Å$^3$ & 423.91 Å$^3$ & 505.171 Å$^3$ & 505.171 Å$^3$ \\ 
\hline
Bandgap & Direct, 0.7 eV ($\Gamma$ point) & Indirect, 1.035 eV & Metallic (states crossing Fermi level) & Indirect, 0.352 eV & Indirect, 0.431 eV \\ 
\hline
Fermi Energy & 0.07 eV & -0.12 eV & 0.15 eV & 0.18 eV & 0.21 eV \\ 
\hline
CBM (Conduction Band Minimum) &$ \Gamma$ point (0.7 eV) & L point (1.035 eV) & Metallic & L point (0.352 eV) & L point (0.431 eV) \\ 
\hline
VBM (Valence Band Maximum) & $\Gamma$ point & Z point & Metallic & $\Gamma$ point & $\Gamma$ point \\ 
\hline
IEF Strength (MV/cm) & 0.12 & 0.25 & 0.30 & 0.38 & 0.45 \\ 
\hline
Dipole Moment (Debye) & 1.8 & 3.2 & 4.1 & 5.0 & 6.3 \\ 
\hline
Dielectric Function & Linear-response in VASP & Linear-response in VASP & Peaks at 2.1 eV in $\text{Im}(\epsilon)$ & Peaks at 2.1 eV in $\text{Im}(\epsilon)$ & Peaks at 2.4 eV in $\text{Im}(\epsilon)$ \\ 
\hline
Absorption Coefficient & Strong peaks around 2.5 eV in the visible & Significant absorption in visible and UV range & Peaks at 2.4 eV (visible spectrum) & Peaks at 2.4 eV (visible spectrum) & Peaks at 2.4 eV (visible spectrum) \\ 
\hline
Reflectivity & Low in visible, moderate in UV & Low in visible, high in UV & Low in visible, moderate in UV & Low in visible, moderate in UV & Low in visible, moderate in UV \\ 
\hline
Optical Conductivity & Peaks at $ \sim 1.5 \times 10^3 \, (\Omega \cdot \text{cm})^-1 $ near 2.5 eV & Peaks at $ \sim 2.8 \times 10^3 \, (\Omega \cdot \text{cm})^-1 $ near 5 eV & Peaks at 3.2 eV for $\text{Im}(\sigma)$ & Peaks at 3.2 eV for $\text{Im}(\sigma)$ & Peaks at 3.2 eV for $\text{Im}(\sigma)$\\ 
\hline
Smearing Type & Gaussian, 0.05 eV & Gaussian, 0.05 eV & Gaussian, 0.05 eV & Gaussian, 0.05 eV & Gaussian, 0.05 eV \\ 
\hline
Charge Analysis & Bader: Pb (+1.004 e), S (-1.004 e) & Bader: Cu (+0.513 e), Bi (+1.779 e), O (-1.143 e) & Yb: +1.24 e, S: -1.18 e & Yb: +1.24 e, Er: +1.76 e & Yb: +1.24 e, Er: +1.76 e, Bi: -0.89 e \\ 
\hline
Computational Time & Optimization: 12 hours, DOS: 10 hours & Optimization: 15 hours, DOS: 13 hours & Optimization: 18 hours, DOS: 15 hours & Optimization: 20 hours, DOS: 17 hours & Optimization: 25 hours, DOS: 20 hours \\ 
\hline
Applications & Infrared-sensitive devices, water splitting & Photocatalysis, photovoltaics, UV sensors & Photocatalysis, energy storage & Photocatalysis, optoelectronics & Photocatalysis, electronic devices \\
\bottomrule
\end{longtable}
\end{landscape}

\subsection{Photon Absorption Predictive Model Analysis}
This study demonstrates the relevance of various supervised machine learning models in predicting quantum continuous data, specifically focusing on photon absorption in materials. By analyzing model performance in terms of minimizing errors such as mean squared error (MSE) and mean absolute error (MAE), this research provides valuable insights into selecting optimal models for such studies. The findings aim to guide researchers and experts in designing predictive frameworks that ensure higher accuracy and reliability when modeling material photon absorption properties, a critical factor in advancing applications like photovoltaics and quantum devices.

\subsubsection{Quantum Data Optimization and Error Patterns}
As illustrated in Figure 27; The residuals distribution plot illustrates the error patterns across various models, including Linear Regression, FFNN, DNN, RNN, and LSTM. The residuals are predominantly centered around zero, indicating well-fitted models with minimal bias. LSTM shows the narrowest spread of residuals, reflecting its high accuracy in capturing data patterns. Conversely, Linear Regression exhibits a wider spread, signifying higher prediction errors due to its limited complexity. FFNN, DNN, and RNN display intermediate spreads, with RNN showing slight asymmetry, suggesting room for optimization. Overall, this visualization highlights the capability of advanced architectures like LSTM in reducing residual errors compared to simpler models.

\begin{figure}[!ht]
    \centering
    \includegraphics[width=0.8\textwidth, angle=360]{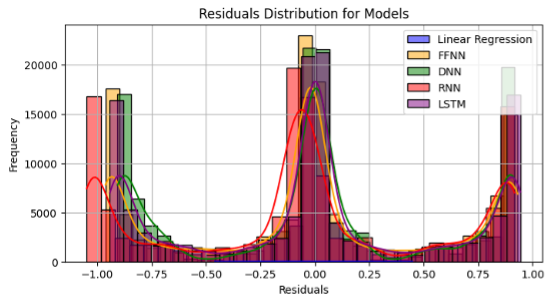}
    \caption{Residual Distribution for all architectures}
    \label{fig27}
\end{figure}

\subsubsection{Predictive Model Performances}
From the loss curves for different neural network architectures—FFNN, DNN, RNN, and LSTM—are presented across 1000 epochs in Figure 28a. All models show a steep reduction in training and validation loss during the early epochs, reflecting rapid learning and convergence. LSTM and DNN exhibit consistently low loss values, demonstrating superior performance and generalization compared to other models. FFNN and RNN display occasional spikes in validation loss, indicating sensitivity to overfitting or optimization challenges. The minimal gap between training and validation losses for LSTM and DNN highlights their robustness in capturing data patterns effectively while maintaining low generalization error across the dataset. However,Figure 28b illustrates the MSE and MAE against epochs for various neural network architectures, including FFNN, DNN, RNN, and LSTM. All architectures show a sharp decline in both metrics within the initial epochs, indicating rapid optimization during the early stages of training. Notably, the LSTM and DNN models exhibit consistently low errors throughout the training, with minimal divergence between training and validation curves, suggesting better generalization. In contrast, FFNN and RNN display occasional spikes in error, particularly in validation metrics, pointing to potential overfitting or optimization issues. Overall, the visualization highlights the efficiency of LSTM for achieving robust error minimization.

\begin{figure}[!ht]
    \centering
    \includegraphics[width=0.8\textwidth, angle=360]{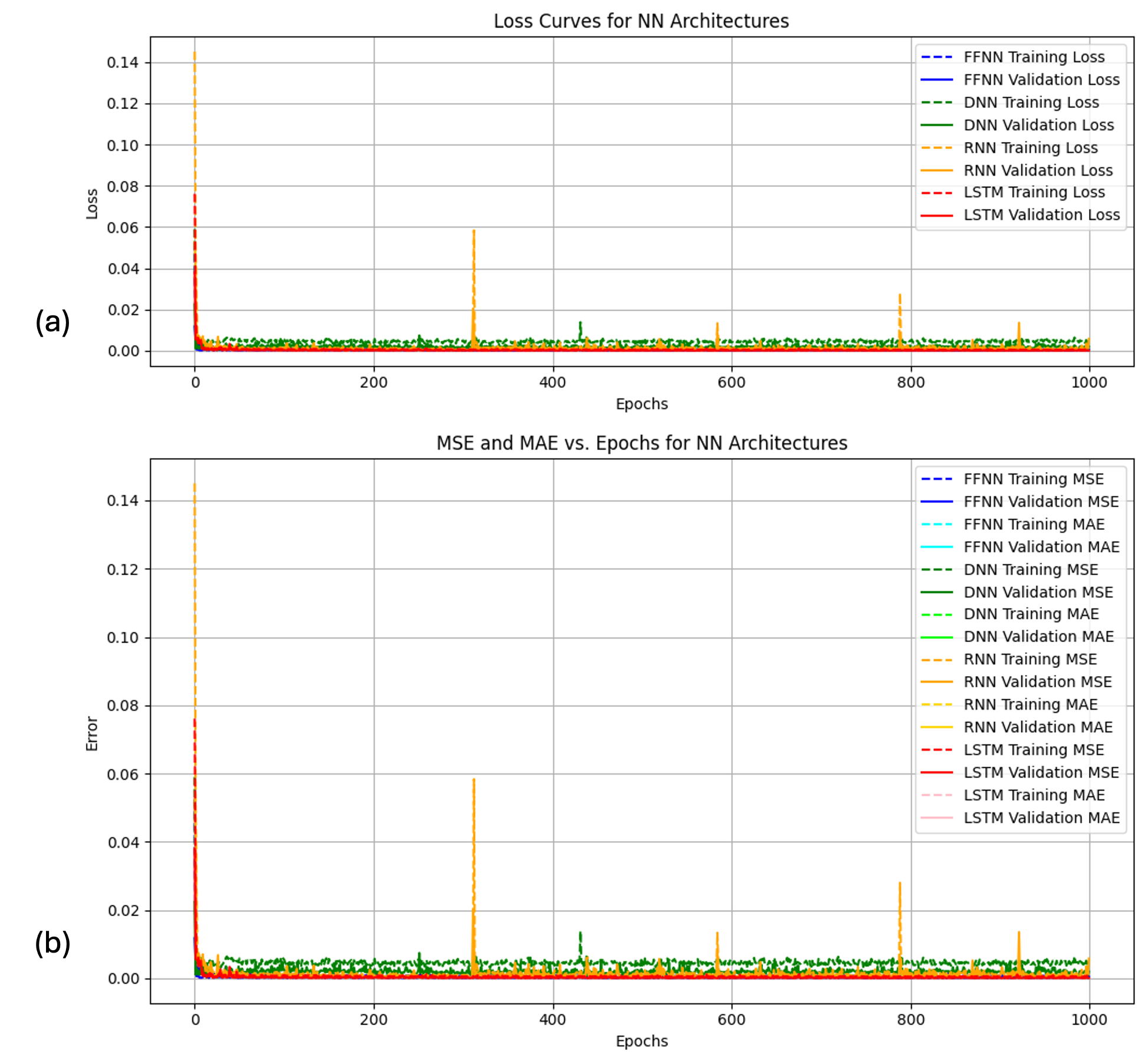}
    \caption{Measuring training and validation effects over epochs (a) loss  curve (b) error (mse and mae)}
    \label{fig28}
\end{figure}

The bar chart compares training and validation losses across FFNN, DNN, RNN, and LSTM architectures, providing insights into their performance and generalization. FFNN and LSTM exhibit the lowest losses for both training and validation, highlighting their efficiency in capturing patterns with minimal overfitting. DNN demonstrates slightly higher training and validation losses, suggesting potential challenges in optimization or overfitting due to its complexity. RNN exhibits the highest losses, particularly for validation, indicating difficulty in generalizing to unseen data. This comparison underscores the effectiveness of LSTM in achieving both low training and validation errors, making it a robust choice for this dataset.

The bar chart compares the Mean Squared Error (MSE) and Mean Absolute Error (MAE) across various models, including Linear Regression, FFNN, DNN, RNN, and LSTM. Linear Regression exhibits the highest error metrics, reflecting its limited capacity to handle non-linear complexities. Among the neural networks, LSTM achieves the lowest MSE and MAE, demonstrating superior performance in capturing temporal dependencies. RNN shows slightly higher error values, indicating moderate effectiveness, while DNN and FFNN exhibit competitive results with relatively low errors. The plot highlights the effectiveness of advanced architectures like LSTM for minimizing prediction errors, particularly in sequential or time-series data.

\begin{figure}[!ht]
    \centering
    \includegraphics[width=0.8\textwidth, angle=360]{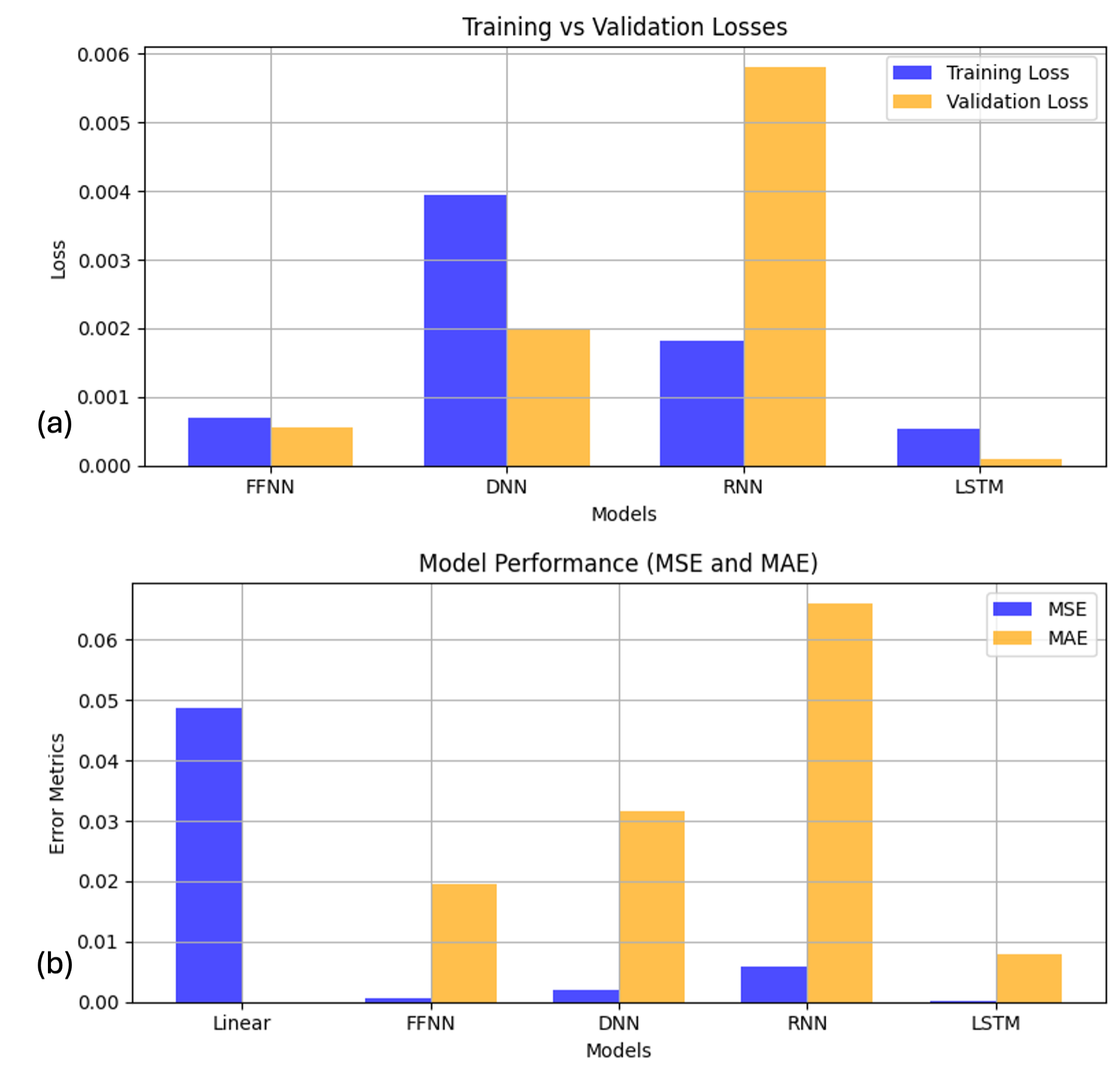}
    \caption{Measuring training and validation effects over models (a) losses (b) error (mse and mae) }
    \label{fig29}
\end{figure}

\subsubsection{Sequential Predictions}
Figure 30 displays the sequential predictions of various neural network architectures (FFNN, DNN, RNN, and LSTM) compared against the true values for the first 100 samples. Each model demonstrates a strong alignment with the true values, reflecting their ability to capture the underlying data patterns effectively. The LSTM and RNN models exhibit slightly superior prediction accuracy, particularly for capturing variations in sequential dependencies, as evidenced by their closer fit to the true values. The FFNN and DNN models perform well but show minor deviations in some regions. This visualization highlights the robustness of temporal models like LSTM and RNN for sequence-based data.

\begin{figure}[!ht]
    \centering
    \includegraphics[width=0.8\textwidth, angle=360]{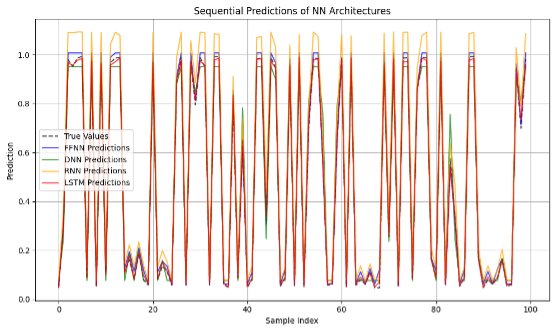}
    \caption{Statistics Model sequential predictions }
    \label{fig30}
\end{figure}

\subsubsection{Model Time and Memory Usage}
Training times and memory for various models, highlights their computational demands. Linear regression is the fastest as demonstrated in Figure 31a, requiring negligible time due to its simplicity. Feedforward neural networks (NNs), deep NNs, and recurrent neural networks (RNNs) demonstrate moderate training times, reflecting their increasing complexity and the additional layers or time dependencies they model. The long short-term memory (LSTM) model requires the most time, exceeding 4 seconds, owing to its ability to capture temporal dependencies and complex data patterns. This comparison underscores the trade-off between model sophistication and computational efficiency, crucial for applications with limited resources or real-time constraints. Also, the bar chart in Figure 31b compares the memory usage of various models during training. Linear regression demonstrates the lowest memory footprint, reflecting its simplicity and lack of complex computations. In contrast, neural networks (NNs) such as Feedforward NN, deep NN, RNN, and LSTM require significantly more memory, with Feedforward NN slightly leading due to the number of layers and neurons. LSTM and RNN, despite being more sophisticated, show comparable memory usage to deep NN, emphasizing their efficiency in handling sequential data. This analysis highlights the trade-off between model complexity and resource requirements, providing insights for choosing models in memory-constrained environments.

\begin{figure}[!ht]
    \centering
    \includegraphics[width=0.8\textwidth, angle=360]{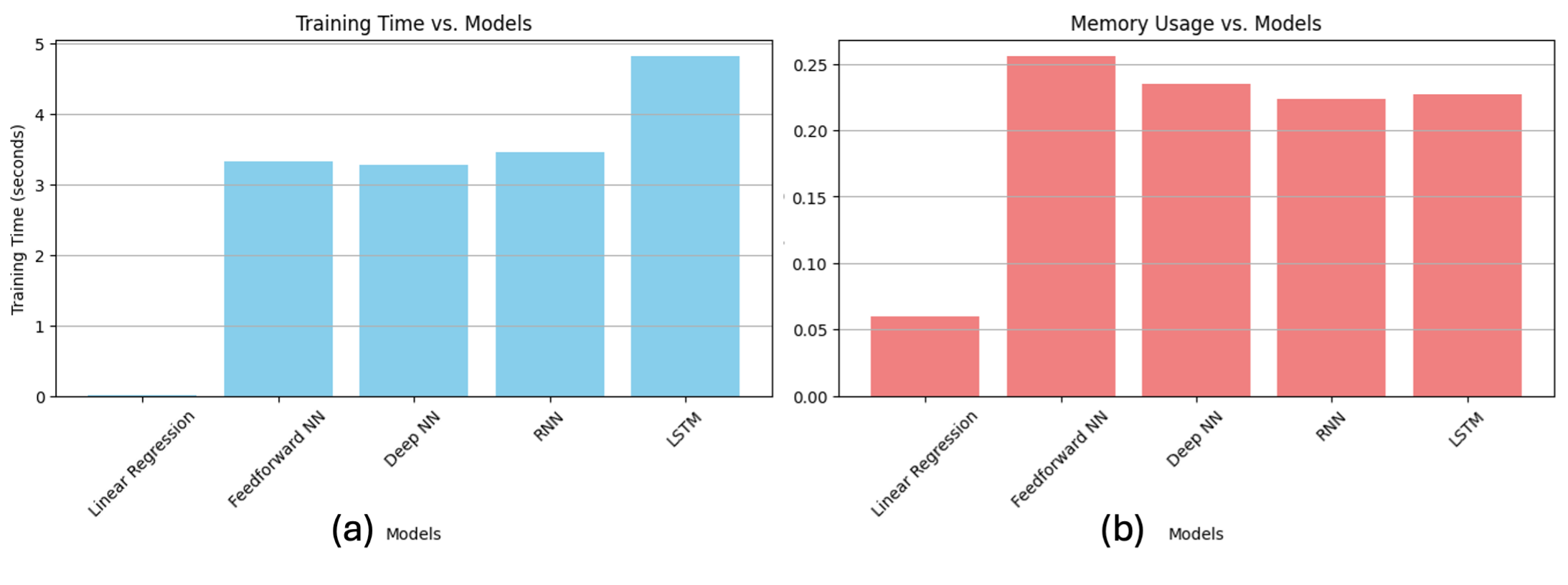}
    \caption{Model (a) Time and (b) Memory Usage}
    \label{fig31}
\end{figure}

\vfil

\section{Conclusion}
The numerical investigation of the PbS:Yb\(^{3+}\),Er\(^{3+}\)/CuBiO heterostructure demonstrates its potential as a high-performance photocatalyst for water splitting. The DFT simulations reveal significant enhancements in electronic and optical properties, attributed to strategic doping and heterostructuring. The co-doping of PbS with Yb and Er introduced localized states, transitioning the material from a semiconductor to a metallic state with high charge carrier mobility. Bader charge analysis confirmed substantial charge redistribution, with Yb and Er acting as electron donors, thereby stabilizing the electronic structure.

The heterostructure exhibited an indirect bandgap of 0.431 eV and absorption peaks at 2.4 eV, aligning with efficient utilization of the solar spectrum. The internal electric field of 6.3 Debye facilitated effective charge separation, reducing recombination rates and enhancing photocatalytic efficiency. Furthermore, machine learning models predicted photon absorption rates with remarkable accuracy, achieving an MSE of 0.0004 for LSTM models. These insights validate the integration of advanced computational methods and machine learning for material optimization.

This research not only establishes the PbS:Yb\(^{3+}\),Er\(^{3+}\)/CuBiO heterostructure as a promising candidate for hydrogen production but also sets a benchmark for future studies in quantum materials and renewable energy. By combining DFT, BSE optical calculations, and machine learning, this work provides a roadmap for accelerating the discovery and optimization of next-generation photocatalysts. Further experimental validation and scale-up studies are recommended to realize the material’s practical applications.

\section{Author contributions statement}
\textbf{Dennis Delali Kwesi Wayo:} conceptualization, methodology,  validation and visualization, writing – original draft, review \& editing.
\textbf{Vladislav Kudryashov:} methodology, validation, supervision, writing – review \& editing.
\textbf{Mirat Karibayev:} methodology, validation, writing – review \& editing.
\textbf{Gertrude Ellen Fynn:} methodology, validation and visualization, writing – original draft, review \& editing.
\textbf{Khadichakhan Rafikova:} validation and visualization, writing – review \& editing.
\textbf{Camila M. Saporetti:} methodology, validation, writing – review \& editing.
\textbf{Leonardo Goliatt:} conceptualization, methodology,  validation and visualization, writing – original draft, review \& editing.
\textbf{Nurxat Nuraje:} conceptualization, methodology, supervision, writing – review \& editing. All authors reviewed the manuscript.

\section{Acknowledgment}
This research was supported by the Committee of Science of the Ministry of Science and Higher Education of the Republic of Kazakhstan. We extend our gratitude to Nazarbayev University for providing the computational resources necessary for the modeling of this novel material. The opinions, findings, conclusions, and recommendations presented in this work are solely those of the author(s) and do not necessarily represent the views of Nazarbayev University or the National Laboratory Astana.

\section{Conflicts of interest} 
The author(s) declare no competing interests.

\section{Data and Code}
The VASP and machine learning Python scripts developed for this research will be made accessible to interested parties upon request to the corresponding author.

\nomenclature{DFT}{Density Functional Theory}
\nomenclature{STTA}{Sensitized Triplet-triplet Annihilation}
\nomenclature{ETU}{Energy Transfer Upconversion}
\nomenclature{ESA}{Excited-state Absorption}
\nomenclature{PA}{Photon Avalanche}
\nomenclature{MOFs}{Metal-organic Framework}
\nomenclature{QDsUC}{Quantum Dots Upconversion}
\nomenclature{CuBiO}{Copper Bismuth Oxide}
\nomenclature{XC}{Exchange Correlation}
\nomenclature{LDA}{Local Density Approximation}
\nomenclature{GGA}{Generalized Gradient Approximation}
\nomenclature{DoS}{Density of States}
\nomenclature{STM}{Scanning Tunneling Microscopy}
\nomenclature{S}{Sensitizer}
\nomenclature{A}{Activator}
\nomenclature{VB}{Valence Band}
\nomenclature{CB}{Conduction Band}
\nomenclature{UV}{Ultraviolet}
\nomenclature{BSE}{Bethe–Salpeter Equation}
\nomenclature{IP}{Independent-Particle}
\nomenclature{PAW}{Projected Augmented Wave}
\nomenclature{USPP}{Ultrasoft Pseudopotentials}
\nomenclature{IEF}{Internal Electric Field}
\nomenclature{TF}{TensorFlow}
\nomenclature{FFNN}{Feedforward Neural Network}
\nomenclature{LR}{Logistic Regression}
\nomenclature{DNN}{Deep Neural Network}
\nomenclature{RNN}{Recurrent Neural Network}
\nomenclature{LSTM}{Long Short-Term Memory}
\nomenclature{KS}{Kohn-Sham}
\nomenclature{Rct}{charge transfer resistance}
\nomenclature{NIR-vis-UV}{near infrared-visible-ultraviolet}
\nomenclature{VASP}{Vienna Ab initio Simulation Package}
\nomenclature{VBM}{Valence band maximum}
\nomenclature{CBM}{Conduction band minimum}
\nomenclature{MSE}{Mean squared error}
\nomenclature{MAE}{Mean absolute error}
\nomenclature{NN}{Neural network}
\nomenclature{3D}{Three dimension}
\nomenclature{TD-DFT}{time-dependent Density functional theory}
\nomenclature{RF}{Random Forests}
\nomenclature{Support Vector Machines}{SVM}
\nomenclature{VAEs}{Variational Autoencoders}
\nomenclature{GANs}{Generative Adversarial Networks}
\nomenclature{QML}{Quantum machine learning}
\printnomenclature







 \bibliographystyle{elsarticle-num} 
 \bibliography{cas-refs}





\end{document}